\RequirePackage{pdf14}

\documentclass[a4paper,11pt]{article}
\pdfoutput=1

\usepackage{array,setspace,mathrsfs,amsfonts,amsmath,yfonts,dsfont,bbm,colonequals,amscd}
\usepackage{relsize,suffix,mathtools,cancel,bbm,tikz-cd}
\usepackage{subcaption}

\usetikzlibrary{positioning}
\usetikzlibrary{chains}
\usetikzlibrary{arrows,fit,decorations.pathreplacing}
\tikzstyle{every picture}+=[remember picture]
\tikzstyle{na} = [baseline=-.5ex]

\usepackage{jheppub} 

\usepackage[T1]{fontenc} 
\usepackage{subcaption}
\graphicspath{ {graphs/} }

\preprint{PUPT-2580, YITP-44}

\newcommand{\be}{\begin{equation}}
\newcommand{\ee}{\end{equation}}

\newcommand{\ba}{\begin{equation}\begin{aligned}}
\newcommand{\ea}{\end{aligned}\end{equation}}

\newcommand{\Disc}{\textup{Disc}}

\newcommand{\Df}{\Delta_{\phi}}
\newcommand{\Gt}{\widetilde{\mathcal{G}}}

\newcommand{\oHat}{\widehat{\mathcal{O}}}
\renewcommand{\o}{\mathcal{O}}
\newcommand{\dHat}{\widehat{\Delta}}
\newcommand{\zTr}{\mbox{$\frac{z}{z-1}$}}

\DeclareMathOperator*{\dDisc}{dDisc}

\abstract{We develop an analytic approach to Boundary Conformal Field Theory (BCFT), focussing on the two-point function of a general pair of scalar primary operators. The resulting crossing equation can be thought of as a vector equation in an infinite-dimensional space ${\cal V}$ of analytic functions of a single complex variable. We argue that in a unitary theory,  functions in ${\cal V}$ satisfy a 
boundedness condition in the Regge limit.  We identify a useful basis for ${\cal V}$, consisting of bulk {\it and} boundary conformal blocks with scaling dimensions which appear in OPEs of the mean field theory correlator. 
Our main achievement is an explicit expression for the action of the {\it dual} basis (the basis of liner functionals on ${\cal V}$) on an arbitrary conformal block. The practical merit of our basis is that it trivializes the study of perturbations around mean field theory. Our results are equivalent to a BCFT version of the Polyakov bootstrap. Our derivation of the expressions for the functionals
relies on the identification of the Polyakov blocks with (suitably improved) boundary and bulk Witten exchange diagrams in $AdS_{d+1}$. We also provide another conceptual perspective on the Polyakov block expansion and the associated functionals, by deriving a new Lorentzian OPE inversion formula for BCFT.

}

\title{\boldmath \LARGE
An Analytic Approach to BCFT$_d$
 }

\author[a,b]{Dalimil Maz\'a\v{c},}
\author[b]{Leonardo Rastelli,}
\author[b,c]{Xinan Zhou}

\affiliation[a]{Simons Center for Geometry and Physics, Stony Brook University, \\Stony Brook, NY 11794, U.S.A.}
\affiliation[b]{C. N. Yang Institute for Theoretical Physics, Stony Brook University, \\Stony Brook, NY 11794, U.S.A.}
\affiliation[c]{Princeton Center for Theoretical Science, Princeton University, \\Princeton, NJ 08544, U.S.A.}

\begin{document}
\maketitle
\flushbottom


\section{Introduction}\label{Introduction}

The study of boundary conditions in conformal field theory has a long and storied history. Boundary CFTs (BCFTs) have manifold physical applications, including surface phenomena in statistical mechanics ({\it e.g}, \cite{Cardy:1984bb})
 the worldsheet description of D-branes in string theory, formal quantum field theory (see {\it e.g.}, \cite{Gaiotto:2008sa} for supersymmetric examples), and holography \cite{Karch:2000gx,Aharony:2003qf}.\footnote{We refer to \cite{Andrei:2018die} for a recent review and guide to the literature.}  They are also a very attractive playground to develop
analytic bootstrap methods \cite{Billo:2016cpy,Gadde:2016fbj,Liendo:2016ymz,Hogervorst:2017kbj,Rastelli:2017ecj,Karch:2017wgy,Lauria:2017wav,Lemos:2017vnx,Liendo:2018ukf,Lauria:2018klo,Bissi:2018mcq}\footnote{In two dimensional rational CFTs, the beautiful work of Cardy  \cite{Cardy:1989ir,Cardy:1991tv}  still provides the prototype of how the study of boundary conditions can lead to analytic insights into the full theory.}.  This comes about because the simplest nontrivial BCFT correlator (the two-point bulk correlator) is a function of a {\it single} conformal cross ratio.\footnote{The positivity property that is essential for the numerical conformal bootstrap \cite{Rattazzi:2008pe} does not hold in this case, making the development of exact methods all the more interesting. See however \cite{Gliozzi:2016cmg} for some approximate results obtained by Gliozzi's truncation method \cite{Gliozzi:2013ysa}, which also does not rely on positivity.}
Some preliminary analytic observations were already made 
in \cite{Liendo:2012hy}, but the time is ripe to revisit this problem. Much analytic control has been achieved recently \cite{Mazac:2016qev,Mazac:2018mdx,Mazac:2018ycv,Mazac:2018qmi} in the study of the CFT$_1$ four-point function, which also depends on a single cross ratio. We can then look forward to analyze a case of similar complexity but greater physical interest.

We will consider the two-point function of a general pair of scalar primary operators in the presence of a conformal boundary condition. The resulting crossing equation can be thought of as a vector equation in an infinite-dimensional space of holomorphic functions of a single complex variable, namely the cross-ratio. Our first goal will be to clarify which vector space we are talking about. In particular, we will argue that it only contains functions satisfying a suitable boundedness condition in the BCFT version of the Regge limit. This is because both individual conformal blocks, and infinite sums of conformal blocks which give rise physical correlators have this property.

Our main achievement in this paper is the construction of a useful basis for this vector space, in the spirit of \cite{Mazac:2018ycv}. The basis consists of bulk \emph{and} boundary conformal blocks with scaling dimensions which appear in the bulk and boundary OPEs of the mean field theory correlator. The two natural boundary conditions for mean field theory, Neumann and Dirichlet, each give rise to its own basis. Expressing the crossing equation in this basis leads to an infinite set of sum rules on the CFT data appearing in the bulk and boundary OPEs. This is equivalent to saying that the sum rules can be derived from the crossing equation by acting on it with elements of the dual basis. The dual basis consists of linear functionals each of which annihilates all but one bulk or boundary conformal block of mean field theory.

Although the sum rules are valid non-perturbatively, the primary practical merit of our basis is that it trivializes the study of perturbations around mean field theory. Indeed, the sum rule associated to a given mean-field operator allows us to solve for the perturbative OPE data of that operator since the contribution of all other mean-field operators is suppressed by one order in perturbation theory in that sum rule. In particular, we can imagine perturbing the mean-field correlator by adding an individual ``single-trace'' conformal block in either channel with a general scaling dimension and small OPE coefficient. On its own, this operation is not consistent with crossing symmetry. We can fix it by giving small anomalous dimensions and anomalous OPE coefficients to the mean-field operators. There is a unique way to do this, provided by expanding the single-trace conformal block in our basis of mean-field conformal blocks.\footnote{In the Neumann case, there is also a single-parameter contact-term ambiguity discussed later.} This result is analogous to the achievements of the analytic bootstrap for the four-point function in higher $d$, which allow one to find the corrections to the OPE data of the mean-field double-trace operators in a given channel due to the presence of individual conformal blocks in the crossed channels \cite{Alday:2007mf,Fitzpatrick:2012yx,Komargodski:2012ek,Alday:2015eya,Alday:2015ewa,Alday:2016njk,Caron-Huot:2017vep}. In the BCFT context, perturbation theory around free theory in the $\epsilon$-expansion was addressed in the nice work \cite{Bissi:2018mcq}. It would be interesting to explore the relation of their methods to ours.

There is an illuminating way of reformulating the idea of last paragraph. We see that for any bulk scaling dimension $\Delta$, there exists a unique function which lies in our vector space, whose bulk OPE only contains the bulk conformal block of dimension $\Delta$ and bulk mean-field conformal blocks and whose boundary OPE only contains boundary mean-field conformal blocks. We will call this function the bulk Polyakov block of dimension $\Delta$. The analogous definition gives also the boundary Polyakov blocks. If we want to deform the mean-field two-point function by a bulk conformal block of dimension $\Delta$, we see that self-consistency requires that we should supplement it by corrections to the mean-field operators so that we are in fact adding the full Polyakov block to the correlator. It is not difficult to argue that Polyakov blocks are computed by exchange Witten diagrams in $AdS_{d+1}$ and Neumann or Dirichlet boundary condition imposed at $AdS_{d}\subset AdS_{d+1}$.

The above line of reasoning leads us to conclude that a general BCFT two-point function can be expanded not only using the bulk or boundary conformal blocks, but also as a sum of bulk \emph{and} boundary Polyakov blocks, with the same spectrum and coefficients as those appearing in the OPEs. The cancellation of spurious mean-field conformal blocks is equivalent to the sum rules discussed above. In other words, we recover a BCFT version of Polyakov's approach to the conformal bootstrap \cite{Polyakov:1974gs}, recently revisited using the language of Mellin space in \cite{Gopakumar:2016wkt,Gopakumar:2016cpb,Gopakumar:2018xqi}.

Reference \cite{Mazac:2018qmi} offered an alternative point of view on the mean-field basis and Polyakov bootstrap for the crossing equation of the CFT$_1$ four-point function using a Lorentzian OPE inversion formula. Inspired by that work, we will sketch the derivation of a Lorentzian inversion formulae for both bulk and boundary OPE of the BCFT two-point function and explain its connection to the rest of our logic.

While the present paper focuses on a rather abstract analysis of the analytic bootstrap program for BCFT, there are a number of possible future applications. As stated above, our basis is particularly useful for performing perturbation theory around mean field theory. Since every CFT with a holographic dual reduces to mean field theory when the bulk is weakly coupled, our results can be used to bootstrap boundary conditions of holographic CFTs. In general, our equations can be employed to explore the conformal manifold of boundary conditions of a fixed CFT$_d$ \cite{Behan:2017mwi,Karch:2018uft}. While conformal manifolds are rare in theories containing a stress-tensor, they are common for nonlocal theories. Conformal boundary conditions are nonlocal theories since they always come with an associated higher-dimensional theory and certainly do not contain a $d-1$-dimensional stress-tensor. Therefore, we expect that a fixed CFT$_d$ generically admits nontrivial conformal manifolds of conformal boundary conditions \cite{Behan:2017dwr, Behan:2017emf}. A very interesting recent example of this set-up is \cite{DavideBCFT}, which considers 3D CFTs with abelian flavour symmetry on the boundary of a free gauge field in 4D half-space. Another important example arises in string theory, where families of boundary conditions correspond to the moduli space of D-branes \cite{Recknagel:1998ih}.

Since our sum rules work for the two-point functions of general pairs of scalar primary operators, it will be interesting to use them to generalize the $\epsilon$-expansion analysis of \cite{Bissi:2018mcq} to more general operators in the Wilson-Fisher theory.

Finally, while the conformal bootstrap of CFT four-point function with external operators with spin is rather complicated due to the large number of tensor-structures involved, the situation for the BCFT two-point function is much more favorable. It would be interesting to generalize our analysis in that direction too.

\bigskip

The rest of the paper is organized as follows. Section \ref{Outline} gives a more detailed overview of our logic. The most important technical work needed, namely the computation of the OPE of exchange Witten diagrams in both channels, is performed in detail Section \ref{Wdiagrams}. In Section \ref{LIFormula}, we derive Lorentzian OPE inversion formulae for the BCFT two-point function and explain their connection to the rest of this paper. In Section \ref{sec:deformation}, we analyze a family of conformal boundary conditions for mean-field theory interpolating between Neumann and Dirichlet and use it to check the consistency of our proposal. In particular, we give a closed formula determining the boundary spectrum for an arbitrary value of the deformation parameter.

\bigskip

{\bf Note:}
While our work was being completed, we became aware of \cite{Kaviraj:2018tfd}, which has substantial overlap with our results in sections \ref{Outline} and \ref{Wdiagrams}. A detailed presentation of this subset of our results was delivered by one of us at Caltech in July 2018 \cite{XinanTalk}.

\section{Outline}\label{Outline}

\subsection{Kinematics}


 Let us first recap some basic kinematics (see, {\it e.g.}, \cite{McAvity:1995zd,Liendo:2012hy} for detailed presentations). 
We consider a Euclidean CFT in $d$ dimensions, in the half-space
 $x_\perp \equiv x_d  \geqslant 0$, with boundary conditions at $x_\perp = 0$ that preserve the appopriate $SO(d, 1)$ subgroup of the original $SO(d+1, 1)$ conformal symmetry. The correlator of two bulk scalar operators
of dimensions $\Delta_1$ and $\Delta_2$ takes the form
\begin{equation} \label{2pt}
G(x,y) \colonequals \langle  {\cal O}_1 (x)   {\cal O}_2 (y)   \rangle = \frac{1}{|2x_\perp|^{\Delta_1}|2y_\perp|^{\Delta_2}}\mathcal{G}(\xi)\; ,
\end{equation}
where $\xi$ is the unique cross ratio,
\begin{equation}
\xi=\frac{(x-y)^2}{4x_\perp y_\perp} =  \frac{(\vec x-\vec y)^2 + (x_\perp - y_\perp)^2}{4x_\perp y_\perp} \;.
\end{equation}
$\xi$ takes positive real values when the two operators live in the Euclidean signature or are spacelike separated in the Lorentzian signature. As familiar, we can expand the two-point function in two inequivalent OPE limits, the bulk limit $\xi \to 0$ and the boundary limit  $\xi \to + \infty$,
\begin{equation} \label{2ptOPEs}
\mathcal{G} (\xi)=\sum_{\mathcal{O}}\lambda_{\mathcal{O}}\, g^B_{\Delta_\mathcal{O}}(\xi)=\sum_{\widehat{\mathcal{O}}} \mu_{\widehat {\cal O}}\, g^b_{ \Delta_{ \widehat {\cal O}}} (\xi) \;.
\end{equation}
Here  $g^B$ and $g^b$ denote  the bulk and boundary conformal blocks (whose well-known expressions will be reviewed in Section 2), and the two sums run  over the set of bulk primary operators $\{ \mathcal{O} \}$
or boundary primary operators $\{ \widehat{\mathcal{O}} \}$ that appear in the respective OPEs.\footnote{As a  general rule, we will use hatted symbols  and a lower case ``b'' for boundary objects, and unhatted symbols and a capital ``B'' for bulk  objects.} 

%

\begin{figure}[h!]
  \centering
  \includegraphics[width=.9\textwidth]{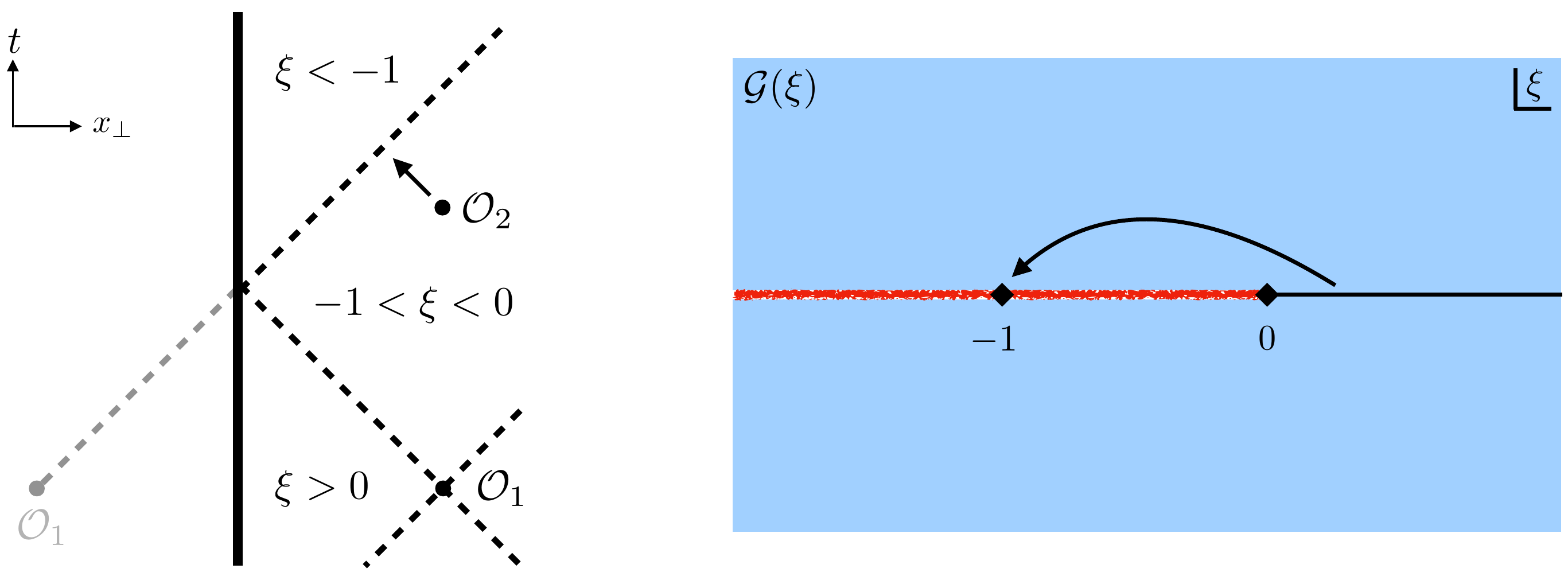}
  \caption{Left: The Regge limit of a two-point function in the presence of a conformal boundary condition. The thick vertical line represents the (timelike) boundary. In the Regge limit, $\mathcal{O}_2$ approaches the lightcone of the mirror reflection of $\mathcal{O}_1$. Right: To reach the Regge limit, we should start with $\mathcal{G}(\xi)$ in the Euclidean regime $\xi>0$ and analytically continue to $\xi\rightarrow -1$. In doing so, we need to go around the branch point $\xi = 0$.}
  \label{fig:Regge}
\end{figure}
\subsection{Regge limit and the spaces ${\cal V}$ and ${\cal U}$}

In any unitary CFT,  ${\cal G}(\xi)$ is a complex analytic function of $\xi$, with branch point singularities at $\xi =0$ and $\xi = \infty$, dictated respectively by the bulk and boundary OPEs. When we continue to the Lorentzian signature such that the time direction runs along the boundary, we find that the configurations where $\mathcal{O}_1$ and $\mathcal{O}_2$ are timelike separated are described by the analytic continuation of ${\cal G}(\xi)$ to $\xi<0$ around the branch-point at $\xi=0$. This regime includes the interesting limit $\xi\rightarrow -1$, which we dub the {\it Regge limit}. Physically, the Regge limit corresponds to configurations where $\mathcal{O}_2$ approaches the light-cone of the mirror reflection of $\mathcal{O}_1$, with the boundary acting as the mirror.

As we show in Appendix \ref{Reggebehavior}, ${\cal G}(\xi)$ obeys a suitable boundedness condition in the BCFT Regge limit, analogously to what happens in the more standard Regge limit of a CFT$_d$ four-point function. This condition takes the form
\be\label{Reggeboundedness}
|G(\xi)| \lesssim (\xi+1)^{-\frac{\Delta_1+\Delta_2}{2}}\quad \textrm{for}\; \xi \to -1^+ \qquad \textrm{(Regge boundedness)}\,.
\ee
For a given choice of bulk external dimensions $\Delta_1$ and $\Delta_2$, we define the space ${\cal V}$ as the space of complex analytic functions $ \{ {\cal G}(\xi) \}$ that have 
(at worst) the same branch point singularities at $\xi \to 0$ and $\xi \to \infty$ of a physical two-point function, and that are ``bounded'' in the Regge limit, {\it i.e.}, they diverge at most as in (\ref{Reggeboundedness}).

We will also find it useful to define a smaller space ${\cal U} \subset {\cal V}$, the space of ``Regge super-bounded'' functions. The function ${\cal F} \in {\cal U}$ if and only if ${\cal F} \in {\cal V}$ and further obeys
\begin{equation}
| {\cal F} (\xi) |  \lesssim (\xi+1)^{-\frac{\Delta_1+\Delta_2-1}{2}+\epsilon}\quad \textrm{for}\; \xi \to -1^+ \qquad \textrm{(Regge super-boundedness)}\,,
\end{equation}
with some $\epsilon>0$. As we show in Appendix \ref{Reggebehavior}, 
both bulk and boundary conformal blocks are Regge bounded, as long as the unitary bound $\Delta_i \geqslant d/2 -1$ is satisfied.

\subsection{Mean field theory}

 To state our main point we first need to recall another bit of kinematics,  the definition of mean field theory.  In mean field theory, the  $n$-point correlator of the ``elementary'' scalar operator $\varphi$ is simply given by its disconnected part,  {\it i.e.} it factorizes into products of two-point functions.  In the presence of a boundary, the form of the mean field theory two-point function depends on a choice of boundary conditions. The two obvious choices
 are Neumann and Dirichlet boundary conditions,\footnote{Both choices preserve the $\mathbb{Z}_2$ symmetry $\varphi \to -\varphi$ so one-point functions are zero.}
 \begin{eqnarray}
 \langle \varphi(x) \varphi(y) \rangle_{\rm Neumann} & = & \frac{1}{(2x_\perp)^{\Delta_\varphi}(2y_\perp)^{\Delta_\varphi}}\left(\xi^{-\Delta_\varphi}+ (\xi+1)^{-\Delta_\varphi}\right) \,  , \\ 
 \langle \varphi(x) \varphi(y) \rangle_{\rm Dirichlet} & = &\frac{1}{(2x_\perp)^{\Delta_\varphi}(2y_\perp)^{\Delta_\varphi}}\left(\xi^{-\Delta_\varphi}- (\xi+1)^{-\Delta_\varphi}\right) \, .
 \end{eqnarray}
  The boundary OPE of $\varphi$ predicts the existence of an infinite tower of boundary modes $\widehat \varphi_n$, with $n \in \mathbb{Z}_{\geq 0}$, of dimensions
  \begin{equation}
 \widehat \Delta_n = \Delta_\varphi + 2n \quad \textrm{(Neumann)} \, , \qquad \widehat \Delta_n = \Delta_\varphi + 2n +1  \quad \textrm{(Dirichlet)} \, .
  \end{equation}
For concreteness, in most of the paper 
we will pick Neumann as our reference boundary condition. In addition to the identity and to $\varphi$, the bulk operator spectrum of the mean field theory consists of the usual  multi-trace composite operators  -- normal ordered products of $\varphi$ sprinkled with derivatives. Similarly, the boundary spectrum consists of the boundary identity, of the single-trace modes $\{ \widehat \varphi_n \}$ and of {\it their} multi-trace composites.  An equivalent and very useful definition of mean field theory is provided by the holographic correspondence. The theory of a single free scalar field in $AdS_{d+1}$ is dual
to CFT$_d$ mean field theory. The BCFT case is obtained by  introducing an $AdS_d$ boundary in $AdS_{d+1}$, with suitable (Neumann or Dirichlet) boundary conditions \cite{DeWolfe:2001pq,Aharony:2003qf}.
The resulting ``half-space'' geometry is what we denote by $hAdS^N_{d+1}$ and $hAdS^D_{d+1}$ for the Neumann and Dirichlet cases, respectively.

\subsection{From functionals to Polyakov blocks, and back}

Our main contention is that the space of super-bounded functions ${\cal U}$ admits a  natural basis ``adapted'' to mean field theory.
The basis functions are bulk and boundary conformal blocks
with quantized dimensions, as dictated by our reference Neumann mean field theory.\footnote{An analogous statement holds if one chooses Dirichlet mean field theory as the reference  BCFT. We will come back to the choice of reference boundary conditions in Section \ref{sec:deformation}.}

 Let us first consider the case $\Delta_1 \neq \Delta_2$.
We claim that
 the following set functions is a basis for ${\cal U}$,
\begin{eqnarray} \label{basis}
g^B_{\Delta_N}  \quad &\textrm{with}&\; \,\Delta_N = \Delta_1 + \Delta_2 + 2 N\, , \quad   N \in \mathbb{Z}_{\geq 0}\, , \\
g^b_{\widehat \Delta^i_n}  \quad &\textrm{with}&\; \, \widehat \Delta^i_N = \widehat \Delta_i + 2n\,, \quad \quad \quad \quad n \in \mathbb{Z}_{\geq 0}\, , \; i = 1,2\,. \nonumber
\end{eqnarray}
 A basis for the dual space ${\cal U}^*$ is given by the set of functionals $\{ \omega_M, \widehat \omega_m^{(j)} \}$, defined by dualizing the primal basis (\ref{basis}),
\begin{eqnarray} \label{dualbasis}
 \omega_M (g^B_{\Delta_N} ) = \delta_{MN},  \quad  &\omega_M (g^b_{  \widehat \Delta^i_n }  )&= 0 \\
 \widehat \omega_m^{(j)}(g^B_{\Delta_N} )=0 \;\;\;\;\,\, , \quad  &\widehat \omega_m^{(j)}  (g^b_{\widehat \Delta^i_n} )&= \delta_{mn} \, \delta^{ij}\, .\nonumber
\end{eqnarray}
Our  main goal is to find  explicit expressions for the action of the functionals $\{ \omega_M, \widehat \omega_m^{(j)} \}$ on bulk and boundary conformal blocks of {\it arbitrary} dimension.
 We will proceed somewhat indirectly. The first step is the definition of the ``Polyakov blocks''.

 \smallskip
 
 {\noindent \it Polyakov blocks}
 
 \smallskip
 
 \noindent
 The bulk Polyakov block of dimension $\Delta$ is defined as the {\it unique} function in ${\cal U}$,
 which admits the following bulk and boundary conformal block expansions,\footnote{The Polyakov blocks depend on the choice of  boundary conditions for the reference mean field theory. To avoid cluttering, we do not indicate explicitly that we are choosing Neumann boundary conditions, but this should be understood throughout.}
\begin{equation}\label{PBexpansion}
\mathfrak{P}^B_\Delta = g^B_\Delta + \sum_N \mathfrak{a}_N \, g^B_{\Delta_N} = \sum_{m, i} \mathfrak{b}_{m}^{(i)} \, 
g^{b}_{\widehat{\Delta}^{(i)}_m}\,.
\end{equation}
for some coefficients  $\mathfrak{a}_N$ and $\mathfrak{b}_{m}^{(i)}$. Acting on this equation with the functionals $\{ \omega_M, \widehat \omega_m^{(j)} \}$ and using the orthonormality relations (\ref{dualbasis}) we immediately find
\begin{equation} \label{PBcoefficients}
\mathfrak{a}_N = - \omega_N(g^B_\Delta)\, , \quad  \mathfrak{b}_{m}^{(i)} =  \widehat{\omega}^{(i)}_m(g^B_\Delta)\,.
\end{equation}
Clearly, if we  were  somehow handed an expression for the bulk Polyakov block of general dimension $\Delta$, we could perform its conformal  block expansion and find the  action of our basis of functionals on the general bulk conformal block. Similarly, the boundary Polyakov block of dimension $\widehat \Delta$ is defined as the unique function in ${\cal U}$ with conformal block expansions
\begin{equation} \label{Pbexpansion}
\mathfrak{P}^b_{\widehat{\Delta}} =  g^b_{\widehat{\Delta}} + \sum_{n, i} \mathfrak{c}^{(i)}_n \,   g^b_{\widehat{\Delta}_n^i} = \sum_N \mathfrak{d}_N \, g^B_{\Delta_N} \;.
\end{equation}
Using (\ref{dualbasis}),  we immediately find
\begin{equation}   \label{Pbcoefficients}
\mathfrak{c}^{(i)}_n= - \widehat \omega_n^{(i)} (g^b_{\widehat\Delta})\,, \quad \mathfrak{d}_N= \omega_N(g^b_{\widehat\Delta})\, ,
\end{equation}
{\it i.e.} the boundary Polyakov block of dimension $\widehat \Delta$ encodes the action of the functionals on the boundary conformal block with the same dimension.

\smallskip

\noindent
{\it Polyakov = Witten}

\smallskip

\noindent
We are now approaching the punchline. There  {\it is}, in fact, an independent way to define the Polyakov blocks. We just need to recall that bulk and boundary exchange Witten diagrams in $hAdS_{d+1}$
have conformal block expansions precisely of the  form (\ref{PBexpansion}) and (\ref{Pbexpansion}), respectively. The boundary exchange Witten diagram is Regge super-bounded, and it must then
{\it coincide} with the boundary Polyakov block,
\begin{equation}  \label{Pb=W}
\mathfrak{P}^b_{\widehat{\Delta}}  \equiv \mathcal{W}^{boundary}_{\widehat{\Delta}}\,.
\end{equation}
On the other hand, the bulk exchange Witten diagram is Regge bounded, but {\it not} super-bounded, {\it i.e.} it belongs to ${\cal V}$ but not to ${\cal U}$. Fortunately, there is a simple fix.
The basic Witten contact diagram in $hAdS_{d+1}$  (with Neumann b.c. and a non-derivative vertex)
belongs to  ${\cal V}$, but not to ${\cal U}$, and admits  bulk and boundary conformal block expansions featuring precisely the set (\ref{basis}) of conformal blocks,
\begin{equation} \label{relation}
{\cal W}^{contact}
 = \sum_N    a_N  \; g^B_{\Delta_N}= \sum_{n, i} \hat  a_{n}^{(i)} \; g^b_{\widehat \Delta^i_n} \,.
\end{equation}
It follows that we can ``improve'' the bulk exchange Witten diagram by adding to it a term proportional to the contact diagram, such that the sum belongs to ${\cal U}$. All in all,
\begin{equation}
\mathfrak{P}^B_{\Delta} \equiv \mathcal{W}^{bulk}_{\Delta} + \theta_\Delta {\cal W}^{contact}\, ,
\end{equation}
where $\theta_\Delta$ is a computable coefficient. This will then be our strategy to find the action of our basis of functionals on general conformal blocks. We will ``just'' need to perform
the conformal block expansion of the (improved) exchange Witten diagrams. To that end, we will have to overcome some technical hurdles, especially
for the crossed channel expansions of the Witten diagrams ({\it i.e.}, the boundary expansion of the bulk exchange diagram, and the bulk expansion of the boundary exchange diagram).
The requisite technical work is performed in Section \ref{Wdiagrams}.

It should now also be becoming clear why we had to introduce the notion of Regge super-boundedness as opposed to mere boundedness. The reason is that the equality of the bulk and boundary OPEs of the contact diagram \eqref{relation} gives a linear relation among our proposed basis vectors in the space $\mathcal{V}$. Since the contact diagram is Regge bounded but not super-bounded, this relation disappears when restricting to $\mathcal{U}$. Had we chosen to work with the Dirichlet boundary condition, we would not be force to introduce $\mathcal{U}$.

\smallskip

\noindent
{\it The Polyakov block expansion}

\smallskip

\noindent
It follows from the definitions (\ref{PBexpansion}, \ref{PBcoefficients}) and (\ref{Pbexpansion}, \ref{Pbcoefficients}) that assuming the BCFT two-point function (\ref{2ptOPEs}) is Regge super-bounded, it admits the following curious representation,
\begin{equation} \label{Polyakovexpansion}
{\cal G} (\xi) = \sum_{\mathcal{O}}\lambda_{\mathcal{O}}\,  \mathfrak{P}^B_{\Delta} (\xi) + \sum_{\widehat{\mathcal{O}}}\mu_{\widehat {\cal O} } \,  \mathfrak{P}^b_{\widehat{\Delta}} (\xi) \;.
\end{equation}
The two sums run over the same bulk and boundary spectra of the usual conformal block expansions  (\ref{2ptOPEs}), and with the {\it same} OPE coefficients. 
To show the validity of this representation,
we can replace the Polyakov blocks by (say) their bulk channel conformal block expansions,
\begin{eqnarray}
{\cal G}   & = & \sum_{\mathcal{O}}\lambda_{\mathcal{O}}\,  g^B_{\Delta}   - \sum_{\mathcal{O}}\lambda_{\mathcal{O}}\,  \sum_N   \,   \omega_N(g^B_\Delta) \, g^B_{\Delta_N}  + \sum_{\widehat{\mathcal{O}}}\mu_{\widehat {\cal O} } \sum_N \omega_N(g^b_{\Delta_{\widehat O}}) \, g^B_{\Delta_N} \\
& = &  \sum_{\mathcal{O}}\lambda_{\mathcal{O}}\,  g^B_{\Delta}   + \sum_N  g^B_{\Delta_N} \,
   \omega_N 
   \left( - \sum_{\mathcal{O}}\lambda_{\mathcal{O}}\, g^B_\Delta    + \sum_{\widehat{\mathcal{O}}}  \mu_{\widehat {\cal O} } \, g^b_{\Delta_{\widehat O}}
\right)\,.
 \nonumber
\end{eqnarray}
The expression in the bracket is identically zero thanks the crossing equation (\ref{2ptOPEs}), and we have thus recovered the bulk conformal block expansion of ${\cal G}$. The only potentially
subtle point in this derivation  is  swapping  the order of the infinite sums (over $N$ and over $\{ {\cal O} \}\, ,\{ \widehat {\cal O} \}$).\footnote{We were not able to prove that this swapping is allowed in general but we are optimistic. It will be important to return to this very important point in future.} One may also prove (\ref{Polyakovexpansion})
by replacing the Polyakov blocks by their expansion in the boundary channels, and follow entirely analogous steps. 

The representation (\ref{Polyakovexpansion}) is somewhat analogous to the one conjectured long ago by Polyakov \cite{Polyakov:1974gs} (and reproposed recently in \cite{Gopakumar:2016wkt,Gopakumar:2016cpb,Gopakumar:2018xqi}) for the four-point function of identical scalar operators in CFT$_d$. In that context,  Polyakov blocks are defined to be fully crossing symmetric, so that one is effectively summing over the $s$, $t$ and $u$ channels, which are all equivalent for identical scalar operators. In our context, the bulk and boundary channels are inequivalent and one necessitates the introduction of two
kinds of ``Polyakov'' blocks. The existence of a Polyakov block representation for the four-point function in CFT$_1$ has been recently demonstrated in \cite{Mazac:2018ycv}, 
following a similar logic as the one we have used here. The Polyakov block turns out to be the crossing-symmetrized sum of Witten exchange diagrams, {\it plus} a multiple of the contact diagram, needed to achieve super-boundedness. The status of a Polyakov block representation in CFT$_d$ is still unclear (at least to us), see \cite{Dey:2017fab,Gopakumar:2018xqi,Mazac:2018ycv} for recent discussions.

\smallskip
{\noindent  \it  The case $\Delta_1 = \Delta_2$}
\smallskip

\noindent
The action of the functionals in the equal dimension case can be obtained by carefully taking the limit $\Delta_1 \to \Delta_2$. Alternatively, we 
recognize that in this limit the two sets of boundary blocks in the basis (\ref{basis}) become degenerate with each other, and to preserve completeness we need to introduce
{\it derivatives} of the boundary blocks with respect to the conformal dimension. The basis then reads (with $\Delta_1 = \Delta_2 \equiv \Delta_\varphi$):
\begin{eqnarray} \label{basisequaldims}
g^B_{\Delta_N}  \quad &\textrm{with}&\; \,\Delta_N = 2 \Delta + 2 N\, , \quad  \quad \quad N \in \mathbb{Z}_{\geq 0}\, , \\
g^b_{\widehat \Delta_n}  \quad &\textrm{with}&\; \, \widehat \Delta_N = \widehat \Delta+ 2n\,, \quad \quad \quad \quad n \in \mathbb{Z}_{\geq 0}\, ,  \nonumber \\
\partial g^b_{\widehat \Delta_n}  \quad &\textrm{with}&\; \, \widehat \Delta_N = \widehat \Delta+ 2n\,, \quad \quad \quad \quad n \in \mathbb{Z}_{\geq 0}\, , \nonumber
\end{eqnarray}
where $\partial g^b_{\widehat \Delta_n}$ stands for the derivative of $g^b_{\widehat \Delta}$ with respect to $\widehat{\Delta}$, evaluated at $\widehat{\Delta}=\widehat{\Delta}_n$. The dual basis $\{ \omega_N \, , \widehat{\omega}_n \,, \widetilde{\omega}_n \}$ is defined by the orthonormality conditions
\begin{equation} \label{orthoequal}
\left[\begin{array}{c}\omega_N \\\widehat{\omega}_n \\\widetilde{\omega}_n\end{array}\right]\, \left[\begin{array}{ccc}g^B_{\Delta_M} & g^b_{\widehat{\Delta}_m} & \partial g^b_{\widehat{\Delta}_m}\end{array}\right]=\left[\begin{array}{ccc}\delta_{NM} &   &   \\  & \delta_{nm} &   \\  &   & \delta_{nm}\end{array}\right]\;.
\end{equation}
The analysis performed above for $\Delta_1 \neq \Delta_2$ can be repeated with the obvious modifications in the equal dimensions case. For example, the 
 boundary conformal block decompositions of the Polyakov blocks now read
\begin{eqnarray}
\mathfrak{P}^B_{\Delta} & = & \sum_m\widehat{\omega}_m(g^B_\Delta)g^{b}_{\widehat{\Delta}_m}+\sum_{m} \widetilde{\omega}_m(g^B_\Delta)\partial g^{b}_{\widehat{\Delta}_m}\; , \\
\mathfrak{P}^b_{\widehat\Delta} & = &  g^b_{\widehat\Delta}-\sum_m\widehat{\omega}_m(g^b_{\widehat\Delta})g^{b}_{{\widehat\Delta}_m}-\sum_{m} \widetilde{\omega}_m(g^b_{\widehat\Delta})\partial g^{b}_{\widehat{\Delta}_m}\; ,
\end{eqnarray}
while (\ref{relation}) gets replaced by
\begin{equation} \label{relationequaldims}
{\cal W}^{contact}
 = \sum_N    \lambda_N  \; g^B_{\Delta_N}= \sum_{n} \widehat  \lambda_{n} \; g^b_{\widehat \Delta_n} +  \sum_{n} \widetilde \lambda_{n} \; \partial g^b_{\widehat \Delta_n}  \,.
\end{equation}

\subsection{From ${\cal U}^*$  to ${\cal V}^*$} 
A generic BCFT two-point function is Regge bounded, but not necessarily super-bounded. 
Our real interest is  in the  space ${\cal V}$ of Regge bounded functions, and in its dual space ${\cal V}^*$. Fortunately, extending the previous analysis to these physically relevant spaces
takes only some minor additional work.   The only complication is that the functions (\ref{basis}) (for $\Delta_1 \neq \Delta_2$) or (\ref{basisequaldims}) (for $\Delta_1 = \Delta_2$) are not quite linearly independent in ${\cal V}$. 
As we have noted, the basic Witten contact diagram ${\cal W}^{contact}$ belongs to ${\cal V}$ but not to ${\cal U}$. Its bulk and boundary conformal block decompositions ((\ref{relation} or  (\ref{relationequaldims}))  imply that the putative basis vectors obey one linear relation.

The inclusion ${\cal U} \subset {\cal V}$ implies  ${\cal V}^* \subset {\cal U}^*$: a  functional acting on the space of super-bounded fun is not necessarily a good functional on the space
of   bounded functions. Indeed (focusing for definiteness on the equal dimension case),  it is easy to see that the basis elements of ${\cal U}^*$  defined by the orthonormality relations (\ref{orthoequal})
are not good functionals on ${\cal V}$. For example, acting with $\omega_N$ on both conformal block expansions (\ref{relationequaldims}) of ${\cal W}^{contact} \in {\cal V}$ we find 
$\lambda_N = 0$  for all $N$, which is an obvious contradiction. The issue is of course that we have not yet taken into account the linear relation. Heuristically, we need to remove one functional from ${\cal U}^*$ in order to obtain ${\cal V}^*$. This is easily accomplished by taking linear combinations of the basis of ${\cal U}^*$, such that the resulting functionals annihilate the difference of the LHS and RHS in ${\cal W}^{contact}$. For example,
we could decide to ``remove'' $\omega_0$ and take as a complete set of functionals in ${\cal V}^*$ the linear combinations
\begin{equation}
\omega_N + \gamma_N \, \omega_0 \, , \qquad \widehat \omega_m + \widehat \gamma_m \, \omega_0 \, , \qquad \widetilde \omega_n + \widetilde \gamma_n \, \omega_0 \, ,\qquad N \geq 1 \, , m, n \geq 0\, ,
\end{equation}
where $\gamma_N$, $\widehat \gamma_m$ and $\widetilde \gamma_n$ are determined by requiring that these linear combinations respect the existence of the contact linear relation.

The bottom-line is that the above linear combinations of functionals lead to sum rules which are valid constraints on the OPE data in the bulk and boundary expansions of a general scalar two-point function in a unitary BCFT.


\subsection{Connection to the Lorentzian inversion formulae}
There is an alternative point of view on the above logic provided by the Lorentzian OPE inversion formula for the BCFT two-point function. We will write down Lorentzian inversion formulae for both the bulk and boundary OPEs. The formulae express the OPE coefficient functions in each channel as weighted integrals of the correlator in Lorentzian configurations. More precisely, both the bulk and boundary formulae involve the double discontinuity around the boundary OPE singularity and a single discontinuity around the bulk OPE singularity. These discontinuities annihilate precisely the set of boundary and bulk mean-field conformal blocks which form the basis for our vector space discussed earlier.

The Lorentzian formulae lead to the Polyakov sum rules in the following way. If we insert a single bulk conformal block of dimension $\Delta$ into the Lorentzian formulae, we get back the coefficient functions of the bulk Polyakov block of dimension $\Delta$, and similarly for the boundary conformal and Polyakov block. When applying the formulae to a general two-point function, the boundary dDisc must be expanded using the boundary OPE and the bulk Disc using the bulk OPE. Combined with the previous sentence, we see that by inserting these OPEs into the Lorentzian formula, we obtain the coefficient function of our correlator as a sum over the coefficient functions of bulk and boundary Polyakov blocks, with the same spectrum and OPE coefficients that appear in the bulk and boundary OPEs.

The construction also gives us a better understanding of the linear functionals which form the dual basis for the primal basis of mean-field bulk and boundary conformal blocks. The functionals arise simply by taking residues of the Lorentzian inversion kernels at the mean-field scaling dimensions.

Our method does not yield an explicit expression for the inversion kernels of the Lorentzian formulae. Instead, consistency with the Euclidean formula dictates that the double discontinuity of Lorentzian inversion kernel is the conformal partial wave. We give an explicit formula for the kernels in the special case $\Delta_1=\Delta_2+1$.


\section{Witten Diagrams with Neumann Boundary Condition}\label{Wdiagrams}
\subsection{Representation in terms of probe-brane diagrams}
In this paper we will consider tree-level Witten diagrams in an $hAdS_{d+1}$ space which is a half of an $AdS_{d+1}$ space. Using the Poincar\'e coordinates, it is defined by 
\begin{equation}
ds^2=\frac{dz_0^2+d\vec{z}^2+dz_\perp^2}{z_0^2}\;,\quad z_\perp\geq0\;.
\end{equation}
In addition to the conformal boundary at $z_0=0$, the $hAdS_{d+1}$ space also has an $AdS_d$ boundary defined by $z_\perp=0$. Let us specify the boundary conditions. For fields $\phi$
 living in the bulk of $hAdS_{d+1}$ we impose the Neumann boundary condition on the $AdS_d$ boundary (and we denote the $hAdS_{d+1}$ space with this choice of boundary condition as $hAdS_{d+1}^N$)
 \begin{equation}\label{Nbc}
\partial_{z_\perp} \phi(z_0,\vec{z},z_\perp)\big|_{z\perp=0}=0\;.
\end{equation}
However for the boundary fields $\widehat{\phi}(z_0,\vec{z})$ that live only on the $AdS_d$ boundary there is no such constraint. The Witten diagrams are then constructed by using the $hAdS_{d+1}^N$ Green's functions which obey the boundary condition (\ref{Nbc}), and the $AdS_d$ Green's functions. More precisely, the $hAdS_{d+1}^N$ bulk-to-bulk propagator  $\widetilde{G}_{BB}^{\Delta}(z,w)$ satisfies the defining equation
\begin{equation}\label{NGBBEOM}
(\square_{d+1}+M^2)\widetilde{G}_{BB}^\Delta(z,w)=\delta^{(d+1)}(z,w)\;,
\end{equation}
with $M^2=\Delta(\Delta-d)$. As a result of (\ref{Nbc}), the propagator $\widetilde{G}_{BB}^\Delta(z,w)$ has to satisfy the boundary condition 
\begin{equation}\label{NGBBbc}
\partial_{z_\perp}\widetilde{G}_{BB}^\Delta(z,w)\big|_{z_\perp\to0}=0\;,\quad \partial_{w_\perp}\widetilde{G}_{BB}^\Delta(z,w)\big|_{w_\perp\to0}=0\;.
\end{equation}
We also have the $AdS_d$  bulk-to-bulk propagator $\widetilde{G}_{BB}^{\widehat{\Delta}}(z,w)$ where both $z$ and $w$ are restricted to be on $AdS_d$. The propagator $\widetilde{G}_{BB}^{\widehat{\Delta}}(z,w)$ satisfies the $AdS_d$ equation of motion
\begin{equation}
(\square_{d}+\widehat{M}^2)\widetilde{G}_{BB}^{\widehat{\Delta}}(z,w)=\delta^{(d)}(z,w)\;
\end{equation}
where 
\begin{equation}
\widehat{M}^2=\widehat{\Delta}(\widehat{\Delta}-d+1)\;.
\end{equation}
We will also need the bulk-to-boundary propagators $\widetilde{G}_{B\partial}^\Delta(z,\vec{x})$. These bulk-to-boundary propagators can be obtained from the bulk-to-bulk propagators by sending one point to the conformal boundary.

\begin{figure}
\begin{subfigure}{.33\textwidth}
  \centering
  \includegraphics[width=.8\linewidth]{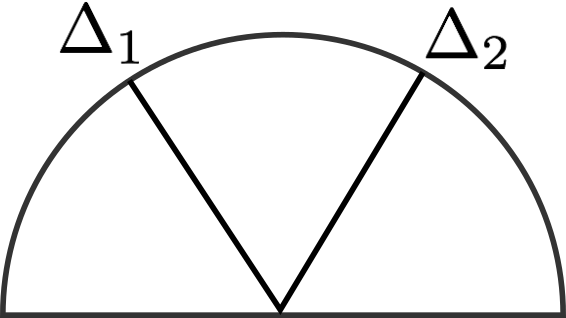}
  \caption{contact diagrams}
  \label{Ncontact}
\end{subfigure}%
\begin{subfigure}{.33\textwidth}
  \centering
  \includegraphics[width=.8\linewidth]{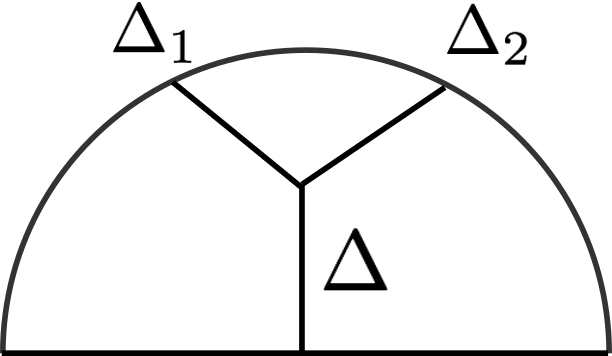}
  \caption{bulk channel exchange diagrams}
  \label{Nbulk}
\end{subfigure}
\begin{subfigure}{.33\textwidth}
  \centering
  \includegraphics[width=.8\linewidth]{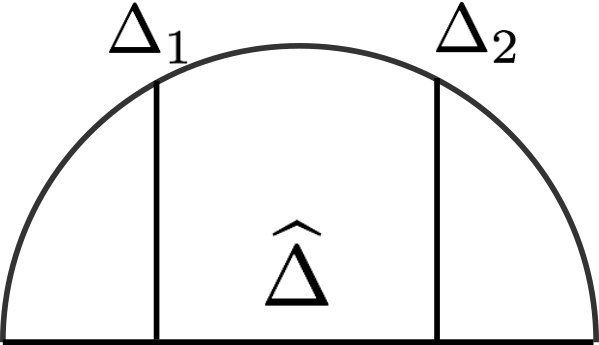}
  \caption{boundary channel exchange diagrams}
  \label{Nboundary}
\end{subfigure}
\caption{Tree level Witten diagrams in $hAdS_{d+1}^N$. The semi-disk represents the $hAdS_{d+1}^N$ space which terminates at an $AdS_d$, represented by the horizontal line.}
\label{Neumanndiagrams}
\end{figure}

In terms of the $hAdS_{d+1}^N$ propagators, we define three types of Witten diagrams with the Neumann boundary condition. The simplest diagram is a contact Witten diagram (Figure \ref{Ncontact}) which comes from a quadratic coupling on $AdS_d$. It is defined to be\footnote{Because we will seldom use contact Witten diagrams with derivatives in the contact vertex, whenever we write $W^{contact}$ we mean by default the zero-derivative contact diagram.}
\begin{equation}
W^{contact}_{\rm Neum}(x,y)=\int_{AdS_d}\frac{d^dw}{w_0^d}\widetilde{G}_{B\partial}^{\Delta_1}(w,x)\widetilde{G}_{B\partial}^{\Delta_2}(w,y)\;.
\end{equation} 
We also define the bulk channel exchange Witten diagram (Figure \ref{Nbulk}) where the external operators exchange a dimension $\Delta$ $hAdS_{d+1}$ field with the $AdS_d$ boundary via a cubic coupling
\begin{equation}
W^{bulk}_{\rm Neum}(x,y)=\int_{AdS_d}\frac{d^dw}{w_0^d}\int_{hAdS_{d+1}^+}\frac{d^{d+1}z}{z_0^{d+1}} \widetilde{G}_{BB}^\Delta(w,z)\widetilde{G}_{B\partial}^{\Delta_1}(z,x)\widetilde{G}_{B\partial}^{\Delta_2}(z,y)\;.
\end{equation}
Finally, we define the boundary channel exchange Witten diagram (Figure \ref{Nboundary}) where an $AdS_d$ field with dimension $\widehat{\Delta}$ is exchanged
\begin{equation}
W^{boundary}_{\rm Neum}(x,y)=\int_{AdS_d}\frac{d^dw_1}{w_{10}^d}\frac{d^dw_2}{w_{20}^d}\widetilde{G}_{BB}^{\widehat{\Delta}}(w_1,w_2)\widetilde{G}_{B\partial}^{\Delta_1}(w_1,x)\widetilde{G}_{B\partial}^{\Delta_2}(w_2,y)\;.
\end{equation}

However it is not the most convenient to work with the $hAdS_{d+1}^N$ propagators because of the nontrivial boundary condition. Instead we will shortly see that we can express the above Witten diagrams in terms the {\it probe-brane} Witten diagrams, where only the usual AdS propagators are used. 

\begin{figure}[htbp]
\begin{center}
\includegraphics[scale=0.4]{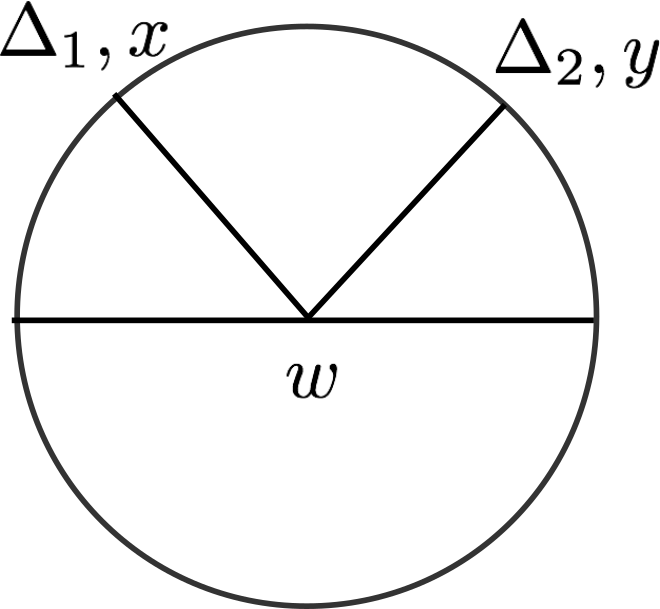}
\caption{A contact Witten diagram $W^{contact}$ in the probe brane setup. The disk represent $AdS_{d+1}$ space and the horizontal line represent the $AdS_d$ interface.}
\label{FigureWcontact}
\end{center}
\end{figure}

\begin{figure}[htbp]
\begin{center}
\includegraphics[scale=0.42]{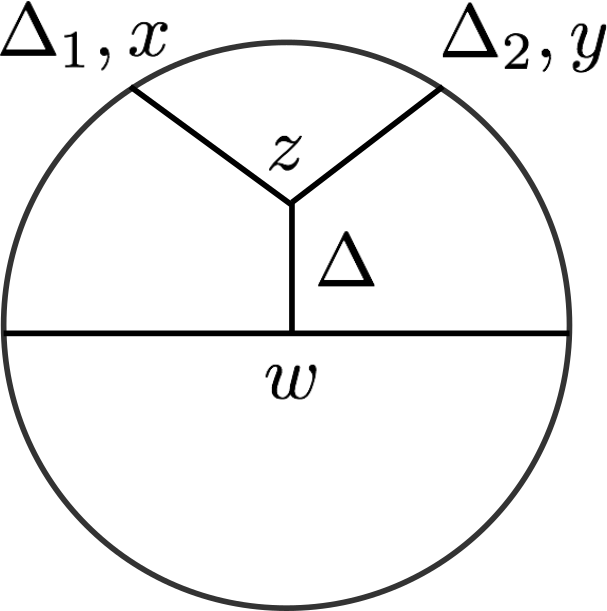}
\caption{A bulk channel exchange Witten diagram $W^{bulk}$ in the probe brane setup.}
\label{FigureWbulk}
\end{center}
\end{figure}

\begin{figure}[htbp]
\begin{center}
\includegraphics[scale=0.42]{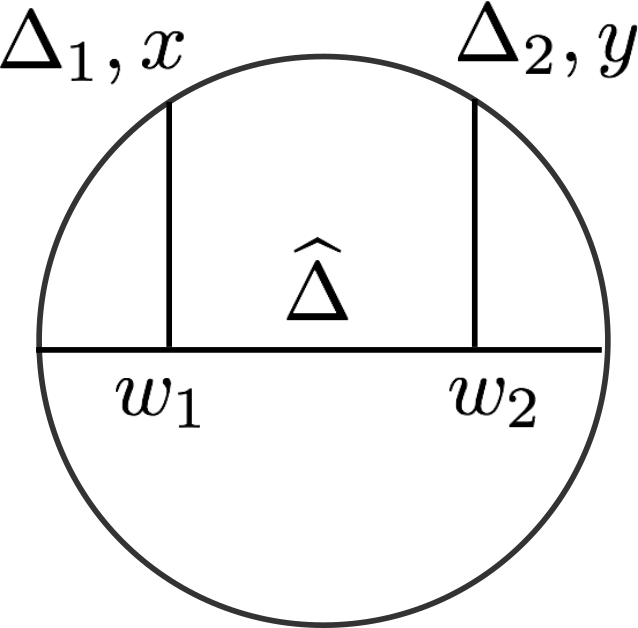}
\caption{A boundary channel exchange Witten diagram $W^{boundary}$ in the probe brane setup.}
\label{FigureWboundary}
\end{center}
\end{figure}

The probe-brane set up is given by an $AdS_{d+1}$ space with an $AdS_d$ brane inserted at $z_\perp=0$ and corresponds to an interface.\footnote{This is the simplest version of the Karch-Randall setup \cite{Karch:2000gx,Karch:2001cw}.} The $AdS_d$ brane coincide with the $z_\perp=0$ slice of $AdS_{d+1}$, and causes no back reaction to the geometry of the latter. In the probe-brane setup the $AdS_{d+1}$ bulk-to-bulk propagator satisfies the equation of motion
\begin{equation}\label{bulkGBBEOM}
(\square_{d+1}+M^2){G}_{BB}^\Delta(z,w)=\delta^{(d+1)}(z,w)\;,
\end{equation}
with no boundary condition when $z_\perp$ or $w_\perp$ goes to zero. The $AdS_d$ bulk-to-bulk propagator $G^{\widehat{\Delta}}_{BB}(z,w)$ is the same as in the $hAdS_{d+1}$ case
\begin{equation}\label{GtildebdreqGbdr}
G^{\widehat{\Delta}}_{BB}(z,w)\equiv\widetilde{G}^{\widehat{\Delta}}_{BB}(z,w)\;.
\end{equation}
The bulk-to-boundary propagators are obtained by taking the boundary limit of the corresponding bulk-to-bulk propagator. For concreteness, let us recall below the explicit expressions of the propagators
\begin{eqnarray}
G^{\Delta}_{BB}(w,z)&=&\frac{\Gamma(\Delta)}{2\pi^{\frac{d}{2}}\Gamma(\Delta-\frac{d}{2}+1)}u^{-\Delta}{}_2F_1\left(\Delta,\frac{2\Delta-d+1}{2},2\Delta-d+1,-\frac{4}{u}\right)\;,\\
G^{\widehat{\Delta}}_{BB}(w,z)&=&\frac{\Gamma(\widehat{\Delta})}{2\pi^{\frac{d-1}{2}}\Gamma(\widehat{\Delta}-\frac{d-1}{2}+1)}u^{-\widehat{\Delta}}{}_2F_1\left(\widehat{\Delta},\frac{2\widehat{\Delta}-d+2}{2},2\widehat{\Delta}-d+2,-\frac{4}{u}\right)\;,\\
G^{\Delta}_{B\partial}(z,x)&=&\left(\frac{z_0}{z_0^2+(\vec{z}-\vec{x})^2+(z_\perp-x_\perp)^2}\right)^\Delta\;
\end{eqnarray}
where we have defined
\begin{equation}\label{defofu}
u=\frac{(\vec{w}-\vec{z})^2+(w_\perp-z_\perp)^2+(w_0-z_0)^2}{w_0z_0}\;.
\end{equation}
We now define the following probe-brane Witten diagrams (Figure \ref{FigureWcontact}, \ref{FigureWbulk}, \ref{FigureWboundary}) similar to those that we have defined in $hAdS_{d+1}$
\begin{eqnarray}
W^{contact}(x,y)&=&\int_{AdS_d}\frac{d^dw}{w_0^d}{G}_{B\partial}^{\Delta_1}(w,x){G}_{B\partial}^{\Delta_2}(w,y)\;,\label{Wcontact}\\
W^{bulk}(x,y)&=&\int_{AdS_d}\frac{d^dw}{w_0^d}\int_{AdS_{d+1}}\frac{d^{d+1}z}{z_0^{d+1}} {G}_{BB}^\Delta(w,z){G}_{B\partial}^{\Delta_1}(z,x){G}_{B\partial}^{\Delta_2}(z,y)\;,\label{Wbulk}\\
W^{boundary}(x,y)&=&\int_{AdS_d}\frac{d^dw_1}{w_{10}^d}\frac{d^dw_2}{w_{20}^d}{G}_{BB}^{\widehat{\Delta}}(w_1,w_2){G}_{B\partial}^{\Delta_1}(w_1,x){G}_{B\partial}^{\Delta_2}(w_2,y)\;.\label{Wboundary}
\end{eqnarray}
Notice that the integration region of $z$ has now been extended to the entire $AdS_{d+1}$.

To express the $hAdS_{d+1}^N$ diagrams in terms of the $AdS_{d+1}$ diagrams, we notice that the $hAdS_{d+1}^N$ propagators can be constructed using the method of images. It is easy to check that the following combination
\begin{equation}\label{NGBBinprobe}
\widetilde{G}^{\Delta}_{BB}(z,w)=G_{BB}^\Delta(z,w)+G_{BB}^\Delta(z,\bar{w})
\end{equation}
satisfies the equation of motion (\ref{NGBBEOM}) and the boundary condition (\ref{NGBBbc}) for $w_\perp\to0$. Here we have defined $\bar{z}=(z_0,\vec{z},-z_\perp)$ to be the mirror point of $z$ with respect to $z_\perp=0$. Note that $G_{BB}^\Delta(z,w)$ only depends on the combination $u$ defined in (\ref{defofu}), and the quantity $u$ is invariant under reflection with respect to $z_\perp=0$.
We can therefore rewrite (\ref{NGBBinprobe}) as
\begin{equation}
\widetilde{G}^{\Delta}_{BB}(z,w)=\frac{1}{2}\left(G_{BB}^\Delta(z,w)+G_{BB}^\Delta(z,\bar{w})+G_{BB}^\Delta(\bar{z},w)+G_{BB}^\Delta(\bar{z},\bar{w})\right)
\end{equation}
which manifests the symmetry between $z$ and $w$. The boundary condition (\ref{NGBBbc}) for $z_\perp\to0$ is then also manifestly satisfied. 

\begin{figure}[htbp]
\begin{center}
\includegraphics[scale=0.4]{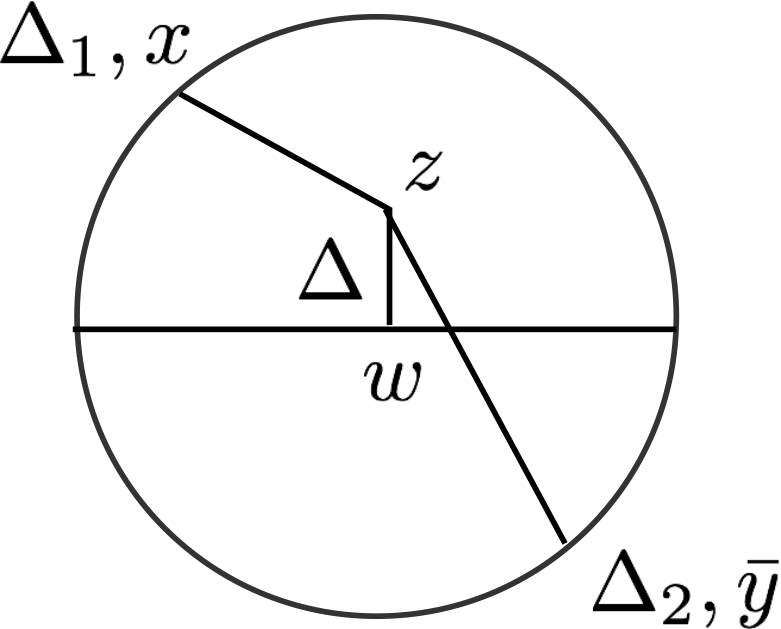}
\caption{A mirror bulk exchange Witten diagram $\bar{W}^{bulk}$ in the probe brane setup. Here $\bar{y}$ is the mirror point of $y$ with respect to the interface, and therefore appears on the other side.}
\label{FigureWbulkmirror}
\end{center}
\end{figure}

From the definitions of Witten diagrams and the relations (\ref{NGBBinprobe}) (\ref{GtildebdreqGbdr}) for the propagators, it is not hard to show that all the Witten diagrams with Neumann boundary condition can be written in terms of the probe-brane Witten diagrams. We have
\begin{equation}
W^{contact}_{\rm Neum}(x,y)=4W^{contact}(x,y)\;,
\end{equation}
\begin{equation}
W^{bulk}_{\rm Neum}(x,y)=2\left(W^{bulk}(x,y)+ W^{bulk}(x,\bar{y})\right)\;,
\end{equation}
\begin{equation}
W^{boundary}_{\rm Neum}(x,y)=8W^{boundary}(x,y)\;.
\end{equation}
We will also denote the diagram $W^{bulk}(x,\bar{y})$, where the boundary point $y$ is replaced by its mirror point $\bar{y}=(\vec{y},-y_\perp)$,  by $\bar{W}^{bulk}(x,y)$ and refer to it as the {\it mirror} bulk exchange Witten diagram (Figure \ref{FigureWbulkmirror}). In the following discussions, we will focus on these probe brane Witten diagrams which are simpler to study. 

\subsection{Relating exchange diagrams and contact diagrams}\label{relatingexchangeandcontact}
\begin{figure}[htbp]
\begin{center}
\includegraphics[scale=0.25]{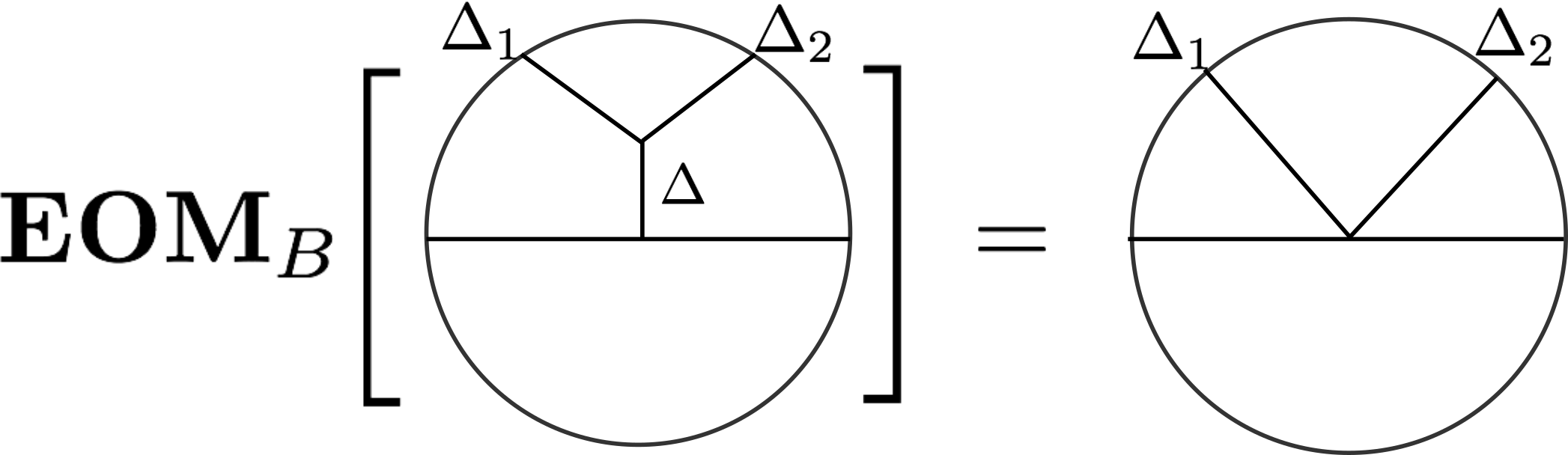}\\
\vspace{7mm}
\includegraphics[scale=0.25]{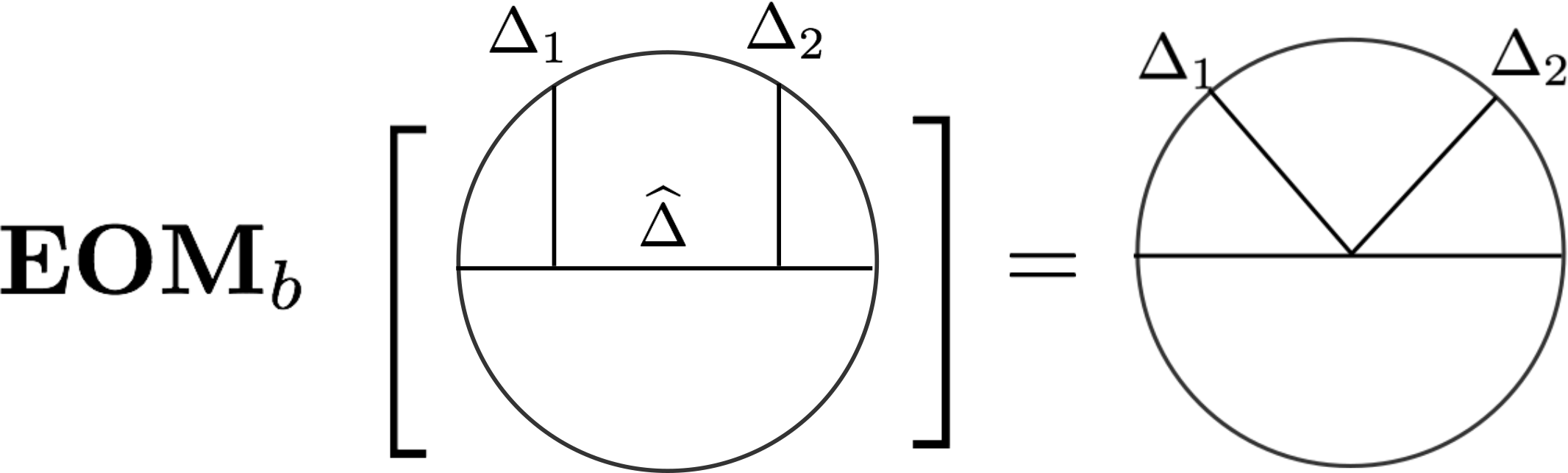}\\
\vspace{8mm}
\includegraphics[scale=0.25]{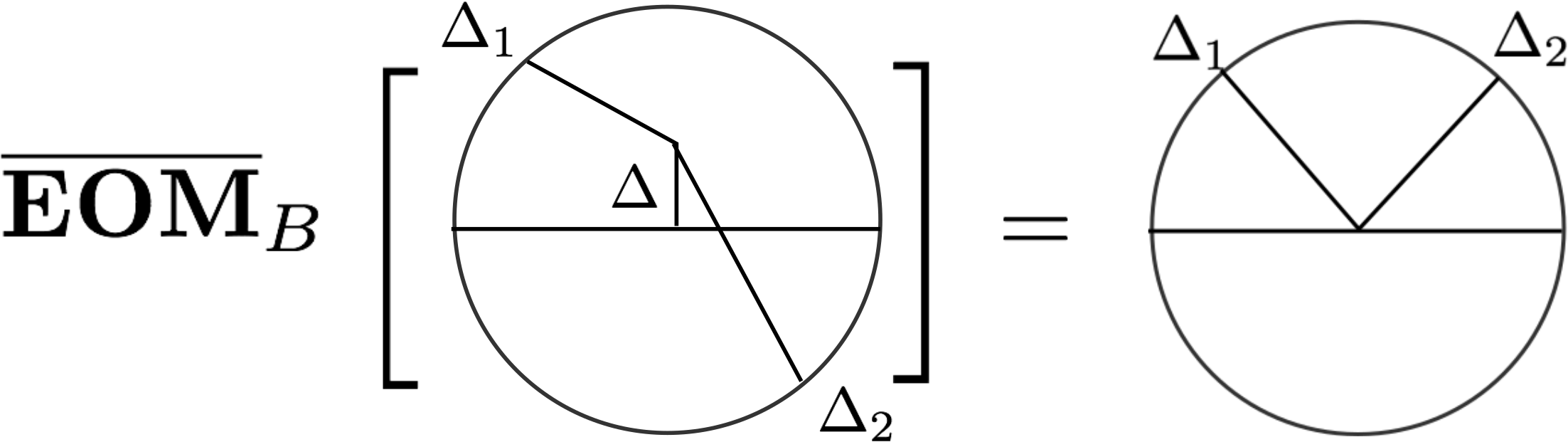}
\caption{The equation of motion identities relating the exchange diagrams to the contact diagram.}
\label{FigEOM}
\end{center}
\end{figure}

The exchange Witten diagrams can be related to the contact Witten diagram by second order differential operators (Figure \ref{FigEOM}). These differential relations exist as a result of the conformal invariance of the integrals that define the exchange diagrams, and the fact that the bulk-to-bulk propagator satisfies the equation of motion in the bulk. As we will see in later sections, these relations play an important role in the conformal block decomposition of the exchange Witten diagrams. In this subsection we give the explicit expressions for these relations. 

We start by considering the bulk exchange Witten diagram $W^{bulk}(x,y)$ in (\ref{Wbulk}), and focus on the $z$-integral
\begin{equation}
I^{bulk}(x,y;w)=\int_{AdS_{d+1}}\frac{d^{d+1}z}{z_0^{d+1}} {G}_{BB}^\Delta(w,z){G}_{B\partial}^{\Delta_1}(z,x){G}_{B\partial}^{\Delta_2}(z,y)
\end{equation}
 By the conformal invariance of the $z$-integral, we have 
 \begin{equation}
 (\mathbf{L}_1+\mathbf{L}_2+\mathfrak{L}_w)_{AB}I^{bulk}(x,y;w)=0\;,
 \end{equation}
where $\mathbf{L}_1$ and $\mathbf{L}_2$ are the generators of the $SO(d,2)$ conformal group for operator 1 and 2, and $\mathfrak{L}_w$ is the $AdS_{d+1}$ isometry generator.  From this identity it follows that
\begin{equation}
\frac{1}{2}(\mathbf{L}_1+\mathbf{L}_2)^{AB}(\mathbf{L}_1+\mathbf{L}_2)_{AB}I^{bulk}(x,y;w)=\frac{1}{2}\mathfrak{L}_w^{AB}\mathfrak{L}_{w,AB}I^{bulk}(x,y;w)=-\square_{d+1}I^{bulk}(x,y;w)\;.
\end{equation}
We now use the equation of motion (\ref{bulkGBBEOM}) for the $AdS_{d+1}$ bulk-to-bulk propagator and perform the remaining $w$-integral. We get the following relation between a bulk exchange Witten diagram and a contact Witten diagram
\begin{equation}
\left(\frac{1}{2}(\mathbf{L}_1+\mathbf{L}_2)^2+\Delta(\Delta-d)\right)W^{bulk}(x,y)=W^{contact}(x,y)\;.
\end{equation}
In terms of functions of the cross ratio, we can write the equation as
\begin{equation}\label{EOMBWbulk}
\mathbf{EOM}_B\,\mathcal{W}^{bulk}(\xi)=\mathcal{W}^{contact}(\xi)
\end{equation}
which defines a differential operator $\mathbf{EOM}_B$. The action of this differential operator is given by
\begin{equation}\label{EOMBdiff}
\begin{split}
\mathbf{EOM}_B\; \mathcal{G}(\xi)={}&-4 (\xi +1) \xi ^2 \mathcal{G}''(\xi )+\xi(2 d-4 (\xi +1) (\Delta_1+\Delta_2+1))  \mathcal{G}'(\xi ) \\{}&+((\Delta -\Delta_1-\Delta_2) (-d+\Delta +\Delta_1+\Delta_2)-4 \Delta_1 \Delta_2 \xi )\mathcal{G}(\xi )\;. 
\end{split}
\end{equation}
Similarly, the mirror bulk exchange Witten diagram satisfies
\begin{equation}
\left(\frac{1}{2}(\mathbf{L}_1+\bar{\mathbf{L}}_2)^2+\Delta(\Delta-d)\right)\bar{W}^{bulk}(x,y)=W^{contact}(x,y)\;.
\end{equation}
The operator $\bar{\mathbf{L}}_2$ is the conformal generator with respect to the mirror point $\bar{y}$. This identity defines a differential operator $\overline{\mathbf{EOM}}_B$ of the cross ratio which turns a bulk exchange Witten diagram into a contact Witten diagram
\begin{equation}\label{EOMBmirror}
\overline{\mathbf{EOM}}_B\,\bar{\mathcal{W}}^{bulk}(\xi)=\mathcal{W}^{contact}(\xi)\;.
\end{equation}
The differential operator has the following explicit expression
\begin{equation}
\begin{split}
\overline{\mathbf{EOM}}_B\; \mathcal{G}(\xi)={}&4 \xi  (\xi +1)^2\mathcal{G}''(\xi)+2 (\xi +1) (d+2 \xi  (\Delta_1+\Delta_2+1))\mathcal{G}'(\xi)\\
{}&+(d (-\Delta +\Delta_1+\Delta_2)+\Delta ^2-(\Delta_1-\Delta_2)^2+4 \Delta_1 \Delta_2 \xi ) \mathcal{G}(\xi)\;.
\end{split}
\end{equation}
Finally let us look at the boundary exchange Witten diagram (\ref{Wboundary}). The $w_1$-integral
\begin{equation}
I^{boundary}(x,w_2)=\int_{AdS_d}\frac{d^dw_1}{w_{10}^d}{G}_{BB}^{\widehat{\Delta}}(w_1,w_2){G}_{B\partial}^{\Delta_1}(w_1,x)
\end{equation}
has $SO(d-2,1)$ invariance, which is the residual conformal symmetry preserved by a conformal boundary in a CFT$_d$. Therefore we have 
\begin{equation}
(\widehat{\mathbf{L}}_1+\widehat{\mathfrak{L}}_{w_2})_{\hat{A}\hat{B}}I^{boundary}(x,w_2)=0
\end{equation}
where $\widehat{\mathbf{L}}_1$ is the $SO(d-2,1)$ conformal generator and $\widehat{\mathfrak{L}}_{w_2}$ is the $AdS_d$ isometry generator. Acting again with these generators, we obtain
\begin{equation}
\frac{1}{2}\widehat{\mathbf{L}}_1^{\hat{A}\hat{B}}\widehat{\mathbf{L}}_{1,\hat{A}\hat{B}}I^{boundary}(x,w_2)=\frac{1}{2}\widehat{\mathfrak{L}}_{w_2}^{\hat{A}\hat{B}}\widehat{\mathfrak{L}}_{w_2,\hat{A}\hat{B}}I^{boundary}(x,w_2)=-\square_d\,I^{boundary}(x,w_2)\;.
\end{equation}
After using the $AdS_d$ equation of motion and integrating over $w_2$, we arrive at the following relation
\begin{equation}
\left(\frac{1}{2}\widehat{\mathbf{L}}_1^2+\widehat{\Delta}(\widehat{\Delta}-d+1)\right)W^{boundary}(x,y)=W^{contact}(x,y)\;.
\end{equation}
Alternatively, we can write it as 
\begin{equation}\label{EOMbboundary}
\mathbf{EOM}_b\,\mathcal{W}^{boundary}(\xi)=\mathcal{W}^{contact}(\xi)\;,
\end{equation}
with $\mathbf{EOM}_b$ being a differential operator defined by
\begin{equation}
\mathbf{EOM}_b\,\mathcal{G}(\xi)=-\frac{1}{2} (2 d \xi +d) \mathcal{G}'(\xi )+\widehat{\Delta}  (\widehat{\Delta}-d +1) \mathcal{G}(\xi )-\xi  (\xi +1) \mathcal{G}''(\xi )\;.
\end{equation}

\subsection{Regge behavior of Witten diagrams}\label{ReggeWitten}
We now look into the behavior of various Witten diagrams in the Regge limit. Our starting point is the contact Witten diagrams for which we have closed form expressions in terms of hypergeometric functions. The simplest contact diagram is the one with no derivatives in the vertex, and we have \cite{Aharony:2003qf,Rastelli:2017ecj}
\begin{equation}\label{Wcontact2F1}
\mathcal{W}^{contact}=\frac{\pi ^{d/2} \Gamma \left(\frac{1}{2} (-(d-1)+\Delta_1+\Delta_2)\right) \, }{\Gamma \left(\frac{1}{2} (\Delta_1+\Delta_2+1)\right)}{}_2F_1\left(\Delta_1,\Delta_2;\frac{1}{2} (\Delta_1+\Delta_2+1);-\xi \right)\;.
\end{equation}
After making the change of the variable into $\rho$, we find the following behavior in the Regge limit
\begin{equation}
\mathcal{W}^{contact}(\rho)\sim (1+\rho)^{1-(\Delta_1+\Delta_2)}\;,\quad \rho\to-1^+\;.
\end{equation}
According to our terminology, the zero-derivative contact Witten diagram is only Regge bounded 
\begin{equation}
\mathcal{W}^{contact}\in \mathcal{V}\;.
\end{equation}
There are also contact diagrams that arise from the contact vertices with derivatives. However, these higher-derivative contact diagrams have more divergent behavior in the Regge limit compared to the zero-derivative one. For example, the Regge behavior of a contact Witten diagram with two derivatives reads
\begin{equation}
\mathcal{W}^{contact}_{2-der}(\rho)\sim (1+\rho)^{-1-(\Delta_1+\Delta_2)}\;,\quad \rho\to-1^+\;.
\end{equation}

To investigate the Regge behavior of the various exchange Witten diagrams, it is advantageous to make use of the equation of motion operators. Since the equation of motion operators turn exchange diagrams into contact diagrams, we only need to know their actions on a power-like singularity at $\rho=-1$. For example, the operator $\mathbf{EOM}_B$ collapses a bulk exchange Witten diagram into a zero-derivative contact diagram
\begin{equation}
\mathbf{EOM}_B[\mathcal{W}^{bulk}]=\mathcal{W}^{contact}\;,
\end{equation}
and turns a singularity $(\rho+1)^{a}$ into a stronger one $(\rho+1)^{a-2}$\;. Since we know that the zero-derivative contact diagram diverges as $(\rho+1)^{1-(\Delta_1+\Delta_2)}$, we find that the bulk exchange Witten diagram has the Regge behavior 
\begin{equation}
\mathcal{W}^{bulk}\sim (\rho+1)^{3-(\Delta_1+\Delta_2)}\;,\quad \rho\to -1^+\;,
\end{equation}
and therefore is Regge super-bounded
\begin{equation}
\mathcal{W}^{bulk}\in \mathcal{U}\;.
\end{equation}
The analysis for the other two exchange  Witten diagrams is analogous and yields the following Regge behavior
\begin{equation}
\bar{\mathcal{W}}^{bulk}\sim (\rho+1)^{1-(\Delta_1+\Delta_2)}\;,\quad \rho\to -1^+\;,
\end{equation}
\begin{equation}
\mathcal{W}^{boundary}\sim (\rho+1)^{3-(\Delta_1+\Delta_2)}\;,\quad \rho\to -1^+\;.
\end{equation}
We have 
\begin{equation}
\bar{\mathcal{W}}^{bulk}\in \mathcal{V}\;,\quad \mathcal{W}^{boundary}\in \mathcal{U}\;.
\end{equation}

\subsection{Conformal block decomposition of the contact Witten diagram}\label{Seccontactdecomposition}
In this subsection, we study the conformal block decomposition of the contact Witten diagram (\ref{Wcontact}). The integral is simple to evaluate and the result can be compactly expressed in terms of a single ${}_2F_1$ function, as we have already mentioned in (\ref{Wcontact2F1}). Using well-known properties of ${}_2F_1$ it is a straightforward exercise to work out the conformal block decomposition of the contact Witten diagram. In the bulk channel, we find double-trace operators with dimensions $\Delta_1+\Delta_2+2N$
\begin{equation}\label{contactinbulk}
\mathcal{W}^{contact}(\xi)=\sum_{N=0}^\infty {a}_N\, g_{\Delta_1+\Delta_2+2N}^B(\xi)
\end{equation}
where 
\begin{equation}
g^B_{\Delta}(\xi)=\xi^{-\frac{\Delta_1+\Delta_2}{2}}\xi^{\frac{\Delta}{2}}{}_2F_1(\frac{\Delta+\Delta_1-\Delta_2}{2},\frac{\Delta+\Delta_2-\Delta_1}{2};\Delta-\frac{d}{2}+1;-\xi)\;,
\end{equation}
is the bulk channel conformal block \cite{McAvity:1995zd}. The decomposition coefficients are given by\footnote{We can also arrive at this result using the BCFT version of the geodesic Witten diagrams \cite{Hijano:2015zsa,Rastelli:2017ecj,Karch:2017wgy}.}
\begin{equation}\small
a_N=\frac{\pi ^{d/2} (-1)^N \Gamma (N+\Delta_1) \Gamma (N+\Delta_2) \Gamma \left(-\frac{d}{2}+N+\Delta_1+\Delta_2\right) \Gamma \left(\frac{1}{2} (-d+2 N+\Delta_1+\Delta_2+1)\right)}{\Gamma (\Delta_1) \Gamma (\Delta_2) \Gamma (N+1) \Gamma \left(\frac{1}{2} (2 N+\Delta_1+\Delta_2+1)\right) \Gamma \left(-\frac{d}{2}+2 N+\Delta_1+\Delta_2\right)}\;.
\end{equation}
In the boundary channel, we have two towers of single-trace operators, with dimensions $\Delta_1+2m$ and $\Delta_2+2m$ respectively
\begin{equation}\label{contactinboundary}
\mathcal{W}^{contact}(\xi)=\sum_{m}\hat{{a}}^{(1)}_mg^b_{\Delta_1+2m}(\xi)+\sum_{m}\hat{{a}}^{(2)}_mg^b_{\Delta_2+2m}(\xi)\;.
\end{equation}
Here
\begin{equation}\label{bchanconfblock}
g^b_{\widehat{\Delta}}(\xi)=\xi^{-\widehat{\Delta}}{}_2F_1(\widehat{\Delta},\widehat{\Delta}-\frac{d}{2}+1;2\widehat{\Delta}+2-d;-\xi^{-1})\;,
\end{equation}
is the boundary channel conformal block \cite{McAvity:1995zd}, and the decomposition coefficients are
\begin{equation}
\begin{split}\label{contactbdrdecompcoe}
{}&\hat{{a}}^{(1)}_m=\frac{\pi ^{\frac{d-1}{2}} 2^{-\Delta_1+\Delta_2-4 m-1} (\Delta_1)_{2 m} \left(\frac{d-2 m-2 \Delta_1+1}{2} \right)_{-m} \Gamma \left(\frac{-2 m-\Delta_1+\Delta_2}{2}\right) \Gamma \left(\frac{-d+2 m+\Delta_1+\Delta_2+1}{2} \right)}{\Gamma (\Delta_2) \Gamma (m+1)}\;,\\
{}&\hat{{a}}^{(2)}_m=\text{replacing $\Delta_1$ with $\Delta_2$ in $\hat{{a}}^{(1)}_m$.}
\end{split}
\end{equation}
So far we have considered  the contact diagrams with generic external dimensions $\Delta_1$ and $\Delta_2$. The special case with equal weights $\Delta_1=\Delta_2=\Delta_\phi$ can be obtained from the above expressions by taking the limit. The bulk decomposition coefficients $a_m$ are regular when $\Delta_1=\Delta_2=\Delta_\phi$. We can therefore straightforwardly take this limit. However the boundary channel coefficients $\hat{{a}}^{(1)}_m$, $\hat{{a}}^{(2)}_m$ contain simple poles in $\Delta_1-\Delta_2\to 0$. This generates additional derivative conformal blocks $\partial_{\widehat{\Delta}}g^b_{\widehat{\Delta}}$ in the boundary channel decomposition
\begin{equation}
\mathcal{W}^{contact}(\xi)=\sum_{n}\hat{a}_{n}g^b_{\Delta_\phi+2n}(\xi)+\sum_{n}\hat{b}_{n}(\partial_{\Delta_\phi}g^b_{\Delta_\phi+2n})(\xi)\;.
\end{equation}
The coefficients $\hat{a}_{n}$, $\hat{b}_{n}$ are given by
\begin{equation}\label{ahatbhat}
\begin{split}
\hat{a}_{n}={}&\frac{\pi ^{\frac{d-1}{2}} \left(-\frac{1}{16}\right)^n \Gamma (2 n+\Delta_\phi ) \Gamma \left(\frac{1}{2} (d-4 n-2 \Delta_\phi +1)\right) \Gamma \left(-\frac{d}{2}+n+\Delta_\phi +\frac{1}{2}\right)}{(n!)^2 \Gamma (\Delta_\phi )^2 \Gamma \left(\frac{1}{2} (d-2 n-2 \Delta_\phi +1)\right)}\\
{}&\times  \left(-H_{\frac{1}{2} (d-2 n-2 \Delta_\phi -1)}+H_{\frac{1}{2} (d-4 n-2 \Delta_\phi -1)}-H_{2 n+\Delta_\phi -1}+H_n+\log (4)\right)\;,\\
\hat{b}_{n}={}&\frac{\pi ^{\frac{d-1}{2}} (-1)^{n+1} 16^{-n} \Gamma (2 n+\Delta_\phi ) \Gamma \left(\frac{1}{2} (d-4 n-2 \Delta_\phi +1)\right) \Gamma \left(-\frac{d}{2}+n+\Delta_\phi +\frac{1}{2}\right)}{(n!)^2 \Gamma (\Delta_\phi )^2 \Gamma \left(\frac{1}{2} (d-2 n-2 \Delta_\phi +1)\right)}\;,
\end{split}
\end{equation}
and $H_n$ denotes the Harmonic number.
\subsection{Direct channel decomposition of exchange Witten diagrams}
In the direct channel, the decomposition of an exchange Witten diagram contains a single-trace conformal block and infinitely many double-trace/single-trace conformal blocks whose dimensions depends on the external dimensions. This is clear from the Mellin representation \cite{Rastelli:2017ecj}. More precisely, a bulk exchange Witten diagram can be written in bulk channel as
\begin{equation}\label{bulkinbulk}
\mathcal{W}^{bulk}(\xi)=A^B g_\Delta^B(\xi)+\sum_{N}A^B_n g^B_{\Delta_1+\Delta_2+2N}(\xi)\;.
\end{equation}
Similarly, decomposing a boundary exchange Witten diagram into the boundary channel gives
\begin{equation}\label{boundaryinboundary}
\mathcal{W}^{boundary}(\xi)=\hat{A}^{b}g^b_{\widehat{\Delta}}(\xi)+ \sum_{n}\hat{A}^{b,(1)}_ng^b_{\Delta_1+2n}(\xi)+\sum_{n}\hat{A}^{b,(2)}_ng^b_{\Delta_2+2n}(\xi)\;.
\end{equation}
The direct channel decompositions can be obtained by starting from the Mellin spectral representations given in \cite{Rastelli:2017ecj}. To proceed one massages the cross ratio dependence in the inverse Mellin transformation into that of a direct channel conformal block. Then one obtains the spectral representations with respect to the direct channel conformal blocks, and the OPE coefficients can be read off from the residues. We give the detailed derivation in Appendix \ref{appspectral}, and only present the final results here. 

From the spectral representation, we find that the single-trace coefficients associated with $g^B_\Delta$ and $g^b_{\widehat{\Delta}}$ are
\begin{equation}\label{AB}\small
\begin{split}
{}&A^B=\frac{\pi ^{d/2} \cos \left(\frac{\pi  \Delta }{2}\right) \Gamma \left(\frac{1-\Delta }{2}\right) \Gamma \left(\frac{-d+\Delta +1}{2}\right)  \Gamma \left(\frac{\Delta +\Delta_1-\Delta_2}{2}\right) \Gamma \left(\frac{-\Delta +\Delta_1+\Delta_2}{2}\right) \Gamma \left(\frac{-d+\Delta +\Delta_1+\Delta_2}{2}\right)}{4 \Gamma (\Delta_1) \Gamma (\Delta_2) \sin \left(\frac{ \pi  (\Delta -\Delta_1+\Delta_2)}{2}\right)\Gamma \left(-\frac{d}{2}+\Delta +1\right) \Gamma \left(\frac{-\Delta +\Delta_1-\Delta_2+2}{2}\right)}\;,
\end{split}
\end{equation}
\begin{equation}\small
\hat{A}^{b}=\frac{\pi ^{\frac{d-1}{2}} \Gamma ({\widehat{\Delta}} ) 2^{-2 {\widehat{\Delta}} +\Delta_1+\Delta_2-3} \Gamma \left(\frac{\Delta_1-{\widehat{\Delta}} }{2}\right) \Gamma \left(\frac{\Delta_2-{\widehat{\Delta}} }{2}\right) \Gamma \left(\frac{-d+{\widehat{\Delta}} +\Delta_1+1}{2} \right) \Gamma \left(\frac{-d+{\widehat{\Delta}} +\Delta_2+1}{2}\right)}{\Gamma (\Delta_1) \Gamma (\Delta_2) \Gamma \left(-\frac{d}{2}+{\widehat{\Delta}} +\frac{3}{2}\right)}\;.
\end{equation}

 The remaining double-trace and single-trace coefficients can be extracted from the spectral function in the same way. But an alternative and faster method is to use the relations discussed in Section \ref{relatingexchangeandcontact}. Let us first look at the bulk exchange Witten diagram. Inserting the bulk channel decompositions (\ref{bulkinbulk}), (\ref{contactinbulk}) into the equation of motion relation (\ref{EOMBWbulk}), we have 
\begin{equation}
\mathbf{EOM}_B\left[A^B g_\Delta^B(\xi)+\sum_{N}A^B_N g^B_{\Delta_1+\Delta_2+2N}(\xi)\right]=\sum_{n}a_N g^B_{\Delta_1+\Delta_2+2N}(\xi)\;.
\end{equation}
Notice that the bulk channel conformal blocks are diagonal under the differential operator $\mathbf{EOM}_B$
\begin{equation}
\begin{split}
\mathbf{EOM}_B[g_\Delta^B(\xi)]={}&0\;,\\
\mathbf{EOM}_B[g^B_{\Delta_1+\Delta_2+2N}(\xi)]={}&(\Delta(\Delta+d)-(\Delta_1+\Delta_2+2N)(\Delta_1+\Delta_2+2N-d))g^B_{\Delta_1+\Delta_2+2N}(\xi)\;.
\end{split}
\end{equation}
Therefore we have the following simple relation
\begin{equation}
A^B_N=\frac{a_N}{\Delta(\Delta-d)-(\Delta_1+\Delta_2+2N)(\Delta_1+\Delta_2+2N-d)}\;.
\end{equation}
Similarly, using the boundary equation of motion (\ref{EOMbboundary}) and the fact that boundary channel conformal blocks are diagonal with respect to $\mathbf{EOM}_b$, we have
\begin{equation}
\hat{A}^{b,(i)}_n=\frac{\hat{a}^{(i)}_n}{{\widehat{\Delta}}({\widehat{\Delta}}-d+1)-(\Delta_i+2n)(\Delta_i+2n-d+1)}\;.
\end{equation}

\subsection{Crossed channel decomposition of the exchange Witten diagram}
To extract all the analytic functionals we also need to obtain the crossed channel decompositions of the exchange Witten diagrams. There are several cases we have to consider. We need to 
decompose the boundary/bulk channel exchange Witten diagram into bulk/boundary channel. We also need to decompose the mirror bulk exchange Witten diagram into both the bulk channel and the boundary channel. Such crossed channel decompositions are in general much more difficult compared to the direct channel decompositions, and no systematic methods exist in the literature. In this subsection we will present a recursive method to obtain the crossed channel decomposition coefficients.

The main idea of our method is to use the contact Witten diagram as a stepping stone.  As we have seen  in \ref{Seccontactdecomposition}, the conformal block decomposition of the contact diagram is very simple in both channels, and the decomposition coefficients were obtained in closed forms. In Section \ref{relatingexchangeandcontact}, we showed that all the exchange Witten diagrams can be related to the same contact Witten diagram by using the equation of motion identity. The action of the equation of motion operators on the crossed channel decompositions of the exchange diagrams should therefore match the decompositions of the contact diagram. Remarkably, the various equation of motion operators admit very simple actions on the crossed channel conformal blocks. The resulting expression can be generally expressed as the linear combination of three conformal blocks with shifted conformal dimensions. This gives rise to infinitely many linear relations among the crossed channel decomposition coefficients. These relations can be recursively solved, and give the crossed channel decomposition. 

Similar techniques for extracting the cross channel decomposition coefficients have also been developed for four-point functions in CFTs without boundaries, see \cite{Zhou:2018sfz}.

\subsubsection{Bulk exchange diagram in the boundary channel}
Let us make the above comments precise by looking at the boundary channel decomposition of a bulk channel exchange Witten diagram. We start with the action of $\mathbf{EOM}_B$ in (\ref{EOMBdiff}) on a boundary conformal block $g^b_{\Delta_1+n}(\xi)$. Using the properties of ${}_2F_1$, it is not hard to verify the following relation\footnote{The relations given here and in the next a few sections are in fact more general. The hypergeometric identities are valid when $n$ is not an integer. Therefore such three-term recursion relations exist for blocks with any conformal dimension.}
\begin{equation}
\mathbf{EOM}_B\; g^b_{\Delta_1+n}(\xi)=\hat{\alpha}_n^{(1)} g^b_{\Delta_1+n-1}(\xi)+\hat{\beta}_n^{(1)} g^b_{\Delta_1+n}(\xi)+\hat{\gamma}_n^{(1)} g^b_{\Delta_1+n+1}(\xi)
\end{equation}
where
\begin{equation}\small
\begin{split}
{}&\hat{\alpha}_n^{(1)}=-4 n (\Delta_1-\Delta_2+n)\;,\\
{}&\hat{\beta}_n^{(1)}=-d (\Delta +\Delta_1-\Delta_2)+n (-2 d+4 \Delta_1+2)+\Delta ^2+\Delta_1 (\Delta_1+2)-\Delta_2^2+2 n^2\;,\\
{}&\hat{\gamma}_n^{(1)}=-\frac{(\Delta_1+n) (d-2 \Delta_1-n-1) (d-\Delta_1-n-2) (-d+\Delta_1+\Delta_2+n+1)}{(d-2 \Delta_1-2 n-3) (d-2 \Delta_1-2 n-1)}\;.
\end{split}
\end{equation}
A similar relation exists for $g^b_{\Delta_2+n}(\xi)$, and can be obtained from the above expressions by swapping $\Delta_1$ with $\Delta_2$.

We apply this relation to the boundary channel decomposition of the bulk exchange Witten diagram
\begin{equation}\label{Wbulkintoboundary}
\mathcal{W}^{bulk}(\xi)=\sum_{n}\hat{A}^{B,(1)}_ng^b_{\Delta_1+n}(\xi)+\sum_{n}\hat{A}^{B,(2)}_ng^b_{\Delta_2+n}(\xi)\;.
\end{equation}
Using the equation of motion identity (\ref{EOMBWbulk})
\begin{equation}
\mathbf{EOM}_B\,\mathcal{W}^{bulk}(\xi)=\mathcal{W}^{contact}(\xi)\;,
\end{equation}
and the boundary channel decomposition of the contact diagram (\ref{contactinboundary}), we get the following recursion relations for $\hat{A}^{B,(1)}_n$ and $\hat{A}^{B,(2)}_n$
\begin{equation}\label{AhatBrecur}
\hat{\gamma}_{n-1}^{(i)}\hat{A}^{B,(i)}_{n-1}+\hat{\beta}_{n}^{(i)}\hat{A}^{B,(i)}_{n}+\hat{\alpha}_{n+1}^{(i)}\hat{A}^{B,(i)}_{n+1}=
\begin{cases}
\hat{a}^{i}_{\frac{n}{2}}\;,\quad n\text{ even}\;,\\
0\;,\quad\quad n\text{ odd}\;.
\end{cases}
\end{equation}
We should also impose the boundary condition $\hat{A}^{B,(i)}_{-1}=0$. These  relations gives us a recursive algorithm for doing the crossed channel decomposition. Our starting point is $n=0$ where we have the identity
\begin{equation}
\hat{\beta}_{0}^{(i)}\hat{A}^{B,(i)}_{0}+\hat{\alpha}_{1}^{(i)}\hat{A}^{B,(i)}_{1}=\hat{a}^{i}_0\;.
\end{equation}
From this equation we can solve $\hat{A}^{B,(i)}_{1}$ in terms of $\hat{A}^{B,(i)}_{0}$. For $n\geq 1$, we can solve $\hat{A}^{B,(i)}_{n+1}$ in terms of $\hat{A}^{B,(i)}_{n}$ and $\hat{A}^{B,(i)}_{n-1}$. The entire decomposition therefore boils down to computing the seed coefficient $\hat{A}^{B,(i)}_{0}$ which we will discuss in Appendix \ref{Secseed}.

We can also consider the decomposition of bulk exchange Witten diagrams with equal weights. Just as in the case of contact Witten diagrams, the equal weight exchange Witten diagrams contain derivative conformal blocks in the boundary channel. The equal weight case can be obtained from the unequal weight results by taking a limit. We will give the  expressions for the decomposition coefficients in Appendix \ref{SecEqualweights}.


\subsubsection{Boundary exchange diagram in the bulk channel}
The same strategy applies to all the other cases. To obtain the bulk channel decomposition of the boundary exchange Witten diagram, we first look at the action of $\mathbf{EOM}_b$ on a bulk channel conformal block $g^B_{\Delta_1+\Delta_2+2N}$. We find
\begin{equation}
\mathbf{EOM}_b\, g^B_{\Delta_1+\Delta_2+2N}=\alpha_N g^B_{\Delta_1+\Delta_2+2N-2}+ \beta_N g^B_{\Delta_1+\Delta_2+2N}+\gamma_N g^B_{\Delta_1+\Delta_2+2N+2}
\end{equation} 
where
\begin{equation}\footnotesize
\begin{split}
{}&\alpha_N=-\frac{1}{2} N (d+2 N-2)\;,\\
{}& \beta_N=\frac{1}{2} \left(-2 d {\widehat{\Delta}} +d \Delta_1+2 {\widehat{\Delta}} ^2+2 {\widehat{\Delta}} +4 N^2+4 \Delta_1 N\right)\\
{}&+\frac{N (d+2 N-2) (\Delta_1+N-1) (d-2 (\Delta_1+N))}{2 (d-2 (\Delta^{12}_N-1))}-\frac{(n+1) (d+2 N) (\Delta_1+N) (d-2 (\Delta_1+N+1))}{2 (d-2 (\Delta^{12}_N+1))}\;,\\
{}& \gamma_N=-\frac{2 (\Delta_1+N) (\Delta_2+N) (2 (\Delta_1+N+1)-d) (2 (\Delta_2+N+1)-d) (-d+\Delta^{12}_N-N+1) (2 (\Delta^{12}_N-N)-d)}{(d-2 \Delta^{12}_N) (d-2 (\Delta^{12}_N+1))^2 (d-2 (\Delta^{12}_N+2))}\;,
\end{split}
\end{equation}
and we have defined the short-hand notation $\Delta^{12}_N\equiv\Delta_1+\Delta_2+2N$.
Apply $\mathbf{EOM}_b$ to the bulk channel decomposition
\begin{equation}
\mathcal{W}^{boundary}(\xi)=\sum_N A^b_n g^B_{\Delta_1+\Delta_2+2N}(\xi)\;,
\end{equation}
and use the equation of motion identity (\ref{EOMbboundary}), we get the following recursion relations for the OPE coefficients $A^b_N$
\begin{equation}
\gamma_{N-1}A^b_{N-1}+\beta_{N}A^b_{n}+\alpha_{N+1}A^b_{N+1}=a_N\;.
\end{equation}
As before, we have defined $A^b_{-1}=0$. The seed coefficient $A^b_0$ is given by (\ref{Abseed}).

\subsubsection{Bulk mirror exchange diagram in the boundary channel}\label{Secbkmirrorinbdr}
We now consider the boundary channel decomposition of the mirror exchange Witten diagram $\bar{\mathcal{W}}^{exchange}(x,y)$. The action of the mirror equation of motion operator $\overline{\mathbf{EOM}}_B$ on a boundary conformal block $g^b_{\Delta_1+n}$ reads
\begin{equation}
\overline{\mathbf{EOM}}_B\; g^b_{\Delta_1+n}(\xi)=\hat{\bar{\alpha}}_n^{(1)} g^b_{\Delta_1+n-1}(\xi)+\hat{\bar{\beta}}_n^{(1)} g^b_{\Delta_1+n}(\xi)+\hat{\bar{\gamma}}_n^{(1)} g^b_{\Delta_1+n+1}(\xi)
\end{equation}
where
\begin{equation}\small
\begin{split}
{}&\hat{\bar{\alpha}}_n^{(1)}=4 n (\Delta_1-\Delta_2+n)\;,\\
{}&\hat{\bar{\beta}}_n^{(1)}=-d (\Delta +\Delta_1-\Delta_2)+n (-2 d+4 \Delta_1+2)+\Delta ^2+\Delta_1 (\Delta_1+2)-\Delta_2^2+2 n^2\;,\\
{}&\hat{\bar{\gamma}}_n^{(1)}=\frac{(\Delta_1+n) (d-2 \Delta_1-n-1) (d-\Delta_1-n-2) (-d+\Delta_1+\Delta_2+n+1)}{(d-2 \Delta_1-2 n-3) (d-2 \Delta_1-2 n-1)}\;.
\end{split}
\end{equation}
Following the same reasoning, the above action leads to recursion relations for the boundary channel decomposition of the mirror exchange Witten diagram
\begin{equation}\label{Wbulkmirrorintoboundary}
\bar{\mathcal{W}}^{bulk}(\xi)=\sum_n\hat{\bar{A}}^{B,(1)}_n g^b_{\Delta_1+n}(\xi)+\sum_n\hat{\bar{A}}^{B,(2)}_n g^b_{\Delta_2+n}(\xi)\;.
\end{equation}
The recursion relations takes the following form
\begin{equation}\label{AbarhatBrecur}
\hat{\bar{\gamma}}_{n-1}^{(i)}\hat{\bar{A}}^{B,(i)}_{n-1}+\hat{\bar{\beta}}_{n}^{(i)}\hat{\bar{A}}^{B,(i)}_{n}+\hat{\bar{\alpha}}_{n+1}^{(i)}\hat{\bar{A}}^{B,(i)}_{n+1}=
\begin{cases}
\hat{a}^{i}_{\frac{n}{2}}\;,\quad n\text{ even}\;,\\
0\;,\quad\quad n\text{ odd}\;,
\end{cases}
\end{equation}
and $\hat{\bar{A}}^{B,(i)}_{-1}=0$. It is important to note that in the above recursion relations
\begin{equation}
\hat{\bar{\alpha}}_n^{(i)}=-\hat{\alpha}_n^{(i)}\;,\quad \hat{\bar{\beta}}_n^{(i)}=\hat{\beta}_n^{(i)}\;,\quad \hat{\bar{\gamma}}_n^{(i)}=-\hat{\gamma}_n^{(i)}\;
\end{equation}
where $\hat{\alpha}_n^{(i)}$, $\hat{\beta}_n^{(i)}$, $\hat{\gamma}_n^{(i)}$ are the recursion coefficients in (\ref{AhatBrecur}) for the bulk exchange Witten diagram. Moreover, as we will show in Appendix \ref{Secseed}, the seed coefficients for the mirror exchange Witten diagram are the same as for the exchange Witten diagram
\begin{equation}
\hat{\bar{A}}^{B,(i)}_{0}=\hat{A}^{B,(i)}_{0}\;.
\end{equation}
This implies that
\begin{equation}
\hat{\bar{A}}^{B,(i)}_{n}=(-1)^n\hat{A}^{B,(i)}_{n}\;.
\end{equation}
Recall that the Neumann boundary condition bulk exchange Witten diagram is a sum of $\mathcal{W}^{bulk}$ and $\bar{\mathcal{W}}^{bulk}$, we therefore find that all the single-trace boundary conformal blocks with conformal dimensions $\Delta_i+2n+1$ are projected out. This is precisely what we expected for the Neumann boundary condition.

\subsubsection{Bulk mirror exchange diagram in the bulk channel}
Finally we discuss the decomposition of the mirror exchange diagram into the bulk channel
\begin{equation}
\bar{\mathcal{W}}^{bulk}(\xi)=\sum_N\bar{A}^B_N g^B_{\Delta_1+\Delta_2+2N}(\xi)\;.
\end{equation}
The action of the operator $\overline{\mathbf{EOM}}_B$ on a bulk channel conformal block $g^B_{\Delta_1+\Delta_2+2N}$ takes the following form
\begin{equation}
\overline{\mathbf{EOM}}_B\; g^B_{\Delta_1+\Delta_2+2N}(\xi)=\bar{\alpha}_N\, g^B_{\Delta_1+\Delta_2+2N-2}+\bar{\beta}_N\, g^B_{\Delta_1+\Delta_2+2N}+\bar{\gamma}_N\, g^B_{\Delta_1+\Delta_2+2N+2}
\end{equation}
where 
\begin{equation}\small
\begin{split}
\bar{\alpha}_N={}&2 N (d+2 N-2)\;,\\
\bar{\beta}_N={}&-(\Delta +\Delta_1-\Delta_2+2 N) (d-\Delta +\Delta_1-\Delta_2+2 N)\\
{}&-\frac{2 N (d+2 N-2) (\Delta_1+N-1) (d-2 (\Delta_1+N))}{d-2 (\Delta^{12}_N-1)}+\frac{2 (N+1) (d+2 N) (\Delta_1+N) (d-2 (\Delta_1+N+1))}{d-2 (\Delta^{12}_N+1)}\;,\\
\bar{\gamma}_N={}&\frac{8 (\Delta_1+N) (\Delta_2+N) (2 (\Delta_1+N+1)-d) (2 (\Delta_2+N+1)-d) (-d+\Delta^{12}_N-N+1) (2 (\Delta^{12}_N-N)-d)}{(d-2 \Delta^{12}_N) (d-2 (\Delta^{12}_N+1))^2 (d-2 (\Delta^{12}_N+2))}\;,
\end{split}
\end{equation}
and $\Delta^{12}_N=\Delta_1+\Delta_2+2N$.  Using the equation of motion (\ref{EOMBmirror}), we get the following recursion relations for the coefficients $\bar{A}^B_N$
\begin{equation}
\bar{\gamma}_{N-1}\bar{A}^B_{N-1}+\bar{\beta}_{N}\bar{A}^B_{N}+\bar{\alpha}_{N+1}\bar{A}^B_{N+1}=a_N\;,
\end{equation}
with $\bar{A}^B_{-1}=0$\;. The seed coefficient $\bar{A}^B_0$ is given by (\ref{AbarBseed}) in Appendix \ref{Secseed}.

\subsection{Relating decomposition coefficients to functional actions}
As we discussed in Section  \ref{ReggeWitten}, the boundary exchange Witten diagram $\mathcal{W}^{boundary}$ and the bulk exchange Witten diagram $\mathcal{W}^{bulk}$ are super-bounded in the Regge limit while the mirror bulk exchange Witten diagram $\bar{\mathcal{W}}^{bulk}$ fails to be. This implies that the $hAdS_{d+1}^N$ exchange diagrams are super-bounded in the boundary channel, but only bounded in the bulk channel. We also notice that the zero-derivative contact diagram $\mathcal{W}^{contact}\in \mathcal{V}$ has the same Regge divergence as $\bar{\mathcal{W}}^{bulk}\in \mathcal{V}$. It implies that a certain linear combination of $\mathcal{W}^{bulk}_{\rm Neum}$ with $\mathcal{W}^{contact}$ can have improved Regge behavior such that it belongs to space $\mathcal{U}$. By construction, these improved $hAdS_{d+1}^N$ exchange diagrams then coincide with the Polyakov blocks up to an overall normalization
\begin{equation}
\mathfrak{P}^B_\Delta=\frac{1}{\mathcal{N}_{bulk}}\mathcal{W}^{bulk}_{\rm Neum}+\theta_\Delta \mathcal{W}^{contact}\;,
\end{equation}
\begin{equation}
\mathfrak{P}^b_{\widehat{\Delta}}=\frac{1}{\mathcal{N}_{boundary}}\mathcal{W}^{boundary}_{\rm Neum}\;.
\end{equation}
Here the normalization factors $\mathcal{N}_{bulk}$ and $\mathcal{N}_{boundary}$ are such that the single-trace conformal blocks have coefficient 1. The coefficient $\lambda$ is fixed by requiring the improved Regge behavior, and is
\begin{equation}
\theta_\Delta=-\frac{(\Delta -1) (\Delta-d+1)}{A^B}\;.
\end{equation}
Acting on $\mathfrak{P}^B_\Delta$ and $\mathfrak{P}^b_{\widehat{\Delta}}$ with the functional basis of $\mathcal{U}^*$, we can then read off from the Witten diagrams the functional action on the generic channel conformal blocks in both channels. Explicitly, the action of these functionals reads
\begin{equation}
\begin{split}
{}&\omega_N(g^B_\Delta)=-\frac{1}{A^B}(A^B_N+\bar{A}^B_N)+\theta_\Delta\, a_N\;,\quad \text{($N$ as in $\Delta_1+\Delta_2+2N$)}\;,\\
{}&\omega_N(g^b_{\widehat{\Delta}})=\frac{1}{\hat{A}^b}A^b_N\;,\quad \text{($N$ as in $\Delta_1+\Delta_2+2N$)}\;,\\
{}&\widehat{\omega}_n^{(i)}(g^B_\Delta)=\frac{1}{A^B}(\hat{A}_{2n}^{B,(i)}+\hat{\bar{A}}_{2n}^{B,(i)})+\theta_\Delta\,\hat{a}^{(i)}_n=\frac{2}{A^B}\hat{A}_{2n}^{B,(i)}+\theta_\Delta\,\hat{a}^{(i)}_n\;, \quad \text{($n$ labels $\Delta_i+2n$)}\;,\\
{}&\widehat{\omega}_n^{(i)}(g^b_{\widehat{\Delta}})=-\frac{1}{\hat{A}^b}\hat{A}_n^{b,(i)}\;, \quad \text{($n$ labels $\Delta_i+2n$)}\;,\\
{}&\widetilde{\omega}(g^B_{\Delta})_n=\frac{1}{A^B}(\hat{B}^B_{2n}+\hat{\bar{B}}^B_{2n})+\theta_\Delta\,\hat{b}_n=\frac{2}{A^B}\hat{B}^B_{2n}+\theta_\Delta\,\hat{b}_n\;,\quad \text{($n$  labels $\Delta_i+2n$)}\;,\\
{}&\widetilde{\omega}(g^b_{\widehat{\Delta}})_n=-\frac{1}{A^b}\hat{B}^b_n\;,\quad \text{($n$ labels $\Delta_i+2n$)}\;.\\
\end{split}
\end{equation}

\section{A Lorentzian Inversion Formula for BCFT}\label{LIFormula}
In this section, we will offer another perspective on our logic by deriving  a Lorentzian OPE inversion formula. We will write down a Lorentzian inversion formula for each channel. The formula for a given channel will express the OPE data in that channel in terms of the two discontinuities across the OPE singularity in both channels. Inserting the OPE expansions into the formulae  immediately leads to the Polyakov expansion of the correlator. Our logic and notation will closely follow \cite{Mazac:2018qmi}, where the analogous construction was made for 1D conformal four-point function.

\subsection{The Euclidean inversion formulae }\label{ssec:euclid}
The starting point of our analysis are the so-called Euclidean inversion formulae . These formulae  express the OPE data in each channel as a weighted integral of the Euclidean correlator. They were derived in \cite{Hogervorst:2017kbj}, where we refer the reader for more details. Below we will review the aspects relevant to our discussion. It will be useful to switch to a different cross-ratio
\be
z\equiv \frac{1}{1+\xi}\quad\Rightarrow\quad \xi =\frac{1-z}{z}\,.
\ee
The Euclidean configurations correspond to $z\in[0,1]$. We have $z\rightarrow 0$ in the boundary OPE limit, and $z\rightarrow 1$ in the bulk OPE limit. By a small change of notation compared to the previous sections, $\mathcal{G}$ will now stand for the four-point function as a function of $z$ rather than $\xi$
\be
\langle\mathcal{O}_1(x)\mathcal{O}_2(y)\rangle = \frac{1}{|2x_{\perp}|^{\Delta_1}|2y_{\perp}|^{\Delta_2}}\mathcal{G}(z)\,.
\ee
Similarly, we use a definition of the conformal blocks adapted to the $z$ variable. We will focus on the boundary channel first. The boundary OPE reads
\be
\mathcal{G}(z) = \sum\limits_{\oHat}\mu_{\oHat}\,\mathfrak{g}^b_{\Delta_{\oHat}}(z)\,,
\ee
where the boundary conformal block in the $z$-variable takes the form
\be
\mathfrak{g}^b_{\dHat}(z) = 
z^{\dHat}{}_2F_1\left(\dHat,\dHat-\frac{d}{2} +1;2 \dHat-d +2;z\right)\,.
\ee
$\mathfrak{g}^b_{\dHat}(z)$ is related to the conformal blocks $g^b_{\dHat}(\xi)$ defined earlier by
\be
\mathfrak{g}^b_{\dHat}(z) = g^b_{\dHat}\left(\mbox{$\frac{1-z}{z}$}\right)\,.
\ee
These conformal blocks are solutions of the boundary-channel Casimir equation, which we can write in the Sturm-Liouville form
\be
\partial_{z}\left[z^{2-d}(1-z)^{\frac{d}{2}}\partial_z f(z)\right]=\dHat(\dHat-d+1) z^{-d}(1-z)^{\frac{d}{2}-1}f(z)\,.
\ee
The idea behind the Euclidean inversion formulae  is that $\mathcal{G}(z)$ can be expanded instead in a complete set of delta-function normalizable eigenfunctions of the Casimir operator. A boundary condition is needed at $z=1$ in order to make the Casimir operator self-adjoint. Analogously to \cite{Hogervorst:2017sfd}, we will choose the boundary condition $f(z)=\textrm{regular}$ at $z=1$. We will call the resulting eigenfunctions conformal partial waves. They take the form
\be
\Psi^{b}_{\dHat}(z) = \frac{\mathfrak{g}^b_{\dHat}(z)}{\widehat{\kappa}(d-1-\dHat)}+\frac{\mathfrak{g}^b_{d-1-\dHat}(z)}{\widehat{\kappa}(\dHat)}\,,
\ee
where
\be
\widehat{\kappa}(\dHat) \equiv \frac{\Gamma (\dHat) \Gamma \left(\dHat-\frac{d}{2} +1\right)}{2 \Gamma (2 \dHat-d +1)}\,.
\ee
One can check a simpler alternative formula holds
\be
\Psi^{b}_{\dHat}(z) = \frac{2}{\Gamma \left(\frac{d}{2}\right)} \,{}_2F_1\left(\dHat ,d-\dHat -1;\frac{d}{2};\frac{z-1}{z}\right)\,.
\ee
The Casimir is self-adjoint with respect to the pairing
\be
\langle f_1,f_2\rangle_b \equiv \int\limits_{0}^{1}\!\!dz\,z^{-d}(1-z)^{\frac{d}{2}-1} f_1(z)f_2(z)\,.
\ee
The complete set of orthogonal, delta-function normalizable conformal partial waves corresponds to the principal series of $SO(1,d)$, {\it i.e.}, $\dHat = \frac{d-1}{2}+ i \widehat{\alpha}$ with $\widehat{\alpha} \geq 0$. Indeed, we find
\be
\langle\Psi^{b}_{\frac{d-1}{2}+i\widehat{\alpha}},\Psi^{b}_{\frac{d-1}{2}+i\widehat{\beta}}\rangle_b = 
\frac{2\pi}{\widehat{\kappa}\!\left(\frac{d-1}{2}+i\widehat{\alpha}\right)\widehat{\kappa}\!\left(\frac{d-1}{2}-i\widehat{\alpha}\right)} \delta(\widehat{\alpha}-\widehat{\beta})\quad\textrm{for }\widehat{\alpha},\widehat{\beta}>0\,.
\ee
Let us now define the \emph{boundary coefficient function} $\widehat{I}_{\dHat}$ of a correlator $\mathcal{G}(z)$ as the overlap
\be
\widehat{I}_{\dHat} \equiv \langle\Psi^{b}_{\dHat},\mathcal{G}\rangle_b = \int\limits_{0}^{1}\!dz\,
z^{-d}(1-z)^{\frac{d}{2}-1}\Psi^{b}_{\dHat}(z)\,\mathcal{G}(z)\,.
\label{eq:EIFb}
\ee
This is the Euclidean inversion formula in the boundary channel. It follows using a standard argument that the correlator can be expressed using $\widehat{I}_{\dHat}$ as
\be
\mathcal{G}(z) =\!\!\! \int\limits_{\frac{d-1}{2}-i\infty}^{\frac{d-1}{2}+i\infty}\!\!\!\frac{d\dHat}{2\pi i}\,
\widehat{\kappa}(\Delta)\,\widehat{I}_{\dHat}\,\mathfrak{g}^b_{\dHat}(z)\,.
\ee
The boundary OPE is recovered by deforming the contour to the right. A primary operator $\oHat$ in the boundary OPE translates to a pole of $\widehat{I}_{\dHat}$ at $\dHat = \Delta_{\oHat}$ with residue fixed in terms of $\mu_{\oHat}$:
\be
\widehat{I}_{\dHat} \sim - \widehat{\kappa}(\Delta_{\oHat})^{-1} \frac{\mu_{\oHat}}{\dHat-\Delta_{\oHat}}\quad\textrm{as}\quad\dHat\rightarrow\Delta_{\oHat}\,.
\ee

The same logic can be applied to the bulk channel. We will write the bulk-channel OPE in the following way
\be
\mathcal{G}(z) = \frac{z^{\Delta_2}}{(1-z)^{\frac{\Delta_1+\Delta_2}{2}}}\sum\limits_{\o}\lambda_{\o}\,\mathfrak{g}^B_{\Delta_{\o}}(1-z)\,,
\ee
where the bulk conformal blocks read
\be
\mathfrak{g}^B_{\Delta}(y) = 
y^{\frac{\Delta}{2}} {}_2F_1\left(\frac{\Delta-\Delta_{12}-d+2}{2},\frac{\Delta-\Delta_{12}}{2};\Delta-\frac{d}{2} +1;y\right)\,,
\ee
where $\Delta_{12}\equiv \Delta_1-\Delta_2$ and we write $y=1-z$ to distinguish the bulk and boundary channels. $\mathfrak{g}^B_{\Delta}(y)$ is related to $g^{B}_{\Delta}(\xi)$ used in the previous sections by
\be
\mathfrak{g}^B_{\Delta}(y) = \frac{y^{\frac{\Delta_1+\Delta_2}{2}}}{(1-y)^{\Delta_2}}g^{B}_{\Delta}\left(\mbox{$\frac{y}{1-y}$}\right)\,.
\ee
The bulk Casimir equation has the following Sturm-Liouville form
\ba
\partial_{y}\left[4y^{1-\frac{d}{2}} (1-y)^{1-\Delta_{12}}\partial_y f(y)\right]
&-y^{-\frac{d}{2}}(1-y)^{-\Delta_{12}}\Delta_{12}(\Delta_{12}+d-2)f(y) = \\
&=\Delta(\Delta-d)y^{-\frac{d}{2}-1}(1-y)^{-\Delta_{12}}f(y)\,.
\ea
Again, we will choose a self-adjoint domain for the Casimir by the boundary condition $f(y)=\textrm{regular}$ at $y=1$. This leads to the following bulk conformal partial waves
\be
\Psi^{B}_{\Delta}(y) = \frac{\mathfrak{g}^B_{\Delta}(y)}{\kappa(d-\Delta)}+\frac{\mathfrak{g}^B_{d-\Delta}(y)}{\kappa(\Delta)}\,,
\ee
where
\be
\kappa(\Delta) \equiv 
\frac{\Gamma \left(\frac{\Delta-\Delta_{12}}{2}\right) \Gamma \left(\frac{\Delta-\Delta_{12}-d+2}{2}\right)}{2 \Gamma \left(\Delta -\frac{d}{2}\right)}
\,.
\ee
Again, there is a simpler useful formula
\be
\Psi^{B}_{\Delta}(y) =
\frac{2}{\Gamma (1-\Delta_{12})} y^{\frac{\text{$\Delta $1}-\text{$\Delta $2}}{2}} {}_2F_1\left(\frac{\Delta-\Delta_{12}}{2},\frac{d-\Delta-\Delta_{12}}{2};1-\Delta_{12};\frac{y-1}{y}\right)\,.
\ee
The complete orthogonal set corresponds to the principal series of $SO(1,d+1)$, {\it i.e.}, $\Delta=\frac{d}{2}+i \alpha$ with $\alpha>0$. The bulk Casimir is self-adjoint with respect to the pairing
\be
\langle f_1,f_2\rangle_B \equiv \int\limits_{0}^{1}\!\!dy\,y^{-\frac{d}{2}-1}(1-y)^{-\Delta_{12}} f_1(y)f_2(y)\,.
\ee
Let us define the \emph{bulk coefficient function} $I_{\Delta}$ as the following overlap
\be
I_{\Delta} \equiv \langle\Psi^{B}_{\Delta},\mathcal{G}_{\textrm{cross}}\rangle_B = \int\limits_{0}^{1}\!dy\,
y^{\frac{\Delta_1+\Delta_2-d-2}{2}}(1-y)^{-\Delta_{1}}\Psi^{B}_{\Delta}(y)\,\mathcal{G}(1-y)\,.
\ee
This is the Euclidean inversion formula for the bulk channel. The correlator can be expanded using $I_{\Delta}$ as follows
\be
\mathcal{G}(z) = \frac{z^{\Delta_2}}{(1-z)^{\frac{\Delta_1+\Delta_2}{2}}}\!\!\! \int\limits_{\frac{d}{2}-i\infty}^{\frac{d}{2}+i\infty}\!\!\!\frac{d\Delta}{2\pi i}
\kappa(\Delta)\,I_{\Delta}\,\mathfrak{g}^B_{\Delta}(1-z)\,.
\ee
The bulk OPE is recovered by deforming the contour to the right. A primary operator $\o$ in the bulk OPE translates to a pole of $I_{\Delta}$ at $\Delta = \Delta_{\o}$ with residue fixed in terms of $\lambda_{\o}$:
\be
I_{\Delta} \sim - \kappa(\Delta_{\o})^{-1} \frac{\lambda_{\o}}{\Delta-\Delta_{\o}}\quad\textrm{as}\quad\Delta\rightarrow\Delta_{\o}\,.
\ee

\subsection{The Lorentzian formulae }
A Lorentzian inversion formula is a formula computing the OPE data from the correlator evaluated in Lorentzian configurations. The first example of such formula was found by Caron-Huot in \cite{Caron-Huot:2017vep} in the context of the four-point function in $D>1$. We will now derive a similar formula for the coefficient functions $\widehat{I}_{\dHat}$ and $I_{\Delta}$ encoding the boundary and bulk OPE data of a two-point function in a BCFT. Since the two-point function depends on a single cross-ratio, our formula will be more closely analogous to the formula of \cite{Mazac:2018qmi} for the four-point function in 1D.

If we consider a boundary condition for a CFT in Euclidean space, all configurations of the two points will have $z\in(0,1)$. When we continue to the Lorentzian signature, we will assume the time direction flows along the boundary. In that case, we still have $z\in(0,1)$ if the two operators in the two-point function are spacelike separated. However, we can now reach also other regions of $z$. The region $z\in(1,\infty)$ corresponds to $\o_1$ and $\o_2$ being timelike separated such that $\o_2$ stays spacelike separated from the mirror image of $\o_1$ on the other side of the boundary. The region $z\in(-\infty,0)$ is reached by making $\o_2$ timelike separated from both $\o_1$ and its mirror image.

The Euclidean two-point function $\mathcal{G}(z)$ can be analytically continued from $z\in(0,1)$ to complex values of $z$. It has a pair of branch cuts at $(-\infty,0]$ and $[1,\infty)$. The branch cuts are present because every time we hit a light-cone, we have to choose an ordering of operators. The Lorentzian formula will depend on appropriate discontinuities across these branch cuts. The discontinuities will be chosen so that they annihilate the contributions of mean-field conformal blocks from our basis.

It will be convenient to switch from $\mathcal{G}(z)$ to
\be
\Gt(z) = z^{-\Delta_2}\mathcal{G}(z)\,.
\ee
In general, adding a tilde over a symbol defined in the previous sections will denote the same object as a function of $z$ and with an extra prefactor $z^{-\Delta_2}$ included. In accordance with this notation, the contribution to  $\Gt(z)$ coming from conformal blocks of dimension $\dHat $, $\Delta$ in the boundary, bulk channel respectively will be denoted
\ba
\widetilde{g}^{b}_{\dHat }(z)&\equiv z^{-\Delta_2}g^{b}_{\dHat}(\xi(z)) = z^{-\Delta_2}\mathfrak{g}^{b}_{\dHat }(z)\\
\widetilde{g}^{B}_{\Delta}(z)&\equiv z^{-\Delta_2}g^{B}_{\Delta}(\xi(z))=(1-z)^{-\frac{\Delta_1+\Delta_2}{2}}\mathfrak{g}^{B}_{\Delta}(1-z)\,,
\label{eq:GTBlockDef}
\ea
so that the boundary and bulk expansions take the following form
\be
\Gt(z) = \sum\limits_{\oHat}\mu_{\oHat}\,\widetilde{g}^b_{\Delta_{\oHat}}(z)
= \sum\limits_{\o}\lambda_{\o}\,\widetilde{g}^B_{\Delta_{\o}}(z)\,.
\ee
For $z\in(1,\infty)$, we consider the discontinuity of $\Gt(z)$ across its branch cut
\be
\Disc[\Gt(z)]\equiv\Gt^{\curvearrowright}(z)-\Gt^{\text{\rotatebox[origin=c]{180}{\reflectbox{$\curvearrowright$}}}}(z)
\ee
Here $\Gt^{\curvearrowright}(z)$ stands for the analytic continuation of $\Gt(z)$ from $z\in(0,1)$ to $z\in(1,\infty)$ passing above the branch point $z=1$, and similarly for $\Gt^{\text{\rotatebox[origin=c]{180}{\reflectbox{$\curvearrowright$}}}}(z)$. $\Disc[\Gt(z)]$ is proportional to the commutator $\left[\o_1,\o_2\right]$ in the Lorentzian configurations where $z>1$. Note that $\Disc[\Gt(z)]$ can be computed from the bulk OPE
\be
\Disc[\Gt(z)] = \sum\limits_{\o}\lambda_{\o}\Disc[\widetilde{g}^{B}_{\Delta_{\o}}(z)]\quad\textrm{for }z>1\,.
\ee
The discontinuity of a bulk block comes purely from its power-law branch cut, so that
\ba
\Disc[&\widetilde{g}^{B}_{\Delta}(z)] = -2 i\sin\left[\frac{\pi}{2}(\Delta-\Delta_1-\Delta_2)\right]\times\\
&\times(z-1)^{\frac{\Delta-\Delta_1-\Delta_2}{2}} {}_2F_1\left(\frac{\Delta-\Delta_{12}-d+2}{2},\frac{\Delta-\Delta_{12}}{2};\Delta-\frac{d}{2} +1;1-z\right)\,.
\ea
In particular, the discontinuity vanishes for the double-trace blocks of mean-field theory
\be
\Delta_N \equiv \Delta_1+\Delta_2 + 2 N\quad N=0,1,\ldots\,.
\ee
Note that one can not compute $\Disc[\Gt(z)]$ for $z>1$ using the boundary OPE since the latter does not converge for $z>1$.

We would like to define a similar quantity which annihilates the contributions of boundary blocks at the mean field dimensions with the Neumann boundary condition
\ba
\dHat ^{(1)}_n&\equiv \Delta_1 + 2n\\
\dHat ^{(2)}_n&\equiv \Delta_2 + 2n\,.
\label{eq:DeltaBoundary}
\ea
This quantity is precisely the double discontinuity around $z=0$, defined by
\be
\dDisc[\Gt(z)] \equiv \cos\!\left(\mbox{$\frac{\pi\Delta_{12}}{2}$}\right)\Gt(z)
-\frac{e^{-\frac{i\pi\Delta_{12}}{2}}}{2}\frac{\Gt^{\curvearrowleft}\!\left(\zTr\right)}{(1-z)^{\Delta_2}}
-\frac{e^{\frac{i\pi\Delta_{12}}{2}}}{2}\frac{\Gt^{\text{\rotatebox[origin=c]{180}{\reflectbox{$\curvearrowleft$}}}}\!\left(\zTr\right)}{(1-z)^{\Delta_2}}
\label{eq:dDiscDef}
\ee
where $z\in(0,1)$. The first term involves just the Euclidean correlator. The second and third term involve the analytic continuation of the correlator from the Euclidean region to the Lorentzian configuration at $\zTr\in(-\infty,0)$. The curved arrows on $\Gt$ indicate how the branch point at $z=0$ should be avoided along the path of analytic continuation. The transformation $z\mapsto\zTr$ is natural since it is a symmetry of the boundary Casimir, and the boundary blocks therefore transform nicely under it. The double discontinuity can be calculated using the boundary OPE since the latter converges in each of the three terms defining $\dDisc$
\be
\dDisc[\Gt(z)] = \sum\limits_{\oHat}\mu_{\oHat}\dDisc[\widetilde{g}^{b}_{\Delta_{\oHat}}(z)]\,.
\ee
To find the $\dDisc$ of an individual boundary block, first note that for $z\in(0,1)$
\be
\frac{\widetilde{g}^{b\curvearrowleft}_{\dHat }\!\left(\zTr\right)}{(1-z)^{\Delta_2}} = 
e^{i\pi(\widehat{\Delta}-\Delta_2)}\widetilde{g}^{b}_{\dHat }(z)\,,\qquad
\frac{\widetilde{g}^{b{\text{\rotatebox[origin=c]{180}{\reflectbox{$\curvearrowleft$}}}}}_{\dHat }\!\left(\zTr\right)}{(1-z)^{\Delta_2}} = 
e^{-i\pi(\widehat{\Delta}-\Delta_2)}\widetilde{g}^{b}_{\dHat }(z)\,.
\ee
It follows that
\be
\dDisc[\widetilde{g}^{b}_{\dHat }(z)] = 2\sin\left[\frac{\pi}{2}(\dHat-\Delta_1)\right]\!\sin\left[\frac{\pi}{2}(\dHat-\Delta_2)\right]\widetilde{g}^{b}_{\dHat }(z)\,.
\ee
As promised, $\dDisc[\widetilde{g}^{b}_{\dHat }(z)]$ has a simple zero whenever $\dHat$ hits one of the mean-field dimensions \eqref{eq:DeltaBoundary}. Pairs of simple zeros coalesce into double zeros whenever $\Delta_1-\Delta_2\in2\mathbb{Z}$. If we were dealing with the Dirichlet boundary condition instead of Neumann, all we would need to do is replace the minus signs in front of the second and third term in \eqref{eq:dDiscDef} with plus signs. Again, $\dDisc[\Gt(z)]$ can not be directly computed using the bulk OPE since the latter does not converge for $z<0$.

Note that the limit $z\rightarrow 1$ of $\dDisc[\Gt(z)]$ and the limit $z\rightarrow\infty$ of $\Disc[\Gt(z)]$ probes the BCFT analog of the Regge limit. Recall that in unitary theories, the two-point function must satisfy a boundedness condition $\mathcal{G}(z) = O(z^{\max(\Delta_1,\Delta_2)})$ as $|z|\rightarrow\infty$. Without loss of generality, we can choose $\Delta_2>\Delta_1$. It follows that $\Gt(z)$ is bounded as $|z|\rightarrow \infty $.

Our strategy for writing down Lorentzian inversion formulae  in the boundary and bulk channel for BCFT will be the reverse of the derivation of \cite{Caron-Huot:2017vep}, which started from the Euclidean formula and derived the Lorentzian one by a contour deformation. First, we will write down a general ansatz for a Lorentzian formula, {\it i.e.}, one depending only on $\dDisc[\Gt(z)]$ evaluated in $z\in(0,1)$ and $\Disc[\Gt(z)]$ evaluated in $z\in(1,\infty)$, each multiplied by yet undetermined inversion kernels. Then we will perform a contour manipulation bringing all integrations into the Euclidean region. We will see that in order for the contour deformation to be allowed, the two kernels multiplying dDisc and Disc need to descend from the same holomorphic function of $z$. Finally, we will constrain this function by imposing that the Euclidean and Lorentzian formula give the same answer.

Let us start with the first step, {\it i.e.}, writing a general ansatz for our Lorentzian inversion formulae . We will have one formula for each channel:
\ba
\widehat{I}_{\dHat} &= 2\!\int\limits_{0}^{1}\!\!dz\,\widehat{K}_{\dHat}(z)\dDisc[\Gt(z)] +
\int\limits_{1}^{\infty}\!\!dz\,\widehat{L}_{\dHat}(z)\frac{\Disc[\Gt(z)]}{i}\\
I_{\Delta} &= 2\!\int\limits_{0}^{1}\!\!dz\,K_{\Delta}(z)\dDisc[\Gt(z)] +
\int\limits_{1}^{\infty}\!\!dz\,L_{\Delta}(z)\frac{\Disc[\Gt(z)]}{i}\,.
\label{eq:LIFs1}
\ea
The factors of $2$ and $1/i$ are a useful convention simplifying several ensuing expressions. It remains to fix the inversion kernels $\widehat{K}_{\dHat}(z)$, $\widehat{L}_{\dHat}(z)$ and $K_{\Delta}(z)$, $L_{\Delta}(z)$. In the next subsection, we will derive a full set of constraints the kernels need to satisfy in order for \eqref{eq:LIFs1} to give the same answers as the Euclidean inversion formulae . We expect the constraints fix the Lorentzian inversion kernels essentially uniquely and it is a very interesting mathematical problem to find the solution. We did not determine the kernels in full generality. We will later discuss the case $\Delta_2-\Delta_1\in2\mathbb{Z}-1$, where all kernels can be found in a closed form.

\subsection{Constraining the inversion kernels}\label{ssec:ContourDef}
\begin{figure}[ht!]%
\begin{center}
\includegraphics[width=\textwidth]{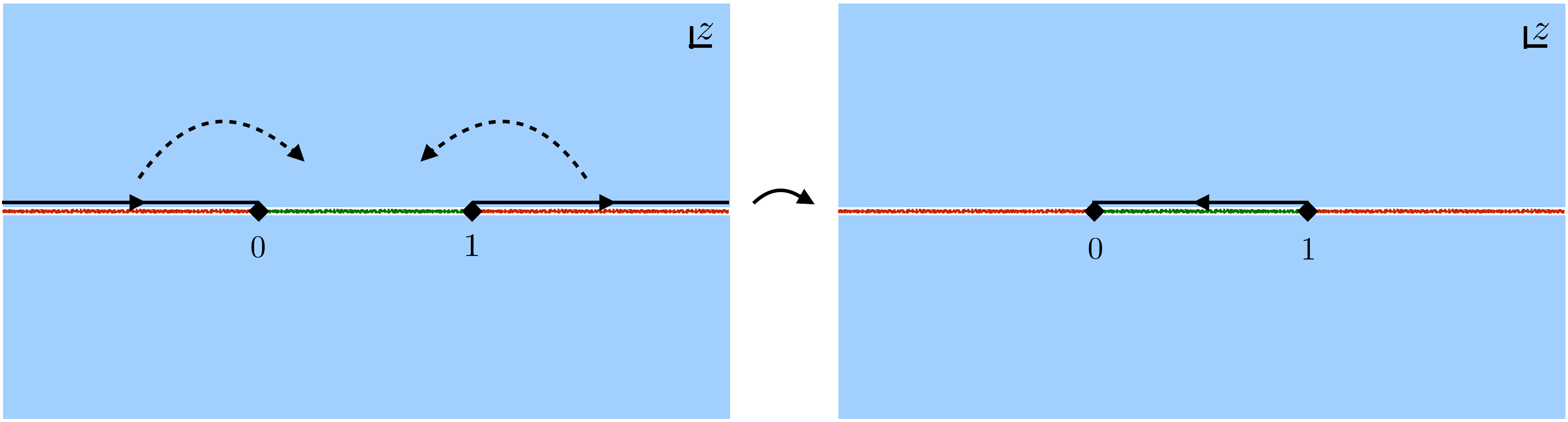}%
\caption{The contour deformation from the Lorentzian to the Euclidean region. The branch cuts of the correlator $\Gt(z)$ are shown in red, while the branch cut of the inversion kernels is shown in green. Figure taken from \cite{Mazac:2018qmi}.}
\label{fig:ContourDef}%
\end{center}
\end{figure}

We will now perform a contour deformation of the Lorentzian formulae  \eqref{eq:LIFs1} which takes all integrations into the Euclidean region $z\in(0,1)$. Since the general form of the boundary and bulk formulae  is  exactly the same, we will work with the bulk formula until the distinction becomes important. We start from \eqref{eq:LIFs1}, write out the definitions of $\dDisc[\Gt(z)]$ and $\Disc[\Gt(z)]$ inside the integrals, and change integration variables in the branch cut contributions to $\dDisc$ to arrive at
\ba
I_{\Delta} &= 2\!\int\limits_{0}^{1}\!\!dz\,K_{\Delta}(z)\dDisc[\Gt(z)] - i
\int\limits_{1}^{\infty}\!\!dz\,L_{\Delta}(z)\Disc[\Gt(z)]\\
& = \int\limits_{0}^{1}\!\!dz\,2\cos\!\left(\mbox{$\frac{\pi\Delta_{12}}{2}$}\right)K_{\Delta}(z)\,\Gt(z)-\\
&-\int\limits_{-\infty}^{0}\!\!dz\,e^{-\frac{i\pi\Delta_{12}}{2}}(1-z)^{\Delta_2-2}K_{\Delta}\left(\zTr \right)\,\Gt^{\curvearrowleft}(z)
- i
\int\limits_{1}^{\infty}\!\!dz\,L_{\Delta}(z)\,\Gt^{\curvearrowright}(z)\\
&-\int\limits_{-\infty}^{0}\!\!dz\,e^{\frac{i\pi\Delta_{12}}{2}}(1-z)^{\Delta_2-2}K_{\Delta}\left(\zTr \right)\,\Gt^{\text{\rotatebox[origin=c]{180}{\reflectbox{$\curvearrowleft$}}}}(z)
+ i
\int\limits_{1}^{\infty}\!\!dz\,L_{\Delta}(z)\,\Gt^{\text{\rotatebox[origin=c]{180}{\reflectbox{$\curvearrowright$}}}}(z)\,.
\ea
The first line on the RHS is already an integral over the Euclidean region. We will now combine the two integrals on the second line and deform the contour to $z\in(0,1)$, as shown in Figure \ref{fig:ContourDef}. We will do the same with the two integrals on the third line except then the contour deformation takes places in the lower-half plane.

For the contour deformation to be admissible, a few conditions must be satisfied. Firstly, the integrands of the two integrals on the second line must define the same analytic function on the upper-half plane. Similarly, the integrands of the two integrals on the third line must define the same analytic function on the lower half-plane. These two constraints together imply that $K_{\Delta}(z)$ and $L_{\Delta}(z)$ arise from the same analytic function. Indeed, it is not difficult to show that the constraints together imply we can always write $K_{\Delta}(z)$ and $L_{\Delta}(z)$ as follows
\ba
K_{\Delta}(z) &= (1-z)^{\frac{\Delta_1+\Delta_2-3}{2}}H_{\Delta}(z)\quad\textrm{for }z\in(0,1)\\
L_{\Delta}(z) &= (z-1)^{-\frac{\Delta_{12}+1}{2}}H_{\Delta}\left(\zTr \right)\quad\textrm{for }z\in(1,\infty)\,,
\ea
where $H_{\Delta}(z)$ is analytic for $z\in\mathbb{C}\backslash(-\infty,0]$. $H_{\Delta}(z)$ should decay sufficiently fast as $z\rightarrow1$ so that we can drop the contribution from the semicircle at infinity. Combining the contributions, we arrive at the following integral over the Euclidean region
\be
I_{\Delta} = 2\!\int\limits_{0}^{1}\!\!dz (1-z)^{\frac{\Delta_1+\Delta_2-3}{2}}S_{\Delta}(z)\Gt(z)\,,
\label{eq:afterDef}
\ee
where $S_{\Delta}(z)$ is essentially the dDisc of the inversion kernel
\be
S_{\Delta}(z) \equiv \cos\!\left(\mbox{$\frac{\pi\Delta_{12}}{2}$}\right)H_{\Delta}(z)
+\frac{e^{\frac{i\pi\Delta_{12}}{2}}}{2}\frac{H_{\Delta}^{\curvearrowleft}\!\left(\zTr\right)}{(1-z)^{\Delta_1-1}}
+\frac{e^{-\frac{i\pi\Delta_{12}}{2}}}{2}\frac{H_{\Delta}^{\text{\rotatebox[origin=c]{180}{\reflectbox{$\curvearrowleft$}}}}\!\left(\zTr\right)}{(1-z)^{\Delta_1-1}}\,.
\ee
Finally, we should impose that \eqref{eq:afterDef} is equivalent to the Euclidean formula in a given channel. This means that besides being holomorphic away from $(-\infty,0]$, the boundary kernel must satisfy
\ba
\cos\!\left(\mbox{$\frac{\pi\Delta_{12}}{2}$}\right)\widehat{H}_{\dHat}(z)
+\frac{e^{\frac{i\pi\Delta_{12}}{2}}}{2}\frac{\widehat{H}_{\dHat}^{\curvearrowleft}\!\left(\zTr\right)}{(1-z)^{\Delta_1-1}}
&+\frac{e^{-\frac{i\pi\Delta_{12}}{2}}}{2}\frac{\widehat{H}_{\dHat}^{\text{\rotatebox[origin=c]{180}{\reflectbox{$\curvearrowleft$}}}}\!\left(\zTr\right)}{(1-z)^{\Delta_1-1}} = \\
&=\frac{z^{\Delta_2-d}(1-z)^{\frac{d-\Delta_1-\Delta_2+1}{2}}}{2}\Psi^{b}_{\dHat}(z)\,.
\label{eq:HbConstraint}
\ea
Similarly, the bulk kernel is also holomorphic away from $(-\infty,0]$ and must satisfy
\ba
\cos\!\left(\mbox{$\frac{\pi\Delta_{12}}{2}$}\right)H_{\Delta}(z)
+\frac{e^{\frac{i\pi\Delta_{12}}{2}}}{2}\frac{H_{\Delta}^{\curvearrowleft}\!\left(\zTr\right)}{(1-z)^{\Delta_1-1}}
&+\frac{e^{-\frac{i\pi\Delta_{12}}{2}}}{2}\frac{H_{\Delta}^{\text{\rotatebox[origin=c]{180}{\reflectbox{$\curvearrowleft$}}}}\!\left(\zTr\right)}{(1-z)^{\Delta_1-1}} =\\
&=\frac{z^{-\Delta_{12}}(1-z)^{\frac{1-d}{2}}}{2}\Psi^{B}_{\Delta}(1-z)
\,,
\label{eq:HBConstraint}
\ea
where $\Psi^{b}_{\Delta}(z),\Psi^{B}_{\Delta}(z)$ are the boundary and bulk conformal partial waves defined in subsection \ref{ssec:euclid}.

Provided we can find kernels satisfying these constraints, the Lorentzian inversion formulae  in the two channels take the following form
\ba
\widehat{I}_{\dHat} &= 2\!\int\limits_{0}^{1}\!\!dz\,(1-z)^{\frac{\Delta_1+\Delta_2-3}{2}}\widehat{H}_{\dHat}(z)\dDisc[\Gt(z)] -\\
&- i\int\limits_{1}^{\infty}\!\!dz\,(z-1)^{-\frac{\Delta_{12}+1}{2}}\widehat{H}_{\dHat}\left(\zTr \right)\Disc[\Gt(z)]
\label{eq:LIFb}
\ea
and
\ba
I_{\Delta} &= 2\!\int\limits_{0}^{1}\!\!dz\,(1-z)^{\frac{\Delta_1+\Delta_2-3}{2}}H_{\Delta}(z)\dDisc[\Gt(z)] -\\
&- i\int\limits_{1}^{\infty}\!\!dz\,(z-1)^{-\frac{\Delta_{12}+1}{2}}H_{\Delta}\left(\zTr \right)\Disc[\Gt(z)]
\label{eq:LIFB}
\ea

We see from \eqref{eq:HbConstraint} and \eqref{eq:HBConstraint} that compatibility of the Lorentzian and Euclidean formulae  fixes the dDisc of $H_{\Delta}(z),\,\widehat{H}_{\dHat}(z)$ in terms of the conformal partial waves. One lesson of the Lorentzian inversion formula is that knowing the dDisc of a function with appropriate analyticity and boundedness properties allows us to reconstruct this function essentially uniquely. Applying the same logic to the inversion kernel itself makes us believe that the equations \eqref{eq:HbConstraint} and \eqref{eq:HBConstraint} go a long way towards fixing $H_{\Delta}(z)$ and $\widehat{H}_{\dHat}(z)$. In the following subsection, we will fix the kernels explicitly in the case $\Delta_1-\Delta_2\in2\mathbb{Z}+1$.

The formulae  \eqref{eq:LIFb}, \eqref{eq:LIFB} tell us that the coefficient functions $\widehat{I}_{\dHat}$, $I_{\Delta}$, and hence the full correlator $\Gt(z)$ can be reconstructed from its dDisc and Disc. This is only true provided the correlator satisfies an appropriate boundedness condition in the Regge limit $z\rightarrow\infty$. Otherwise, there can be additional contributions when going from the Euclidean to the Lorentzian formula coming from the semicircle at infinity, which we neglected in the above. By analogy with what happens in the case of the bosonic four-point function in 1D, we expect there always exist distinguished inversion kernels $\widehat{H}_{\dHat}(z)$, $H_{\Delta}(z)$ such that the formulae  \eqref{eq:LIFb}, \eqref{eq:LIFB} apply to all Euclidean-normalizable and Regge super-bounded functions. Once the formula for superbounded functions is derived, one can ``improve'' the kernel by subtractions to derive a formula for functions with less strict boundedness properties in the Regge limit, along the lines of section 6.5 of reference \cite{Mazac:2018qmi}.

\subsection{A solvable example}
Our life simplifies when $\Delta_1-\Delta_2\in2\mathbb{Z}+1$. In this case, the boundary spectrum $\{\dHat ^{(1)}_n\}\cup\{\dHat ^{(2)}_n\}$ has integer spacing, and the double discontinuity \eqref{eq:dDiscDef} reduces to a simple discontinuity. Indeed, let us write $\Delta_2 = \Delta_1 + 2M+1$, where $M\in\mathbb{Z}_{\geq 0}$. We find
\be
\dDisc[\Gt(z)] = \frac{(-1)^M}{2i(1-z)^{\Delta_2}}
\left[\Gt^{\curvearrowleft}\left(\zTr\right)-\Gt^{\text{\rotatebox[origin=c]{180}{\reflectbox{$\curvearrowleft$}}}}\left(\zTr\right)\right]\,.
\ee
Similarly, the constraints on the inversion kernels \eqref{eq:HbConstraint}, \eqref{eq:HBConstraint} become formulae  for their single discontinuity across the branch cut $z\in(-\infty,0]$. The actual kernel then can be recovered from
\be
H_{\Delta}(z) = \int\limits_{-\infty}^{0}\!\!\frac{dw}{2\pi i}\,\frac{H^{\curvearrowleft}_{\Delta}(w)-H^{\text{\rotatebox[origin=c]{180}{\reflectbox{$\curvearrowleft$}}}}_{\Delta}(w)}{w-z}\,,
\ee
which follows from Cauchy's integral formula and analyticity of $H_{\Delta}(z)$ by a contour deformation. Equivalently, we can derive the Lorentzian formulae  directly by first writing the standard dispersion relation for $\Gt(z)$, inserting it into the Euclidean inversion formulae , and interchange the two integrations. We are ignoring various subtleties coming from possible boundary contributions at $z=0,1,\infty$.

The above procedure gives the following formula for the Lorentzian inversion kernel for the boundary OPE data
\ba
\widehat{H}_{\dHat}(z) =
&\frac{\Gamma \left(\Delta _2-\dHat \right) \Gamma \left(\Delta _2+\dHat-d+1\right)}{\pi  \Gamma \left(\Delta _2\right) \Gamma \left(\Delta _2-\frac{d}{2}+1\right)}z^{-1}(z-1)^{\frac{\Delta_2-\Delta_1+1}{2}}\times\\
&\times{}_3F_2\left( {\begin{array}{*{20}{c}}
{1,\Delta _2-\dHat ,\Delta _2+\dHat-d+1}\\
{\Delta _2,\Delta _2-\frac{d}{2}+1}
\end{array};\frac{z-1}{z}} \right)\,.
\label{eq:HbInt}
\ea
Note that it is indeed holomorphic for $z\in\mathbb{C}\backslash(-\infty,0]$ since $\Delta_2-\Delta_1\in2\mathbb{Z}_{\geq 0}+1$. The same procedure leads to a formula for the bulk Lorentzian inversion kernel in the form of an infinite series around $z=1$
\be
H_{\Delta}(z) = z^{-1}(z-1)^{\frac{\Delta_2-\Delta_1+1}{2}}\sum\limits_{j=0}^{\infty}c_j\left(\mbox{$\frac{z-1}{z}$}\right)^j\,,
\ee
where
\ba
c_j = 
&\frac{\Gamma \left(\frac{\Delta _1+\Delta _2-\Delta }{2}\right) \Gamma \left(j+1-\Delta _1+\Delta _2\right)}
{\pi\Gamma \left(\frac{2 j-\Delta _1+3 \Delta _2+2-\Delta}{2}\right)\Gamma \left(1-\Delta _1+\Delta _2\right)}\times\\
&\times{}_3F_2\left( {\begin{array}{*{20}{c}}
{\frac{\Delta_2-\Delta_1+2-\Delta }{2},\frac{\Delta_2-\Delta_1+d-\Delta}{2},\Delta _2-\Delta _1+j+1}\\
{\frac{3 \Delta _2-\Delta_1-\Delta+2j+2}{2},\Delta _2-\Delta _1+1}
\end{array};1}\right)\,.
\label{eq:HBIntC}
\ea

\subsection{Polyakov blocks from the inversion formula}
The Lorentzian inversion formulae  \eqref{eq:LIFb}, \eqref{eq:LIFB} very directly encode the boundary and bulk OPEs of the Polyakov blocks. Since we are including an extra factor of $z^{-\Delta_2}$ in the correlators, we will work with Polyakov blocks including the same factor, {\it i.e.}, we define
\ba
\widetilde{\mathfrak{P}}^{b}_{\dHat}(z) &= z^{-\Delta_2}\mathfrak{P}^{b}_{\dHat}(\xi(z))\\
\widetilde{\mathfrak{P}}^{B}_{\Delta}(z) &= z^{-\Delta_2}\mathfrak{P}^{B}_{\Delta}(\xi(z))\,,
\ea
where $\mathfrak{P}^{b}_{\dHat}(\xi)$, $\mathfrak{P}^{B}_{\Delta}(\xi)$ are the Polyakov blocks used in the previous sections and $\xi(z) = \mbox{$\frac{1-z}{z}$}$. Recall that the Polyakov blocks admit the following bulk and boundary OPEs
\ba
\widetilde{\mathfrak{P}}^{B}_{\Delta}(z) &=
\widetilde{g}^{B}_{\Delta}(z)+\sum\limits_{N}\mathfrak{a}_N\widetilde{g}^{B}_{\Delta_N}(z)=
\sum\limits_{n,i}\mathfrak{b}_n^{(i)}\widetilde{g}^{b}_{\widehat{\Delta}^{(i)}_n}(z)\\
\widetilde{\mathfrak{P}}^{b}_{\dHat}(z) &=
\widetilde{g}^{b}_{\dHat}(z) + \sum\limits_{n,i}\mathfrak{c}_n^{(i)}\widetilde{g}^{b}_{\widehat{\Delta}^{(i)}_n}(z) =
\sum\limits_{N}\mathfrak{d}_N\widetilde{g}^{B}_{\Delta_N}(z)\,,
\label{eq:PolyakovOPEs}
\ea
where $\widehat{\Delta}^{(1)}_n = \Delta_1+2n$, $\widehat{\Delta}^{(2)}_n = \Delta_2+2n$ and $\Delta_N = \Delta_1+\Delta_2+2N$. It follows from these OPEs that the dDisc of the boundary Polyakov block of dimension $\widehat{\Delta}$ equals the dDisc of a single boundary conformal block of dimension $\widehat{\Delta}$, and its Disc vanishes. In fact, the boundary Polyakov block is the unique function satisfying these properties which is also super-bounded in the Regge limit. Similarly, the bulk Polyakov block of dimension $\Delta$ is the unique Regge super-bounded function whose dDisc vanishes and whose Disc equals the Disc of a single bulk conformal block of dimension $\Delta$.

Since the Polyakov blocks are super-bounded, the Lorentzian inversion formulae  apply to them. Furthermore, when inserting the OPEs of the Polyakov blocks into the inversion formula, only a single term survives. We conclude that the boundary and bulk coefficient functions of the boundary Polyakov block of dimension $\Delta_{\oHat}$ are given by respectively
\ba
\widehat{\mathcal{I}}_{b}(\dHat,\Delta_{\oHat}|\Delta_1,\Delta_2)&\equiv\widehat{I}_{\dHat}[\mathfrak{P}^{b}_{\Delta_{\oHat}}]
=
2\!\int\limits_{0}^{1}\!\!dz\,(1-z)^{\frac{\Delta_1+\Delta_2-3}{2}}\widehat{H}_{\dHat}(z)\dDisc[\widetilde{g}^{b}_{\Delta_{\oHat}}(z)]\\
\mathcal{I}_{b}(\Delta,\Delta_{\oHat}|\Delta_1,\Delta_2)&\equiv I_{\Delta}[\mathfrak{P}^{b}_{\Delta_{\oHat}}]
=
2\!\int\limits_{0}^{1}\!\!dz\,(1-z)^{\frac{\Delta_1+\Delta_2-3}{2}}H_{\Delta}(z)\dDisc[\widetilde{g}^{b}_{\Delta_{\oHat}}(z)]\,.
\label{eq:ICalb}
\ea
Similarly, the boundary and bulk coefficient functions of the bulk Polyakov block of dimension $\Delta_{\o}$ are given by
\ba
\widehat{\mathcal{I}}_{B}(\dHat,\Delta_{\o}|\Delta_1,\Delta_2) &\equiv
\widehat{I}_{\dHat}[\mathfrak{P}^{B}_{\Delta_{\o}}]
=
- i
\int\limits_{1}^{\infty}\!\!dz\,(z-1)^{-\frac{\Delta_{12}+1}{2}}\widehat{H}_{\dHat}\left(\zTr \right)\Disc[\widetilde{g}^{B}_{\Delta_{\o}}(z)]
\\
\mathcal{I}_{B}(\Delta,\Delta_{\o}|\Delta_1,\Delta_2) &\equiv
I_{\Delta}[\mathfrak{P}^{B}_{\Delta_{\o}}]=
- i
\int\limits_{1}^{\infty}\!\!dz\,(z-1)^{-\frac{\Delta_{12}+1}{2}}H_{\Delta}\left(\zTr \right)\Disc[\widetilde{g}^{B}_{\Delta_{\o}}(z)]\,.
\label{eq:ICalB}
\ea
Let us explain how these expressions give rise to the OPE of Polyakov blocks \eqref{eq:PolyakovOPEs}. In order to reproduce the first term, we should find that $\widehat{\mathcal{I}}_{b}(\dHat,\Delta_{\oHat}|\Delta_1,\Delta_2)$ contains a pole at $\dHat = \Delta_{\oHat}$ with residue $-\widehat{\kappa}(\Delta_{\oHat})^{-1}$. This pole can only come from a $z\rightarrow 0$ singularity of the first integral in \eqref{eq:ICalb}. In the specific case discussed in the previous subsection, {\it i.e.}, $\Delta_2-\Delta_1 \in 2\mathbb{Z}+1$, we indeed find
\be
\widehat{H}_{\dHat}(z) \sim \frac{z^{\Delta_2-\dHat-1}}{4\sin\left[\frac{\pi}{2}(\dHat-\Delta_1)\right]\sin\left[\frac{\pi}{2}(\dHat-\Delta_2)\right]\widehat{\kappa}(\dHat)}\quad\textrm{as }z\rightarrow 0\,,
\ee
which produces precisely the right pole and residue in $\widehat{\mathcal{I}}_{b}(\dHat,\Delta_{\oHat}|\Delta_1,\Delta_2)$. Similarly, the coefficient function $\mathcal{I}_{B}(\Delta,\Delta_{\o}|\Delta_1,\Delta_2)$ should have a pole at $\Delta = \Delta_{\o}$ with residue $-\kappa(\Delta_{\o})^{-1}$, which requires $H_{\Delta}(z)$ to have a specific singularity as $z\rightarrow\infty$.

It remains to understand the origin of the poles corresponding to the mean-field operator contributions to the Polyakov blocks. We claim these arise entirely from corresponding poles of $\widehat{H}_{\dHat}(z)$ and $H_{\Delta}(z)$ themselves, rather than from the integration over $z$. For this to be the case, $\widehat{H}_{\dHat}(z)$ should have simple poles at $\dHat = \widehat{\Delta}^{(1,2)}_n$ and $H_{\Delta}(z)$ should have simple poles at $\Delta = \Delta_N$. Let us denote the residues of the inversion kernels at these locations as follows
\ba
\widehat{H}_{\dHat}(z) &\sim \frac{\widehat{h}^{(i)}_n(z)}{\dHat-\widehat{\Delta}^{(i)}_n}\quad\textrm{as }\dHat\rightarrow\widehat{\Delta}^{(i)}_n\\
H_{\Delta}(z) &\sim \frac{h_{N}(z)}{\Delta-\widehat{\Delta}_N}\quad\textrm{as }\Delta\rightarrow\Delta_N\,.
\ea
We can see these poles are indeed present for the example of the previous section. $\widehat{H}_{\dHat}(z)$ of equation \eqref{eq:HbInt} has simple poles at $\dHat=\Delta_2 + \mathbb{Z}_{\geq 0}$ coming from the prefactor $\Gamma(\Delta_2-\dHat)$. These corresponding to alternating $\widehat{\Delta}^{(2)}_{n}$ and $\widehat{\Delta}^{(1)}_{n}$. Similarly, the expected poles of $H_{\Delta}(z)$ are manifest in the prefactor $\Gamma\left(\frac{\Delta_1+\Delta_2-\Delta}{2}\right)$ visible in \eqref{eq:HBIntC}.

Starting from the residues $\widehat{h}^{(i)}_n(z)$, $h_{N}(z)$, we can compute the OPE coefficients $\mathfrak{a}_N$, $ \mathfrak{b}^{(i)}_{n}$, $\mathfrak{c}^{(i)}_n$, $\mathfrak{d}_N$ from equations \eqref{eq:ICalb} and \eqref{eq:ICalB}
\ba
\mathfrak{a}_N &= i\kappa(\Delta_N)
\int\limits_{1}^{\infty}\!\!dz\,(z-1)^{-\frac{\Delta_{12}+1}{2}}h_{N}\left(\zTr \right)\Disc[\widetilde{g}^{B}_{\Delta}(z)]\\
\mathfrak{b}_{n}^{(i)} &= i\widehat{\kappa}(\widehat{\Delta}_{n}^{(i)})
\int\limits_{1}^{\infty}\!\!dz\,(z-1)^{-\frac{\Delta_{12}+1}{2}}\widehat{h}^{(i)}_n\left(\zTr \right)\Disc[\widetilde{g}^{B}_{\Delta}(z)]\\
\mathfrak{c}_{n}^{(i)} &= -2\widehat{\kappa}(\widehat{\Delta}_{n}^{(i)})\!\int\limits_{0}^{1}\!\!dz\,(1-z)^{\frac{\Delta_1+\Delta_2-3}{2}}\widehat{h}^{(i)}_n(z)\dDisc[\widetilde{g}^{b}_{\widehat{\Delta}}(z)]\\
\mathfrak{d}_N &= -2\kappa(\Delta_N)\!\int\limits_{0}^{1}\!\!dz\,(1-z)^{\frac{\Delta_1+\Delta_2-3}{2}}h_{N}(z)\dDisc[\widetilde{g}^{b}_{\widehat{\Delta}}(z)]\,.
\label{eq:PolOPECs}
\ea
In the cases where the inversion kernels are known, these formulae  provide a useful alternative for computing the OPE coefficients of the mean-field operators in Witten exchange diagrams for BCFT.

\subsection{Polyakov expansion of the correlator}
We are now ready to explain how the Lorentzian inversion formulae  \eqref{eq:LIFb} and \eqref{eq:LIFB} lead to the expansion of the correlator into Polyakov blocks. The expansion is obtained by inserting the boundary and bulk OPEs into the Lorentzian inversion formulae . We will write the boundary and bulk OPEs as follows
\be
\Gt(z) = \sum\limits_{\oHat}\mu_{\oHat}\,\widetilde{g}^b_{\Delta_{\oHat}}(z)
= \sum\limits_{\o}\lambda_{\o}\,\widetilde{g}^B_{\Delta_{\o}}(z)\,,
\ee
where $\widetilde{g}^b_{\dHat}(z)$ and $\widetilde{g}^B_{\Delta}(z)$ were given in \eqref{eq:GTBlockDef}. The Lorentzian inversion formulae  involve $\dDisc[\Gt(z)]$ with $z\in(0,1)$ and $\Disc[\Gt(z)]$ with $z\in(1,\infty)$. As discussed earlier, $\dDisc[\Gt(z)]$ can be expanded using the boundary OPE, but not using the bulk OPE, and vice versa for $\Disc[\Gt(z)]$.
\ba
\dDisc[\Gt(z)] &= \sum\limits_{\oHat}\mu_{\oHat}\dDisc[\widetilde{g}^{b}_{\Delta_{\oHat}}(z)]
\quad\textrm{for }z\in(0,1)\\
\Disc[\Gt(z)] &= \sum\limits_{\o}\lambda_{\o}\,\Disc[\widetilde{g}^{B}_{\Delta_{\o}}(z)]
\quad\textrm{for }z\in(1,\infty)\,.
\ea
Inserting these expansions into the inversion formulae  \eqref{eq:LIFb} and \eqref{eq:LIFB} gives
\be
\widehat{I}_{\dHat} = \sum\limits_{\oHat}\mu_{\oHat}\,\widehat{\mathcal{I}}_{b}(\dHat,\Delta_{\oHat}|\Delta_1,\Delta_2)+
 \sum\limits_{\o}\lambda_{\o}\,\widehat{\mathcal{I}}_{B}(\dHat,\Delta_{\o}|\Delta_1,\Delta_2)
 \label{eq:Pb}
\ee
and
\be
I_{\Delta} = \sum\limits_{\oHat}\mu_{\oHat}\,\mathcal{I}_{b}(\Delta,\Delta_{\oHat}|\Delta_1,\Delta_2)+
 \sum\limits_{\o}\lambda_{\o}\,\mathcal{I}_{B}(\Delta,\Delta_{\o}|\Delta_1,\Delta_2)\,,
 \label{eq:PB}
\ee
where $\mathcal{I}_{b,B}$, $\widehat{\mathcal{I}}_{b,B}$ are the coefficient functions of Polyakov blocks discussed in the previous subsection. In going from \eqref{eq:LIFb}, \eqref{eq:LIFB} to \eqref{eq:Pb}, \eqref{eq:PB}, we assumed that we can commute the OPE sums and the $z$-integration. This is definitely allowed for the boundary channel sum in the case of identical external operators, for $\dHat$ on the principal series. This is because in that case the OPE gives $\dDisc[\Gt(z)]$ as a sum of positive terms and the swapping of integration and summation follows from the dominated convergence theorem. By analogy with the situation for the 1D four-point function, we expect the sums over operators in \eqref{eq:Pb} (and \eqref{eq:PB}) to converge uniformly in $\dHat$ (and $\Delta$), in any compact region of the complex plane away from the poles of the individual terms. If that is the case, it follows that $\widehat{I}_{\dHat}$ and $I_{\Delta}$ are meromorphic, with poles only at poles of the individual terms in the sums over operators. In the following, we will assume this is indeed what happens.

Provided the last assertion holds, equations \eqref{eq:Pb} and \eqref{eq:PB} say precisely that the correlator can be expanded in Polyakov blocks, at least at the level of the coefficient functions.

\subsection{Sum rules and functionals}
For equations \eqref{eq:Pb} and \eqref{eq:PB} to be consistent with the boundary and bulk OPEs, the spurious poles of $\mathcal{I}_{b,B}$, $\widehat{\mathcal{I}}_{b,B}$ at mean-field operators must cancel out in $I_{\Delta}$, $\widehat{I}_{\dHat}$ after performing the sum over $\oHat$ and $\o$. These are the Polyakov sum rules. As explained before, they can be alternatively derived by acting with suitable linear functionals on the standard crossing equation
\be
\sum\limits_{\oHat}\mu_{\oHat}\,\widetilde{g}^b_{\Delta_{\oHat}}(z)
= \sum\limits_{\o}\lambda_{\o}\,\widetilde{g}^B_{\Delta_{\o}}(z)\,.
\ee
We have one functional associated to every mean-field operator, namely functionals $\widehat{\omega}_n^{(i)}$ for the boundary mean-field operators and $\omega_N$ for the bulk ones. Recall that the OPE coefficients of mean-field operators in Polyakov blocks are given by the action of functionals on conformal blocks according to\footnote{Recall that the functionals were defined to act on $\mathcal{G}$, rather than on $\Gt = z^{-\Delta_2}\mathcal{G}$. This will lead to a few extra $z^{-\Delta_2}$ factors in the following formulae .}
\ba
\mathfrak{a}_N &= -\omega_N(g^{B}_{\Delta})\\
\mathfrak{b}_{n}^{(i)} &= \widehat{\omega}^{(i)}_n(g^{B}_{\Delta})\\
\mathfrak{c}_{n}^{(i)} &= -\widehat{\omega}^{(i)}_n(g^{b}_{\widehat{\Delta}})\\
\mathfrak{d}_N &= \omega_N(g^{b}_{\widehat{\Delta}})\,.
\ea
We can compare these expressions with \eqref{eq:PolOPECs}, where the same coefficients are computed from the inversion formulae . We see that those equations indeed compute $\mathfrak{a}_N$ and $\mathfrak{b}_n^{(i)}$ from a linear action on the bulk conformal block $\widetilde{g}^{B}_{\Delta}$. Similarly, $\mathfrak{c}_n^{(i)}$ and $\mathfrak{d}_N$ are computed by a linear action on the boundary block $\widetilde{g}^{b}_{\widehat{\Delta}}$. The only issue seems to be that the definitions of the functionals provided by $\mathfrak{a}_N$ and $\mathfrak{b}_n^{(i)}$ do not appear equivalent to those provided by $\mathfrak{c}_n^{(i)}$ and $\mathfrak{d}_N$. To see that in fact they are equivalent, let us focus on the bulk functional $\omega_N$. The formulae  for $\mathfrak{a}_N$ and $\mathfrak{d}_N$ provided by \eqref{eq:PolOPECs} lead to the following two ways to define the action of functional $\omega_N$ on a general test-function $\mathcal{F}(z)$ ($\mathcal{F}(z)$ should have the same complex-analytic properties as a generic physical two-point function $\mathcal{G}(z)$):
\ba
\mathfrak{a}_N:\quad \omega_N(\mathcal{F})&\equiv
-i\kappa(\Delta_N)
\int\limits_{1}^{\infty}\!\!dz\,(z-1)^{-\frac{\Delta_{12}+1}{2}}h_{N}\left(\zTr \right)\Disc[z^{-\Delta_2}\mathcal{F}(z)]\\
\mathfrak{d}_N:\quad \omega_N(\mathcal{F})&\equiv
-2\kappa(\Delta_N)\!\int\limits_{0}^{1}\!\!dz\,(1-z)^{\frac{\Delta_1+\Delta_2-3}{2}}h_{N}(z)\dDisc[z^{-\Delta_2}\mathcal{F}(z)]\,.
\ea
We claim that these two formulae  are in fact completely equivalent, that is for every Regge super-bounded function $\mathcal{F}(z)$, there is an exact identity
\ba
&2\!\int\limits_{0}^{1}\!\!dz\,(1-z)^{\frac{\Delta_1+\Delta_2-3}{2}}h_{N}(z)\dDisc[z^{-\Delta_2}\mathcal{F}(z)]-\\
-&i\int\limits_{1}^{\infty}\!\!dz\,(z-1)^{-\frac{\Delta_{12}+1}{2}}h_{N}\left(\zTr \right)\Disc[z^{-\Delta_2}\mathcal{F}(z)] = 0\,.
\label{eq:FunId}
\ea
We can prove this identity by the same contour deformation as used in subsection \ref{ssec:ContourDef} to take us from the Lorentzian to the Euclidean inversion formula. After all contributions are placed in $z\in(0,1)$, we use the identity
\be
\cos\!\left(\mbox{$\frac{\pi\Delta_{12}}{2}$}\right)h_{N}(z)
+\frac{e^{\frac{i\pi\Delta_{12}}{2}}}{2}\frac{h_{N}^{\curvearrowleft}\!\left(\zTr\right)}{(1-z)^{\Delta_1-1}}
+\frac{e^{-\frac{i\pi\Delta_{12}}{2}}}{2}\frac{h_{N}^{\text{\rotatebox[origin=c]{180}{\reflectbox{$\curvearrowleft$}}}}\!\left(\zTr\right)}{(1-z)^{\Delta_1-1}} = 0
\ee
to show the RHS of \eqref{eq:FunId} indeed vanishes. This identity follows by taking the residue of \eqref{eq:HBConstraint} at $\Delta=\Delta_N$ since the RHS of \eqref{eq:HBConstraint} has no pole at that location.

We can use identical logic to show that the following two definitions of the boundary functionals $\widehat{\omega}_n^{(i)}$ are completely equivalent:
\ba
\mathfrak{b}^{(i)}_n:\quad \widehat{\omega}_n^{(i)}(\mathcal{F})&\equiv
i\widehat{\kappa}(\widehat{\Delta}_{n}^{(i)})
\int\limits_{1}^{\infty}\!\!dz\,(z-1)^{-\frac{\Delta_{12}+1}{2}}\widehat{h}^{(i)}_n\left(\zTr \right)\Disc[z^{-\Delta_2}\mathcal{F}(z)]
\\
\mathfrak{c}^{(i)}_n:\quad \widehat{\omega}_n^{(i)}(\mathcal{F})&\equiv
2\widehat{\kappa}(\widehat{\Delta}_{n}^{(i)})\!\int\limits_{0}^{1}\!\!dz\,(1-z)^{\frac{\Delta_1+\Delta_2-3}{2}}\widehat{h}^{(i)}_n(z)\dDisc[z^{-\Delta_2}\mathcal{F}(z)]\,.
\ea

This concludes the explanation of how functionals $\omega_N$ and $\omega_n^{(i)}$ arise from the Lorentzian inversion formula.

\subsection{Open questions}
In this section, we have given a sketch of the Lorentzian inversion formula BCFT and its connection to the Polyakov bootstrap and the associated functionals. The discussion was incomplete in several important aspects, which should be addressed:
\begin{itemize}
\item Find an explicit formula for the inversion kernels $H_\Delta(z)$, $\widehat{H}_{\dHat}(z)$ for general $\Delta_{1,2}$ and $d$ starting from their analyticity properties and constraints \eqref{eq:HbConstraint}, \eqref{eq:HBConstraint}, or otherwise.
\item Construct improved inversion formulae  which apply also to correlators which are Regge bounded but not necessarily super-bounded. This point should be trivialized in the Dirichlet case, where there is no Regge-bounded contact diagram.
\item Show that one can swap the OPE sums for dDisc and Disc with the integration over $z$, which leads to \eqref{eq:Pb} and \eqref{eq:PB}, for $\Delta,\dHat$ on the principal series.
\item Prove that the OPE sums in \eqref{eq:Pb} and \eqref{eq:PB} converge to a meromorphic function of $\Delta,\dHat$ whose only poles are those of the individual terms in the sums.
\end{itemize}
The last two points were proven in \cite{Mazac:2018qmi} by appealing to the positivity of the coefficients in the conformal block expansion. While this condition is not generally present in the BCFT context, we are optimistic that the claims of this section are correct and can eventually be proven rigorously.

\section{A Deformation of the Mean Field Theory}\label{sec:deformation}
\subsection{The set-up}
In this section, we will consider an interesting family of conformal boundary conditions for the mean field theory. The family smoothly interpolates between the Neumann and Dirichlet boundary condition and thus provides a useful test of our logic.

Throughout this section, we take our CFT$_{d}$ to be the mean field theory of a scalar operator $\phi$. This has an $AdS$ bulk description as the theory of a free massive scalar field $\Phi$ in $AdS_{d+1}$, with the mass $M$ related to the scaling dimension of $\phi$ by $(M R)^2 = \Df(\Df-d)$, where $R$ is the $AdS$ radius. We can now define a family of conformal boundary conditions in the following way. First, we restrict the theory in $AdS_{d+1}$ to half of $AdS_{d+1}$ defined by $z_{\perp}>0$, and denoted $hAdS_{d+1}$. Second, we add to the action a ``coupling'' given by integrating $ t  \Phi^2$ over the $AdS_d$ boundary of $hAdS_{d+1}$ at $z_{\perp}=0$. The total bulk action then reads
\be
S = \frac{1}{2}\int_{hAdS_{d+1}}d^{d+1}z\sqrt{g_{d+1}}\left[g^{\mu\nu}\partial_\mu\Phi\partial_\nu\Phi+M^2\Phi^2\right]+\frac{ t }{2}\int_{AdS_{d}}d^{d}z\sqrt{g_d}\,\Phi^2\,,
\ee
where $ t $ is a positive real parameter. The variation principle leads to the following boundary term on $AdS_d$
\begin{equation}
\int_{AdS_d}d^{d}z\sqrt{g_d}\left( t \Phi - \frac{R}{z_0}\partial_{\perp}\Phi\right)\delta\Phi
\end{equation}
which has to vanish for any $\delta\Phi$. This gives the boundary condition
\begin{equation}
\left. t \Phi - \frac{R}{z_0}\partial_{\perp}\Phi\right|_{z_{\perp}=0}=0\;.
\end{equation}
This interpolates smoothly between the Neumann boundary condition, for which $ t =0$, and the Dirichlet boundary condition, for which $ t =\infty$.

The CFT data intrinsic to the CFT$_{d}$ stay fixed and equal to mean field theory for any $ t $. We are interested in the CFT data of the BCFT$_{d-1}$ as a function of $ t $. Most importantly, these are the scaling dimensions of the primary operators $\oHat_n$ appearing in the bulk-to-boundary OPE of $\phi$, as well as the corresponding bulk-to-boundary OPE coefficients. Since the deformation of the action is quadratic in $\Phi$, we can find a closed solution for general $ t $.

\subsection{Boundary Scaling Dimensions From Holography}
To work out the boundary spectrum, we need to solve for the wavefunction of the field in $hAdS_{d+1}$. We set $R=1$. It is convenient to use the following coordinates that foliates $AdS_{d+1}$ into $AdS_d$ slices \cite{Aharony:2003qf}
\begin{equation}
ds_{AdS_{d+1}}^2=\cosh^2r(e^{2w}d\vec{z}^2+dw^2)+dr^2\;.
\end{equation}
This set of coordinates is related to the Poincar\'e coordinates by
\begin{equation}
z_\perp=e^{-w}\tanh r\;,\quad z_0=e^{-w}\cosh^{-1}r\;.
\end{equation}
We then write an $AdS_{d+1}$ field as
\begin{equation}
\Phi=\sum_n\psi_n(r)\Phi_n(\vec{z},w)
\end{equation}
where $\Phi_n$ satisfies the $AdS_d$ equation of motion
\begin{equation}
\square_{AdS_d}\Phi_n+\widehat{M}_n^2\Phi_n=0\;,
\end{equation}
and is dual to a boundary operator $\oHat_n$ living in the BCFT \cite{Aharony:2003qf}. We would like to find the scaling dimensions $\dHat_n$ of $\oHat_n$.

From the equations of motion of $\Phi$ and $\Phi_n$ we find a second order differential equation for $\psi_n$
\begin{equation}\label{eqnpsi}
\partial_r^2\psi_n+d \tanh r \partial_r \psi_n +(\cosh r)^{-2} \widehat{M}_n^2\psi_n-M^2\psi_n=0\;.
\end{equation}
This differential equation is also supplemented by the following boundary conditions for $\psi_n$. First of all, we note that we can approach $AdS_d$ by taking $r\to0$ and keeping $w$ fixed. Recall that in the new coordinates
\begin{equation}
\partial_\perp\Phi=e^w(-\tanh r\psi_n\partial_w\Phi_n+\partial_r \psi \Phi_n)\;,
\end{equation}
and the first term becomes zero as $r\to0$. Therefore the boundary condition at $r=0$ is simply
\begin{equation}
\partial_r \psi_n- t \psi_n=0\;.
\end{equation}
The other boundary condition comes from studying the behavior at $r\to\infty$. Note that this has the effect of keeping $z_\perp$ and $\vec{z}$ finite, while sending $z_0\to 0$. From the expectation that the bulk field $\Phi=\sum_n\psi_n\Phi_n$ is dual to an operator with conformal dimension $\Df$ on $\partial AdS_{d+1}$, we expect that at large $r$
\begin{equation}\label{psinbcatinfty}
\psi_n(r)\sim (e^{r})^{-\Df}
\end{equation}
where 
\begin{equation}
\Df(\Df-d)=M^2\;.
\end{equation}
We now solve for $\psi_n$. It is useful to make a change of variables into $x\equiv e^r$\;, after which the equation (\ref{eqnpsi}) becomes
\begin{equation}\label{eqnpsib}
\frac{d \left(x^2-1\right) x \psi '(x)}{x^2+1}+x^2 \psi ''(x)+x \psi '(x)+  \left(\Df  (d-\Df)+\frac{4 \dHat_n x^2 (-d+\dHat_n+1)}{\left(x^2+1\right)^2}\right)\psi(x)=0\;.
\end{equation}
Here we have written the squared mass of $\psi_n$ as 
\begin{equation}
\widehat{M}_n^2=\dHat_n(\dHat_n-d+1)\;.
\end{equation}
The general solution to equation  (\ref{eqnpsib}) takes following form
\begin{equation}
\begin{split}
\psi_n(x)={}&C_1\frac{x^{\Df}}{(1+x^2)^{\dHat_n}}\, _2F_1\left(\frac{1}{2} (d-2 \dHat_n),\Df -\dHat_n;-\frac{d}{2}+\Df +1;-x^2\right)\\
{}&+C_2\frac{x^{d-\Df}}{(1+x^2)^{\dHat_n}}\, _2F_1\left(\frac{d}{2}-\dHat_n,d-\Df -\dHat_n;\frac{d}{2}-\Df +1;-x^2\right)\;.
\end{split}
\end{equation}
When $x\to\infty$, the boundary condition (\ref{psinbcatinfty}) fixes the ratio
\begin{equation}
\frac{C_1}{C_2}=-\frac{ \Gamma \left(\frac{d}{2}-\Df +1\right) \sin \left(\frac{1}{2} \pi  (d-2 \dHat_n)\right) \Gamma (\Df -\dHat_n) \Gamma (-d+\Df +\dHat_n+1)}{\pi  \Gamma \left(-\frac{d}{2}+\Df +1\right)}\;.
\end{equation}
Further imposing the condition that at $x=1$ ($r=0$)
\begin{equation}
x\partial_x \psi_n- t \psi_n=0\;,
\end{equation}
we find the following equation relating $ t $ and $\dHat_n$:
\begin{equation}\label{gDeltan}
 t =-\frac{2 \Gamma \left(\frac{\Df -\dHat_n+1}{2}\right) \Gamma \left(\frac{-d+\Df +\dHat_n+2}{2}\right)}{\Gamma \left(\frac{\Df -\dHat_n}{2}\right) \Gamma \left(\frac{-d+\Df +\dHat_n+1}{2}\right)}\;.
\end{equation}
This equation gives the spectrum of the boundary operators $\oHat_n$ as an implicit function of $ t $. When $ t =0$, the spectrum consists of the zeros of the RHS, which lie at $\dHat_n = \Df+2n$ with $n=0,1,\ldots$, agreeing with the spectrum of the Neumann boundary condition. When $ t =+\infty$, we need the poles of the RHS, which lie at $\dHat_n = \Df+2n+1$, this time agreeing with the Dirichlet spectrum. As $ t $ is varied from $0$ to $+\infty$, the spectrum smoothly interpolates between the Neumann and Dirichlet case.

\subsection{Coefficient function in the boundary channel at finite coupling}
Having found the boundary spectrum, we will now determine the boundary conformal block expansion of the two-point function of $\phi$ for general $ t $. We write the two-point function as follows
\be
\langle\phi(x_1)\phi(x_2)\rangle = \frac{1}{|2x_{\perp}|^{\Df}|2y_{\perp}|^{\Df}}\mathcal{G}_{ t }(z)\,,
\ee
where $z = \frac{1}{1+\xi}$. We write the boundary OPE as
\be
\mathcal{G}_{ t }(z) = \sum\limits_{n=0}^{\infty}\mu_n(t)g^b_{\dHat_n(t)}(z)\,.
\ee
In practice, we will determine the coefficient function $\widehat{I}_{\dHat}( t )$ of $\mathcal{G}_{ t }(z)$, as defined by the Euclidean inversion formula \eqref{eq:EIFb}. The boundary OPE then can be read off from the poles and residues of $\widehat{I}_{\dHat}(t)$ according to
\be
\widehat{I}_{\dHat}(t) \sim - \widehat{\kappa}(\dHat_n(t))^{-1} \frac{\mu_{n}(t)}{\dHat-\dHat_n(t)}\quad\textrm{as}\quad\dHat\rightarrow\dHat_n(t)\,,
\ee
First, let us consider the two-point functions for $ t =0$ and $ t =\infty$, which correspond to the Neumann and Dirichlet boundary conditions
\ba
\mathcal{G}_{ t =0}(z) &= \left(\mbox{$\frac{z}{1-z}$}\right)^{\Df}+z^{\Df}\\
\mathcal{G}_{ t =\infty}(z) &= \left(\mbox{$\frac{z}{1-z}$}\right)^{\Df}-z^{\Df}\,.
\ea
The scalar product with the partial waves can be evaluated with the result
\ba
\widehat{I}_{\dHat}(t=0) = 
\frac{4\pi
\Gamma \left(\frac{d}{2}-\Delta _{\phi }\right)}
{ \Gamma \left(\Delta _{\phi }\right)}
\frac{\Gamma \left(\Delta_{\phi }-\dHat \right)
\Gamma \left(\Delta _{\phi }-d+1+\dHat\right)}
{\Gamma \left(\frac{\dHat+1-\Delta _{\phi }}{2}\right)
\Gamma \left(\frac{d-\dHat -\Delta _{\phi }}{2}\right)
\Gamma \left(\frac{\Delta _{\phi }+1-\dHat}{2}\right)
\Gamma \left(\frac{\Delta _{\phi }-d+2+\dHat }{2}\right)}\\
\widehat{I}_{\dHat}(t=\infty) = 
\frac{4\pi
\Gamma \left(\frac{d}{2}-\Delta _{\phi }\right)}
{ \Gamma \left(\Delta _{\phi }\right)}
\frac{\Gamma \left(\Delta_{\phi }-\dHat \right)
\Gamma \left(\Delta _{\phi }-d+1+\dHat\right)}
{\Gamma \left(\frac{\dHat+2-\Delta _{\phi }}{2}\right)
\Gamma \left(\frac{d+1-\dHat -\Delta _{\phi }}{2}\right)
\Gamma \left(\frac{\Delta _{\phi }-\dHat}{2}\right)
\Gamma \left(\frac{\Delta _{\phi }-d+1+\dHat }{2}\right)}
\label{eq:IHatND}
\ea
$\widehat{I}_{\dHat}(0)$ and $\widehat{I}_{\dHat}(\infty)$ are shadow-symmetric as they should be. Now, let us consider $\widehat{I}_{\dHat}(t)$ in perturbation theory for small $ t $
\be
\widehat{I}_{\dHat}(t) = \sum\limits_{j=0}^{\infty}\widehat{I}^{(j)}_{\dHat} t ^j\,.
\ee
There is only one Witten diagram contributing to $\widehat{I}^{(j)}_{\dHat}$ for each $j$, involving $j$ integrated insertions of $\Phi^2$ on $AdS_d$ connected in a linear chain. Since the deformation is gaussian, the diagrams for $j\geq1$ form a geometric sequence
\be
\widehat{I}^{(j)}_{\dHat} = \left[ t  K(\dHat)\right]^{j-1} \widehat{I}^{(1)}_{\dHat}\,,
\ee
where
\be
K(\dHat)=
-\frac{\Gamma \left(\frac{\Df-\dHat }{2}\right) \Gamma \left(\frac{\Df-d+1+\dHat}{2}\right)}{2\Gamma\left(\frac{\Df+1-\dHat}{2}\right)\Gamma \left(\frac{\Df+2-d+\dHat }{2}\right)}\,.
\ee
We can now sum the geometric series
\be
\widehat{I}_{\dHat}(t) = \widehat{I}^{(0)}_{\dHat}+\frac{\widehat{I}^{(1)}_{\dHat}}{1- t  K(\dHat)}\,.
\ee
$\widehat{I}^{(0)}_{\dHat}$ and $\widehat{I}^{(1)}_{\dHat}$ can be fixed by using the known values at $t=0,\infty$ in \eqref{eq:IHatND}, the final result being
\be
\widehat{I}_{\dHat}(t) = \frac{\widehat{I}_{\dHat}(0) - t  K(\dHat) \widehat{I}_{\dHat}(\infty)}{1- t  K(\dHat)}\,.
\label{eq:cDefAn}
\ee
The primary operators exchanged in the boundary OPE correspond to poles of $\widehat{I}_{\dHat}(t)$. These come from the zeros of the denominator
\be
 t  K(\dHat) = 1\,,
 \label{eq:spectrum2}
\ee
in agreement with the earlier result \eqref{gDeltan}.

\subsection{Comparison with the Polyakov Bootstrap}
We can use the presented family of boundary conditions to test the consistency of the Polyakov sum rules. We will do so by expanding the boundary OPE data at small $t$ and check that the resulting sum rules are satisfied up to $O(t^2)$. Recall the form of the boundary and bulk OPEs along the deformation
\be
\mathcal{G}_t = \sum\limits_{n=0}^{\infty}\mu_n(t)g^b_{\dHat_n(t)} =g^B_0+
 \sum\limits_{N=0}^{\infty}\lambda_N(t)g^B_{\Delta_{N}}\,.
\ee
Crucially, in the bulk channel the spectrum $\Delta_N = 2\Delta_{\phi}+2N$ is independent of $t$, and the only $t$-dependence comes from the one-point functions in the presence of the boundary. On the other hand, in the boundary channel both scaling dimensions and OPE coefficients are nontrivial functions of $t$. We will write the perturbative expansion of the boundary OPE data as follows
\ba
\dHat_n(t) &= 2\Df+2n + \sum\limits_{j=1}^{\infty}\widehat{\gamma}_n^{(j)}t^j\\
\mu_n(t) &= \sum\limits_{j=0}^{\infty}\mu_n^{(j)}t^j\,.
\ea
It is not hard to use \eqref{eq:spectrum2} to find the anomalous dimensions up to $O(t^2)$:
\begin{align}
\widehat{\gamma}_n^{(1)} &= \frac{\Gamma \left(n+\frac{1}{2}\right) \Gamma \left(n+\Delta _{\phi }-\frac{d}{2}+\frac{1}{2}\right)}{\pi\,\Gamma(n+1) \Gamma \left(n+\Delta _{\phi }-\frac{d}{2}+1\right)}\\
\widehat{\gamma}_n^{(2)} &=
\frac{\Gamma \left(n+\frac{1}{2}\right)^2 \Gamma \left(n+\Delta _{\phi }-\frac{d}{2}+\frac{1}{2}\right)^2}{2\pi^2\,\Gamma(n+1)^2 \Gamma \left(n+\Delta_{\phi }-\frac{d}{2}+1\right)^2}
\left[H_{n-\frac{1}{2}}-H_n+H_{n+\Delta_{\phi}-\frac{d}{2} -\frac{1}{2}}
-H_{n+\Delta_{\phi}-\frac{d}{2}}
\right]\,,
\end{align}
where $H_z$ is the harmonic number. We will also need the OPE coefficients up to $O(t)$:
\ba
\mu^{(0)}_n &= 
\frac{\left(\Delta _{\phi }\right)_{2 n} \left(\Delta _{\phi }-\frac{d}{2}+1\right)_n}{2^{2 n-1}(2 n)! \left(\Delta _{\phi }+n-\frac{d}{2}+\frac{1}{2}\right)_n}\\
\mu^{(1)}_n &= \frac{1}{2}\partial_n(\mu^{(0)}_n\widehat{\gamma}_n^{(1)})\,.
\ea
where $(a)_b = \frac{\Gamma(a+b)}{\Gamma(a)}$ is the Pochhammer symbol. Correspondingly, the two-point function can be expanded in perturbation theory
\be
\mathcal{G}_t(z) = \sum\limits_{j=0}^{\infty}\mathcal{G}^{(j)}(z)t^j\,.
\ee
Since $\mathcal{G}_t(z)$ is a two-point function in a unitary theory, it must be bounded in the Regge limit, i.e. $\mathcal{G}_t(z) = O(z^{\Df})$ as $z\rightarrow\infty$. Furthermore, the perturbation $\Phi^2$ generating the deformation is a relevant operator on $AdS_d$. Therefore, we expect that also the individual terms $\mathcal{G}^{(j)}(z)$ are bounded in the Regge limit. This means that the sum rules resulting from functionals living in $\mathcal{V}^*$ apply to each $\mathcal{G}^{(j)}(z)$. We will focus on $\mathcal{G}^{(j)}(z)$ for $j=0,1,2$. Their boundary OPEs take the form
\ba
\mathcal{G}^{(0)}(z) &= \sum\limits_{n=0}^{\infty}\mu_n^{(0)} g^{b}_{2\Df+2n}(z)\\
\mathcal{G}^{(1)}(z) &= \sum\limits_{n=0}^{\infty}\left[\mu_n^{(1)} g^{b}_{2\Df+2n}(z)+
\mu_n^{(0)}\widehat{\gamma}^{(1)}_n\partial_{\dHat}g^{b}_{2\Df+2n}(z)\right]\\
\mathcal{G}^{(2)}(z) &= \sum\limits_{n=0}^{\infty}\left[\mu_n^{(2)} g^{b}_{2\Df+2n}(z)+
\left(\mu_n^{(0)}\widehat{\gamma}^{(2)}_n+\mu_n^{(1)}\widehat{\gamma}^{(1)}_n\right)\partial_{\dHat}g^{b}_{2\Df+2n}(z)+\right.\\
&\qquad\quad\,+\left.\frac{1}{2}\mu_n^{(0)}(\widehat{\gamma}^{(1)}_n)^2\partial^2_{\dHat}g^{b}_{2\Df+2n}(z)\right]\,.
\label{eq:gDefOPEs}
\ea
We can now test the BCFT Polyakov sum rules. We will focus on the sum rules coming from the functionals $\widetilde{\omega}_n$. Recall that these functionals have the following properties when acting on the bulk and boundary conformal blocks, and $\widehat{\Delta}$-derivatives of the boundary blocks
\ba
&\widetilde{\omega}_n(g^{B}_{\Delta_M}) = 0\\
&\widetilde{\omega}_n(g^{b}_{\widehat{\Delta}_m}) = 0\\
&\widetilde{\omega}_n(\partial_{\dHat}g^{b}_{\widehat{\Delta}_m})=\delta_{nm}\,.
\ea
We can not derive sum rules by directly applying $\widetilde{\omega}_n$ to the expressions in \eqref{eq:gDefOPEs}, since $\widetilde{\omega}_n$ live in $\mathcal{U}^*$, but not in $\mathcal{V}^{*}$. We can construct functionals $\widetilde{\omega}^{\textrm{s}}_n$ belonging to $\mathcal{V}^*$ by subtracting a suitable multiple of $\widetilde{\omega}_0$ from the remaining $\widetilde{\omega}_n$
\be
\widetilde{\omega}^{\textrm{s}}_n = \widetilde{\omega}_n - q_n\widetilde{\omega}_0\,.
\ee
The coefficients $q_n$ can be fixed as follows. Firstly, note that each $\widetilde{\omega}^{\textrm{s}}_n$ annihilates each $\mathcal{G}^{(j)}$. This is because we can expand the latter in the bulk channel, where we find only double-trace conformal blocks $g^{B}_{\Delta_N}$ for $j\geq1$, and with the addition of an identity conformal block for $j=0$. All these conformal blocks are annihilated by $\widetilde{\omega}^{\textrm{s}}_n$. Applying $\widetilde{\omega}^{\textrm{s}}_n$ to the OPE of $\mathcal{G}^{(0)}(z)$ shown on the first line of \eqref{eq:gDefOPEs} leads to the trivially correct equation $0=0$ since each $\widetilde{\omega}^{\textrm{s}}_n$ also annihilates each $g^{b}_{2\Df+2n}(z)$. Applying $\widetilde{\omega}^{\textrm{s}}_n$ to $\mathcal{G}^{(1)}(z)$ on the second line of \eqref{eq:gDefOPEs}, we see that the resulting sum rules are satisfied if and only if
\be
q_n = \frac{\mu_n^{(0)}\widehat{\gamma}^{(1)}_n}{\mu_0^{(0)}\widehat{\gamma}^{(1)}_0}\,,
\ee
which therefore fixes all $a_n$. This is of course just the constraint alluded to in the introduction that $\widetilde{\omega}^{\textrm{s}}_n$ must annihilate the boundary expansion of the contact diagram. A nontrivially check of the logic comes from applying $\widetilde{\omega}^{\textrm{s}}_n$ to the boundary OPE of $\mathcal{G}^{(2)}(z)$, which gives
\be
\left(\mu_n^{(0)}\widehat{\gamma}^{(2)}_n+\mu_n^{(1)}\widehat{\gamma}^{(1)}_n\right) -
q_n\left(\mu_0^{(0)}\widehat{\gamma}^{(2)}_0+\mu_0^{(1)}\widehat{\gamma}^{(1)}_0\right) + s_n - q_n s_0 = 0\,,
\label{eq:sumRuleTilde}
\ee
where we defined the following infinite sums
\be
s_n\equiv \frac{1}{2}\sum\limits_{m=0}^{\infty}
\mu_m^{(0)}(\widehat{\gamma}^{(1)}_m)^2
\widetilde{\omega}_n(\partial^2_{\dHat}g^{b}_{2\Df+2m})\,.
\ee
$\widetilde{\omega}_n(\partial^2_{\dHat}g^{b}_{2\Df+2m})$ can be computed since thanks to results of Section \ref{Wdiagrams}, we know $\widetilde{\omega}_n(g^{b}_{\dHat})$ for arbitrary $\dHat$. The infinite sum over $m$ can then be performed analytically in special cases and numerically in the generic case. We checked that the sum rules \eqref{eq:sumRuleTilde} are indeed satisfied, providing a check of the consistency of our proposal. We stress that the Polyakov sum rules can be used to fix OPE data even in cases where, unlike here, no other analytic solution is available.

\acknowledgments
We thank the California Institute of Technology for the great hospitality during the Bootstrap 2018 workshop where part of this work was written up, and the participants for helpful conversations and comments. The work of LR and XZ is supported in part by NSF Grant PHY-162062.

\appendix

\section{Regge Behavior}\label{Reggebehavior}
The discussion of the two-point function Regge limit is facilitated by introducing the $\rho$ coordinate which is related to $\xi$ via  \cite{Hogervorst:2013sma}
\begin{equation}
\xi=\frac{(1-\rho)^2}{4\rho}\;.
\end{equation}
This change of the variable maps the $\xi$-plane into the unit disc $|\rho|<1$. The physical regime of the cross ratio $\xi\in(0,\infty)$ corresponds to the interval $\rho\in(0,1)$ along the real axis where $\xi=0$ and $\xi=\infty$ are mapped $\rho=1$ and $\rho=0$ respectively. The point $\rho=-1$ ($\xi=-1$) is another special point of interest which lies outside of the physical regime. The analytic continuation of the two-point function from within the unit $\rho$-disc to $\rho=-1$ is  what we will refer to as the {\it Regge limit} for BCFT. In the following we will prove two general statements about BCFT two-point functions regarding their boundedness.

\vspace{2mm}
\noindent\textbf{\textit{Statement 1:}} the two-point function of different operators $\langle \mathcal{O}_1\mathcal{O}_2\rangle$ is bounded by the square root of the two-point functions $\langle \mathcal{O}_1\mathcal{O}_1\rangle$ or $\langle \mathcal{O}_2\mathcal{O}_2\rangle$ of identical operators . 
\vspace{2mm}

\noindent The proof of this statement follows simply from the application of the Cauthy-Schwarz inequality, and the unitarity of a physical two-point function
\begin{equation}
\begin{split}
|\mathcal{G}_{\langle O_1O_2\rangle}|{}&=|\sum_k \hat{a}_{1k}\hat{a}_{2k} g^b_{\widehat{\Delta}_k}(\xi)|\leq \sum_k |\hat{a}_{1k}\hat{a}_{2k}| |g^b_{\widehat{\Delta}_k}(\xi)|\\
{}&\leq \sqrt{\sum_k (\hat{a}_{1k}^2|g^b_{\widehat{\Delta}_k}(\xi)|)\sum_k (\hat{a}_{2k}^2|g^b_{\widehat{\Delta}_k}(\xi)|)}\\
{}&\leq \sqrt{|\mathcal{G}_{\langle O_1O_1\rangle}||\mathcal{G}_{\langle O_2O_2\rangle}|}\;.
\end{split}
\end{equation}

\vspace{2mm}
\noindent\textbf{\textit{Statement 2:}} the growth of a physical two-point function $\mathcal{G}(\rho)$ near the unit circle $|\rho|=1$ is bounded by its behavior at $\rho=1$, {\it i.e.}, the bulk channel limit. 
\vspace{2mm}

\noindent To present the proof, let us first point out a useful fact of the boundary channel conformal block. In the $\rho$ coordinate, we can write the boundary channel conformal block (\ref{bchanconfblock}) as
 \begin{equation}
\begin{split}
g^b_{\widehat{\Delta}}(\rho)={}&\left(\frac{4\rho}{(1-\rho)^2}\right)^{\widehat{\Delta}}{}_2F_1(\widehat{\Delta},\widehat{\Delta}-\frac{d}{2}+1;2\widehat{\Delta}+2-d;\frac{-4\rho}{(1-\rho)^2})\\
={}& (4\rho)^{\widehat{\Delta}}{}_2F_1(\widehat{\Delta},\frac{d-1}{2};\widehat{\Delta}-\frac{d}{2}+\frac{3}{2};\rho^2)
\end{split}
\end{equation}
where in the second line we have used a quadratic transformation.\footnote{The identity we used is
\begin{equation}
\, {}_2F_1(a,b;a-b+1;z)=\frac{\, {}_2F_1\left(a,a-b+\frac{1}{2};2 a-2 b+1;-\frac{4 \sqrt{z}}{\left(1-\sqrt{z}\right)^2}\right)}{\left(1-\sqrt{z}\right)^{2 a}}\;, \quad |z|<1\;.
\end{equation}} From the second expression we see that the boundary channel conformal block admits a power expansion in $\rho$ around the origin, and the expansion coefficients are positive for $\widehat{\Delta}>\frac{d-3}{2}$. Let us first look at a two-point function $\langle \mathcal{O}_1\mathcal{O}_1\rangle$ of identical operators, for which the boundary channel decomposition reads
\begin{equation}
\mathcal{G}_{\langle \mathcal{O}_1\mathcal{O}_1\rangle}=\sum_k \hat{a}_{1k}^2 g^b_{\widehat{\Delta}_k}(\rho)\;.
\end{equation}
We expand the RHS in powers of $\rho$ and assume that no operators with dimension $\widehat{\Delta}<\frac{d-3}{2}$ appear in the boundary channel OPE. From the positivity of $\hat{a}_{1k}^2$ and the expansion coefficients of the boundary channel conformal blocks, we conclude that $\mathcal{G}_{\langle \mathcal{O}_1\mathcal{O}_1\rangle}$ has a power expansion in $\rho$ at $\rho=0$ with {\it all positive} coefficients, to wit  
\begin{equation}
\mathcal{G}_{\langle \mathcal{O}_1\mathcal{O}_1\rangle}=\sum_{k}\sum_{n}\hat{c}_{kn}\rho^{\widehat{\Delta}_k+2n}\;, \quad \hat{c}_{kn}>0\;.
\end{equation}
This immediately implies the following inequality: on the circle $|\rho|=r$, $r<1$, the physical two-point function is bounded by its value at $\rho=r$ ({\it i.e.}, the intersection of the circle with the positive real axis)
\begin{equation}\label{boundOOr}
|\mathcal{G}_{\langle \mathcal{O}_1\mathcal{O}_1\rangle}|_{|\rho|=r}\leq \mathcal{G}_{\langle \mathcal{O}_1\mathcal{O}_1\rangle}\big|_{\rho=r}\;.
\end{equation}
The expansion in $\rho$ converges inside the unit disc $|\rho|<1$ but diverges at the boundary of the disc due to the exchange of the identity operator in the bulk channel ($\rho=1$). As $\rho\to1^-$ along the real axis, the two-point function diverges as $\delta r^{-2\Delta_1}$ where $\delta r=1-r$ is the difference of radii between the unit circle and $|\rho|=r$ circle. The inequality (\ref{boundOOr}) then implies the following bound on the Regge behavior 
\begin{equation}
\mathcal{G}_{\langle \mathcal{O}_1\mathcal{O}_1\rangle}(\rho)\lesssim (1+\rho)^{-2\Delta_1}\;, \quad \rho\to -1^+\;.
\end{equation}
Furthermore, when combined with our first statement, we have proved the following bound on the Regge behavior for a general physical two-point function 
\begin{equation}
\mathcal{G}_{\langle \mathcal{O}_1\mathcal{O}_2\rangle}(\rho)\lesssim (1+\rho)^{-(\Delta_1+\Delta_2)}\;, \quad \rho\to -1^+\;.
\end{equation}

As we can straightforwardly verify, each individual conformal block has the following Regge behavior 
\begin{equation}
g^B_\Delta(\rho)\sim (\rho+1)^{2-d}\;,\quad g^b_{\widehat{\Delta}}(\rho)\sim (\rho+1)^{2-d}\;,\quad \rho\to -1^+\;.
\end{equation}
Therefore, as long as $\Delta_{1,2}\geq (d-2)/2$, both the bulk channel conformal block and the boundary channel conformal block are bounded in the Regge limit
\begin{equation}
g^B_\Delta(\rho)\lesssim (1+\rho)^{-(\Delta_1+\Delta_2)}\;, \quad g^b_{\widehat{\Delta}}(\rho)\lesssim (1+\rho)^{-(\Delta_1+\Delta_2)}\;,\quad \rho\to-1^+\;.
\end{equation}


\section{Direct Channel Decomposition From the Spectral Representation}\label{appspectral}
In this appendix, we give the details of deriving the direct channel decomposition from the spectral representation of exchange Witten diagrams. This makes use of the Mellin representation formalism for BCFT correlators \cite{Rastelli:2017ecj}.\footnote{See also \cite{Goncalves:2018fwx} for the Mellin formalism for defect CFTs.}

We start with the bulk exchange Witten diagram (\ref{Wbulk}). In \cite{Rastelli:2017ecj}, the bulk exchange Witten diagram was shown to have the following spectral representation
\begin{equation}\small
\begin{split}
\mathcal{W}^{bulk}={}&\xi^{-\frac{\Delta_1+\Delta_2}{2}}\int_{-i\infty}^{i\infty}\frac{dc}{2\pi i} \int_{\mathcal{C}}\frac{d\tau}{2\pi i}  \xi^{\tau}\bigg\{4^{\tau} \frac{\Gamma(\tau)\Gamma(\tau+\frac{\Delta_1-\Delta_2}{2})\Gamma(\tau+\frac{\Delta_2-\Delta_1}{2})\Gamma(\frac{h+c}{2}-\tau)\Gamma(\frac{h-c}{2}-\tau)}{\Gamma(\frac{1}{2}-\tau)\Gamma(2\tau)}\bigg\}\\
{}&\times \frac{1}{(\Delta-h)^2-c^2}\frac{\Gamma(\frac{\Delta_1+\Delta_2-h+c}{2})\Gamma(\frac{1+c-h}{2})\Gamma(\frac{\Delta_1+\Delta_2-h-c}{2})\Gamma(\frac{1-c-h}{2})}{8\pi^{\frac{1}{2}-h}\Gamma(\Delta_1)\Gamma(\Delta_2)\Gamma(c)\Gamma(-c)}
\end{split}
\end{equation}
where $h=\frac{d}{2}$ and the contour $\mathcal{C}$ for $\tau$ is parallel to the imaginary axis. One can show that the term in the brackets can be rewritten as
\begin{equation}
f^{bulk}(c)M[g^B_{h+c}](\tau)+f^{bulk}(-c)M[g^B_{h-c}](\tau)
\end{equation}
with
\begin{equation}\small
f^{bulk}(c)=\frac{2 \sqrt{\pi } \Gamma (-c) \cos \left(\frac{1}{2} \pi  (c+h)\right) \csc \left(\frac{1}{2} \pi  (c-\Delta_1+\Delta_2+h)\right) \Gamma \left(\frac{1}{2} (c+h+\Delta_1-\Delta_2)\right)}{\Gamma \left(\frac{1}{2} (-c-h+\Delta_1-\Delta_2+2)\right)}\;.
\end{equation}
Here $M[g^B_{\Delta}](\tau)$ is the {\it reduced Mellin amplitude}\footnote{See \cite{Rastelli:2017ecj} for definitions and details of the Mellin representation formalism of BCFT. } for a bulk channel conformal block
\begin{equation}\label{gBMellin}
g^B_{\Delta}(\xi)=\xi^{-\frac{\Delta_1+\Delta_2}{2}}\int_{\mathcal{C}} \frac{d\tau}{2\pi i} \xi^{\tau} M[g^B_{\Delta}](\tau)
\end{equation} 
\begin{equation}
M[g^B_{\Delta}](\tau)=\frac{\Gamma(\tau+\frac{\Delta_1-\Delta_2}{2})\Gamma(\tau+\frac{\Delta_2-\Delta_1}{2})\Gamma(-\tau+\frac{\Delta}{2})}{\Gamma(\tau+\frac{\Delta}{2}-h+1)} \frac{\Gamma(\Delta-h+1)}{\Gamma(\frac{\Delta+\Delta_1-\Delta_2}{2})\Gamma(\frac{\Delta+\Delta_2-\Delta_1}{2})}\;.
\end{equation}
Therefore we can write 
\begin{equation}
\mathcal{W}^{bulk}(\xi)=\xi^{-\frac{\Delta_1+\Delta_2}{2}}\int_{-i\infty}^{i\infty}\frac{dc}{2\pi i}  \int_{\mathcal{C}}\frac{d\tau}{2\pi i}  \xi^{\tau}\bigg(\mathcal{F}^{bulk}(c) M[g^B_{h+c}](\tau)+\mathcal{F}^{bulk}(-c)M[g^B_{h-c}](\tau)\bigg)
\end{equation}
where
\begin{equation}
\mathcal{F}^{bulk}(c)=f^{bulk}(c) \times \frac{1}{(\Delta-h)^2-c^2}\frac{\Gamma(\frac{\Delta_1+\Delta_2-h+c}{2})\Gamma(\frac{1+c-h}{2})\Gamma(\frac{\Delta_1+\Delta_2-h-c}{2})\Gamma(\frac{1-c-h}{2})}{8\pi^{\frac{1}{2}-h}\Gamma(\Delta_1)\Gamma(\Delta_2)\Gamma(c)\Gamma(-c)}\;.
\end{equation}
We can perform the $\tau$ integral using (\ref{gBMellin}), and the result is written as a spectral representation with respect to the bulk channel conformal blocks
\begin{equation}
\mathcal{W}^{bulk}(\xi)=\int_{-i\infty}^{i\infty}\frac{dc}{2\pi i}  \mathcal{F}^{bulk}(c)g^B_{h+c}(\xi)+\mathcal{F}^{bulk}(-c)g^B_{h-c}(\xi)=\int_{-i\infty}^{i\infty}\frac{dc}{2\pi i}  2\mathcal{F}^{bulk}(c)g^B_{h+c}(\xi)
\end{equation}
where in the second equality we have used the $c\to-c$ shadow symmetry. Closing the contour to the right and pick up the poles we get the OPE coefficients. The residue at $c=-h+\Delta$ gives the OPE coefficient of the single-trace operator. The residues at $c=-h+\Delta_1+\Delta_2+2N$, with non negative integers $N$ give the OPE coefficients for the double-trace operators.

The analysis for the boundary exchange Witten diagram is completely analogous. The spectral representation in Mellin space was given in \cite{Rastelli:2017ecj}
\begin{equation}\small
\begin{split}
\mathcal{W}^{boundary}{}&=\int_{-i\infty}^{i\infty}\frac{dc}{2\pi i}\int_{\mathcal{C}}\frac{d\tau}{2\pi i} \xi^{-\tau}\bigg\{\frac{4^{-\tau }\Gamma (\tau ) \Gamma \left(h-\tau -\frac{1}{2}\right)\Gamma \left(h+c -\tau -\frac{1}{2}\right) \Gamma \left(h-c -\tau -\frac{1}{2}\right)}{\Gamma (2 h-2 \tau -1)}\bigg\}\\
{}&\times \frac{\pi ^{\frac{1}{2} (2 h-1)} 2^{\Delta_1+\Delta_2-3} \Gamma \left(\frac{c-h+\Delta_1+\frac{1}{2}}{2}\right) \Gamma \left(\frac{- c- h+ \Delta_1+\frac{1}{2}}{2}\right) \Gamma \left(\frac{c-h+\Delta_2+\frac{1}{2}}{2}\right) \Gamma \left(\frac{-c-h+ \Delta_2+\frac{1}{2}}{2}\right)}{\Gamma (-c) \Gamma (c) \Gamma (\Delta_1) \Gamma (\Delta_2) \left(\left({\widehat{\Delta}} -h+\frac{1}{2}\right)^2-c^2\right)}\;.
\end{split}
\end{equation}
The term in the brackets can be rewritten in the following form 
\begin{equation}
f^{boundary}(c)M[g^b_{h-\frac{1}{2}+c}](\tau)+f^{boundary}(-c)M[g^b_{h-\frac{1}{2}-c}](\tau)
\end{equation}
where 
\begin{equation}
f^{boundary}(c)=-\frac{\pi  2^{-2 c-2 h+1} \csc (\pi  c) \Gamma \left(c+h-\frac{1}{2}\right)}{\Gamma (c+1)}\;,
\end{equation}
and $M[g^b_{\widehat{\Delta}}](\tau)$ is the reduced Mellin amplitude for a boundary channel conformal block
\begin{equation}
g^b_{\widehat{\Delta}}=\int_{\mathcal{C}}\frac{d\tau}{2\pi i} \xi^{-\tau}M[g^b_{\widehat{\Delta}}](\tau)\;,
\end{equation}
\begin{equation}
M[g^b_{\widehat{\Delta}}](\tau)=\frac{\Gamma (\tau ) \Gamma ({\widehat{\Delta}} -\tau ) \Gamma (-2 h+2 {\widehat{\Delta}} +2) \Gamma (-h+\tau +1)}{\Gamma ({\widehat{\Delta}} ) \Gamma (-h+{\widehat{\Delta}} +1) \Gamma (-2 h+{\widehat{\Delta}} +\tau +2)}\;.
\end{equation}
The boundary exchange Witten diagram can therefore be written as
\begin{equation}
\begin{split}
\mathcal{W}^{boundary}(\xi)={}&\int_{-i\infty}^{i\infty}\frac{dc}{2\pi i} \,2 \mathcal{F}^{boundary}(c)g^b_{h-\frac{1}{2}+c}(\xi)
\end{split}
\end{equation}
where 
\begin{equation}
\begin{split}
\mathcal{F}^{boundary}(c)={}&f^{boundary}(c)\times\frac{\pi ^{\frac{1}{2} (2 h-1)} 2^{\Delta_1+\Delta_2-3} }{\Gamma (\Delta_1) \Gamma (\Delta_2) \left(\left({\widehat{\Delta}} -h+\frac{1}{2}\right)^2-c^2\right)}\\
{}&\times \frac{\Gamma \left(\frac{c-h+\Delta_1+\frac{1}{2}}{2}\right) \Gamma \left(\frac{- c- h+ \Delta_1+\frac{1}{2}}{2}\right) \Gamma \left(\frac{c-h+\Delta_2+\frac{1}{2}}{2}\right) \Gamma \left(\frac{-c-h+ \Delta_2+\frac{1}{2}}{2}\right)}{\Gamma (-c) \Gamma (c) }\;.
\end{split}
\end{equation}
Closing the contour to the right, the residue at $c=-h+\frac{1}{2}+{\widehat{\Delta}}$ gives the OPE coefficient for the  single-trace operator with dimension $\widehat{\Delta}$, and the residues at $c=-h+\frac{1}{2}+\Delta_i+2m$ give the rest boundary channel OPE coefficients for  single-trace operators with dimensions $\Delta_i+2m$.


\section{Computing the Seed Coefficients}\label{Secseed}
In this section we compute the seed coefficients needed for the crossed channel decompositions of exchange Witten diagrams. The main strategy is to use and generalize the method of \cite{DHoker:1999aa} to write an exchange Witten diagram as an infinite sum of contact Witten diagrams. The seed OPE coefficients are extracted from this representation by taking certain limit and then performing resummation.

\subsection{Bulk exchange Witten diagrams}
We start by reviewing the method in \cite{DHoker:1999aa} of computing the three-point integral that appear in the bulk channel exchange Witten diagrams
\begin{equation}
I^{bulk}(x_1,x_2;w)=\int_{AdS_{d+1}}\frac{d^{d+1}z}{z_0^{d+1}} {G}_{BB}^\Delta(w,z){G}_{B\partial}^{\Delta_1}(z,x_1){G}_{B\partial}^{\Delta_2}(z,x_2)
\end{equation}
It is convenient to perform a translation such that
\begin{equation}
x_1\to 0\;,\;\;\;x_2\to x_{21}\equiv x_2-x_1\;.
\end{equation}
This is followed by a conformal inversion,
\begin{equation}
x_{12}'=\frac{x_{12}}{(x_{12})^2}\;,\;\;\; z'=\frac{z}{z^2}\;,\;\;\; w'=\frac{w}{w^2}.
\end{equation}
After these transformations the integral becomes,
\begin{equation}
I^{bulk}(x_1,x_2;w)=(x_{12})^{-2\Delta_2}J(w'-x_{12}')
\end{equation}
where
\begin{equation}
J(w)=\int\frac{d^{d+1}z}{z_0^{d+1}}\;G_{BB}^{\Delta}(u)\;z_0^{\Delta_1}\left(\frac{z_0}{z^2}\right)^{\Delta_2}\;.
\end{equation}
The scaling behavior of $J(w)$ under $w\to\lambda w$ together with the Poincar\'e symmetry dictates that $J(w)$ takes the form
\begin{equation}
J(w)=w_0^{\Delta_1-\Delta_2} f(t)
\end{equation}
where 
\begin{equation}
t=\frac{w_0^2}{w^2}\;,
\end{equation}
and the physical region of $t$ is $[0,1]$. The function $f(t)$ is constrained by the following differential equation,
\begin{equation}\label{bulkdiffeqn}
4t^2(t-1)f''+4t[(\Delta_1-\Delta_2+1)t-\Delta_1+\Delta_2+\frac{d}{2}-1]f'+[(\Delta_1-\Delta_2)(d-\Delta_1+\Delta_2)+M^2]f=t^{\Delta_2}
\end{equation}
where $M^2=\Delta(\Delta-d)$. This equation comes from acting with the equation of motion of the field in the bulk-to-bulk propagator. The function $f$ is further subject to two boundary conditions: 
\begin{enumerate}
\item[1)] From the OPE limit, we know that $f(t)$ should behave like
\begin{equation}
f(t)\sim t^{\frac{\Delta-\Delta_1+\Delta_2}{2}}\;,\quad t\to 0\;.
\end{equation}
\item[2)] 
From its integral definition, $f(t)$ has to be smooth at $t=1$ \cite{DHoker:1998ecp}. 
\end{enumerate}

Let us now look at the solutions to this equation. When $\Delta=\Delta_1+\Delta_2-2m$ with $m\in \mathbb{Z}$ and $m>0$ it is easy to find a polynomial special solutions for $f(z)$. This solution was first given in \cite{DHoker:1999aa} and takes the following form
\begin{equation}\label{bulkpolysol}
f(t)=\sum_{k=k_{\rm min}}^{k_{\rm max}} a_k t^k
\end{equation} 
with
\begin{equation}\label{bulkpolycoe}
\begin{split}
{}&k_{\rm min}=(\Delta-\Delta_{1}+\Delta_2)/2\;,\;\;\;\;\; k_{\rm max}=\Delta_2-1\;,\\
{}&a_{k-1}=a_k \frac{(k-\frac{\Delta}{2}+\frac{\Delta_{1}-\Delta_2}{2})(k-\frac{d}{2}+\frac{\Delta}{2}+\frac{\Delta_{1}-\Delta_2}{2})}{(k-1)(k-1-\Delta_{1}+\Delta_2)}\;,\\
{}& a_{\Delta_2-1}=\frac{1}{4(\Delta_1-1)(\Delta_2-1)}\;.
\end{split}
\end{equation}
The equation (\ref{bulkdiffeqn}) also admits the follwoing homogeneous solutions 
\begin{equation}\small
f_1(t)=t^{\frac{1}{2} (\Delta -\Delta_1+\Delta_2)} \, _2F_1\left(\frac{1}{2} (\Delta -\Delta_1+\Delta_2),\frac{1}{2} (\Delta +\Delta_1-\Delta_2);-\frac{d}{2}+\Delta +1;t\right)\;,
\end{equation}
and
\begin{equation}\small
f_2(t)=\, _2F_1\left(\frac{1}{2} (\Delta -\Delta_1+\Delta_2),\frac{1}{2} (d-\Delta -\Delta_1+\Delta_2);\frac{d}{2};1-\frac{1}{t}\right)\;.
\end{equation}
However, neither solutions can be added to the special solution as they would spoil the boundary conditions. More precisely, this is because $f_1(t)$ has a branch point at $t=1$ while $f_2(t)$ is smooth; while $f_2(t)$ has the wrong asymptotic behavior for $t\to 0$ which grows like $t^{\frac{d-\Delta-\Delta_1+\Delta_2}{2}}$. 

When the truncation condition $\Delta=\Delta_1+\Delta2-2m$ is not satisfied, one can still get a special solution from (\ref{bulkpolysol}). The series  (\ref{bulkpolycoe}) now does not truncate, and can be written in terms of a ${}_3F_2$ function. However it has the wrong boundary behavior. It implies that the solution should be accompanied with homogenous solutions. By studying its behavior near $t=0$ and $t=1$, we find the correct combination of solutions is
\begin{equation}\small
f(t)=C_s^{bulk} t^{\Delta_2} {}_3F_2\left(1,\Delta_1,\Delta_2;-\frac{\Delta }{2}+\frac{\Delta_1}{2}+\frac{\Delta_2}{2}+1,-\frac{d}{2}+\frac{\Delta }{2}+\frac{\Delta_1}{2}+\frac{\Delta_2}{2}+1;t\right)+C_{h,1}^{bulk}f_1(t)
\end{equation}
where
\begin{equation}\small
\begin{split}
C_s^{bulk}={}&-\frac{1 \, }{(-\Delta +\Delta_1+\Delta_2) (-d+\Delta +\Delta_1+\Delta_2)}\;,\\
C_{h1}^{bulk}={}&\frac{\Gamma \left(\frac{1}{2} (\Delta +\Delta_1-\Delta_2)\right) \Gamma \left(\frac{1}{2} (\Delta -\Delta_1+\Delta_2)\right) \Gamma \left(\frac{1}{2} (-\Delta +\Delta_1+\Delta_2)\right) \Gamma \left(\frac{1}{2} (-d+\Delta +\Delta_1+\Delta_2)\right)}{4 \Gamma (\Delta_1) \Gamma (\Delta_2) \Gamma \left(-\frac{d}{2}+\Delta +1\right)}\;.
\end{split}
\end{equation}
It will also appear to be useful to write the above solution as power series
\begin{equation}
f(t)=t^{\Delta_2}\sum_{i=0}^\infty P_i\,t^i+t^{\frac{\Delta-\Delta_1+\Delta_2}{2}}\sum_{i=0}^\infty Q_i\,t^i
\end{equation}
where
\begin{equation}\small
P_i=\frac{(\Delta_1)_i (\Delta_2)_i}{(\Delta -\Delta_1-\Delta_2) (-d+\Delta +\Delta_1+\Delta_2) \left(\frac{-\Delta +\Delta_1+\Delta_2+2}{2}\right)_i \left(\frac{-d+\Delta +\Delta_1+\Delta_2+2}{2}\right)_i}
\end{equation}
and
\begin{equation}\small
\begin{split}
Q_i={}&\frac{(-1)^i \Gamma \left(\frac{d-2 i-2\Delta}{2}\right)\sin \left(\frac{\pi  (d-2 \Delta )}{2} \right)\Gamma \left(\frac{-d+\Delta +\Delta_1+\Delta_2}{2}\right)}{4 \pi  \Gamma (i+1)\Gamma (\Delta_1) \Gamma (\Delta_2)}\\
{}&\times \frac{\Gamma \left(\frac{\Delta -\Delta_1+\Delta_2}{2}\right) \Gamma \left(\frac{\Delta +\Delta_1-\Delta_2}{2}\right) \Gamma \left(\frac{-\Delta +\Delta_1+\Delta_2}{2}\right)\Gamma \left(\frac{-\Delta +\Delta_1-\Delta_2+2}{2}\right)  \Gamma \left(\frac{-\Delta -\Delta_1+\Delta_2+2}{2}\right) }{\Gamma \left(\frac{-\Delta +\Delta_1-\Delta_2-2 i+2}{2}\right)\Gamma \left(\frac{-\Delta -\Delta_1+\Delta_2-2 i+2}{2}\right)}\;.
\end{split}
\end{equation}
After obtaining this solution, we can undo the inversion and translation and the upshot is that each power $t^a$ becomes a contact vertex at $w$,
\begin{equation}
(x_1-x_2)^{2(a-\Delta_2)}G_{B\partial}^{a+\Delta_1-\Delta_2}(x_1,w)\;G_{B\partial}^{a}(x_2,w)\;.
\end{equation}
Therefore the bulk exchange Witten diagram can be written as an infinite sum of contact Witten diagrams
\begin{equation}\small\label{WbulkconB}
W^{bulk}=\sum_{i=0}^\infty (x_1-x_2)^{2i} P_i\, W^{contact}_{\Delta_1+i,\Delta_2+i}+\sum_{i=0}^\infty (x_1-x_2)^{\Delta-\Delta_1-\Delta_2+2i} Q_i\, W^{contact}_{\frac{\Delta+\Delta_1-\Delta_2}{2}+i,\frac{\Delta-\Delta_1+\Delta_2}{2}+i}
\end{equation}
where we have labelled the the contact Witten diagram by two external dimensions. Written in terms of the cross ratio,
\begin{equation}\label{Wbulkcon}
\mathcal{W}^{bulk}(\xi)=\sum_{i=0}^\infty P_i\, \xi^i  \mathcal{W}^{contact}_{\Delta_1+i,\Delta_2+i}(\xi)+\sum_{i=0}^\infty Q_i\, \xi^{\frac{\Delta-\Delta_1-\Delta_2}{2}+i}  \mathcal{W}^{contact}_{\frac{\Delta+\Delta_1-\Delta_2}{2}+i,\frac{\Delta-\Delta_1+\Delta_2}{2}+i}(\xi)
\end{equation}
Similarly, in replacing $x_2$ by $\bar{x}_2$, and use the fact that
\begin{equation}
W^{contact}(x_1,x_2)=W^{contact}(x_1,\bar{x}_2)\;,
\end{equation}
we can also write the mirror exchange Witten diagram as an infinite sum of contact Witten diagrams
\begin{equation}\small\label{WmirrorbulkconB}
\bar{W}^{bulk}=\sum_{i=0}^\infty (x_1-\bar{x}_2)^{2i} P_i\, W^{contact}_{\Delta_1+i,\Delta_2+i}+\sum_{i=0}^\infty (x_1-\bar{x}_2)^{\Delta-\Delta_1-\Delta_2+2i} Q_i\, W^{contact}_{\frac{\Delta+\Delta_1-\Delta_2}{2}+i,\frac{\Delta-\Delta_1+\Delta_2}{2}+i}\;.
\end{equation}
In terms of the cross ratio, it reads
\begin{equation}\label{Wmirrorbulkcon}
\bar{\mathcal{W}}^{bulk}(\xi)=\sum_{i=0}^\infty P_i\, (\xi+1)^i  \mathcal{W}^{contact}_{\Delta_1+i,\Delta_2+i}(\xi)+\sum_{i=0}^\infty Q_i\, (\xi+1)^{\frac{\Delta-\Delta_1-\Delta_2+i}{2}}  \mathcal{W}^{contact}_{\frac{\Delta+\Delta_1-\Delta_2}{2}+i,\frac{\Delta-\Delta_1+\Delta_2}{2}+i}(\xi)\;.
\end{equation}

\subsection{Boundary exchange Witten diagrams}
Let us now use the same strategy for the boundary exchange Witten diagrams. We focus on the two-point integral
\begin{equation}
I^{boundary}(x_1,w_2)=\int_{AdS_d}\frac{d^dw_1}{w_{10}^d}{G}_{BB}^{\widehat{\Delta}}(w_1,w_2){G}_{B\partial}^{\Delta_1}(w_1,x_1)
\end{equation}
This integral has $AdS_{d}$ isometry and should depend on a single variable $t$ invariant under the scaling $w_2\to \lambda w_2$, $x_1\to \lambda x_1$
\begin{equation}\label{tboundary}
t\equiv\frac{w_{2,0}^2+x_{1,\perp}^2+(\vec{w}_2-\vec{x}_1)^2}{w_{2,0} x_{1,\perp}}\;.
\end{equation}
and its physical region of is $[2,\infty)$. The function $I^{boundary}(x_1,w_2)$ then takes the form
\begin{equation}
I^{boundary}(x_1,w_2)=x_{1,\perp}^{-\Delta_1}p(t)\;.
\end{equation}
To work out $f(t)$, we use the equation of motion for the bulk-to-bulk propagator inside $AdS_d$. It leads to the following equation
\begin{equation}
-(t^2-4)p''(t)-d tp'(t)+\widehat{M}^2p(t)=t^{-\Delta_1}
\end{equation}
where $\widehat{M}^2={\widehat{\Delta}}({\widehat{\Delta}}-(d-1))$. This second order differential equation is also supplemented by the following two boundary conditions: 
\begin{enumerate}
\item[1)] as $t\to\infty$, $p(t)\sim t^{-\Delta}$;
\item[2)] at $t=2$, the function $p(t)$ is smooth. 
\end{enumerate}

We now consider the solutions to this equation. When $\Delta_1={\widehat{\Delta}}-2m$ with $m\in\mathbb{Z}$, $m>0$, one finds a polynomial solution \cite{Rastelli:2017ecj}
\begin{equation}\label{boundarypolysol}
p(t)=\sum_{k_{\rm min}}^{k_{\rm max}}b_k t^k \, ,
\end{equation}
where
\begin{equation}\label{boundarypolycoe}
\begin{split}
{}& b_{k+2}=\frac{(k+{\widehat{\Delta}})(k-({\widehat{\Delta}}-(d-1)))}{4(k+1)(k+2)}b_k\;,\\
{}& {k_{\rm min}}=-\Delta_1+2\;,\\
{}& {k_{\rm max}}= -{\widehat{\Delta}}\;,\\
{}& b_{k_{\rm min}}=\frac{1}{4(-\Delta_1+2)(-\Delta_1+1)}\;.
\end{split}
\end{equation}
It is easy to check that further adding the homogeneous solutions
\begin{equation}\small
p_1(t)= t^{-{\widehat{\Delta}} }  {}_2F_1\left(\frac{{\widehat{\Delta}} }{2},\frac{{\widehat{\Delta}} +1}{2};-\frac{d}{2}+{\widehat{\Delta}} +\frac{3}{2};\frac{4}{t^2}\right)\;,
\end{equation}
and
\begin{equation}\small
p_2(t)= t^{-d+{\widehat{\Delta}} +1} {}_2F_1\left(\frac{1}{2} (d-{\widehat{\Delta}} -1),\frac{d-{\widehat{\Delta}} }{2};\frac{1}{2} (d-2 {\widehat{\Delta}} +1);\frac{4}{t^2}\right)\;,
\end{equation}
will spoil the boundary behavior. When ${\widehat{\Delta}}$ and $\Delta_1$ takes generic values, the series (\ref{boundarypolysol}) no longer terminates. But the special solution we get from the resumed series does not satisfy the boundary conditions. Instead, we should use the following solution for $p(t)$
\begin{equation}\small
\begin{split}
p(t)={}&-\frac{t^{-\Delta_1} \, _3F_2\left(1,\frac{\Delta_1}{2}+\frac{1}{2},\frac{\Delta_1}{2};-\frac{{\widehat{\Delta}} }{2}+\frac{\Delta_1}{2}+1,-\frac{d}{2}+\frac{{\widehat{\Delta}} }{2}+\frac{\Delta_1}{2}+\frac{3}{2};\frac{4}{t^2}\right)}{(\Delta_1-{\widehat{\Delta}} ) (-d+{\widehat{\Delta}} +\Delta_1+1)}\\
{}&+\frac{\left(\pi  \csc (\pi  {\widehat{\Delta}} ) \sin (\pi  \Delta_1) \Gamma (1-\Delta_1) \Gamma \left(\frac{\Delta_1}{2}-\frac{{\widehat{\Delta}} }{2}\right) \csc \left(\frac{1}{2} \pi  (-d+{\widehat{\Delta}} +\Delta_1+1)\right) \right)}{4 \Gamma (1-{\widehat{\Delta}} ) \Gamma \left(-\frac{d}{2}+{\widehat{\Delta}} +\frac{3}{2}\right) \Gamma \left(\frac{d}{2}-\frac{{\widehat{\Delta}} }{2}-\frac{\Delta_1}{2}+\frac{1}{2}\right)}p_1(t)\;.
\end{split}
\end{equation}
It is also convenient to write this solution as power series in $1/t$
\begin{equation}
p(t)=t^{-\Delta_1}\sum_{i=0}^\infty R_i\, t^{-2i}+t^{-{\widehat{\Delta}}}\sum_{i=0}^\infty S_i t^{-2i}
\end{equation}
where
\begin{equation}\small
R_i=\frac{(\Delta_1)_{2 i}}{2 (\Delta -\Delta_1) \left(\frac{1}{2} (-{\widehat{\Delta}} +\Delta_1+2)\right)_i \left(\frac{1}{2} (-d+{\widehat{\Delta}} +\Delta_1+1)\right)_{i+1}}\;,
\end{equation}
\begin{equation}\small
S_i=\frac{(-1)^{i+1} \sin (\pi  \Delta_1) \Gamma (1-\Delta_1) \cos \left(\frac{\pi  (d-2 {\widehat{\Delta}} )}{2}\right) \Gamma \left(\frac{\Delta_1-{\widehat{\Delta}} }{2}\right) \Gamma \left(\frac{d-2 i-2 {\widehat{\Delta}} -1}{2}\right)}{4 i! \sin (\pi  {\widehat{\Delta}} )\sin \left(\frac{\pi  (-d+{\widehat{\Delta}} +\Delta_1+1)}{2}\right)\Gamma (-2 i-{\widehat{\Delta}} +1) \Gamma \left(\frac{d-{\widehat{\Delta}} -\Delta_1+1}{2}\right)}\;.
\end{equation}
It is not difficult to find from the definition that each $t^{-a}$ corresponds to a contact vertex
\begin{equation}
x_{1,\perp}^{-\Delta_1+a}\; G_{B\partial}^{a}(x_1,w_2)\;G_{B\partial}^{\Delta_2}(x_2,w_2)\;.
\end{equation}
Therefore we can write the boundary exchange Witten diagram as an infinite sum of contact Witten diagrams
\begin{equation}
W^{boundary}=\sum_{i=0}^\infty (x_{1,\perp})^{2i} R_i\, W^{contact}_{\Delta_1+2i,\Delta_2}+\sum_{i=0}^\infty (x_{1,\perp})^{{\widehat{\Delta}}-\Delta_1+2i} S_i\, W^{contact}_{{\widehat{\Delta}}+2i,\Delta_2}\;.
\end{equation}
Written in terms of the cross ratio, we have
\begin{equation}\label{Wbdrcon}
\mathcal{W}^{boundary}(\xi)=\sum_{i=0}^\infty 2^{-2i} R_i\, \mathcal{W}^{contact}_{\Delta_1+2i,\Delta_2}(\xi)+\sum_{i=0}^\infty 2^{-{\widehat{\Delta}}+\Delta_1-2i} S_i\, \mathcal{W}^{contact}_{{\widehat{\Delta}}+2i,\Delta_2}(\xi)\;.
\end{equation}

\subsection{Extracting seed coefficients}
Building on our previous results (\ref{Wbulkcon}), (\ref{Wmirrorbulkcon}) and (\ref{Wbdrcon}), we now extract the various OPE coefficients. 

We start with the bulk exchange Witten diagram (\ref{Wbulkcon}). To compute the seed OPE coefficient $\hat{A}^{B,(1)}_0$ for boundary channel decomposition we note that $\mathcal{W}^{contact}$ has the following expansion around $\xi=\infty$
\begin{equation}\label{Wbulkcontactaroundinfty}\small
\begin{split}
{}& \mathcal{W}^{contact}_{\Delta_1+i,\Delta_2+i}(\xi)= c_1\,\xi^{-\Delta_1-i}\left(1+\mathcal{O}(\xi^{-1})\right)+c_2\,\xi^{-\Delta_2-i}\left(1+\mathcal{O}(\xi^{-1})\right)\;,\\
{}& \mathcal{W}^{contact}_{\frac{\Delta+\Delta_1-\Delta_2}{2}+i,\frac{\Delta-\Delta_1+\Delta_2}{2}+i}(\xi)=c'_1\,\xi^{-\frac{\Delta+\Delta_1-\Delta_2}{2}-i}\left(1+\mathcal{O}(\xi^{-1})\right)+c'_2\,\xi^{-\frac{\Delta-\Delta_1+\Delta_2}{2}-i}\left(1+\mathcal{O}(\xi^{-1})\right)\;.
 \end{split}
\end{equation}
It is easy to see that $\xi^{-\Delta_1-i}\left(1+\mathcal{O}(\xi^{-1})\right)$ and $\xi^{-\frac{\Delta+\Delta_1-\Delta_2}{2}-i}\left(1+\mathcal{O}(\xi^{-1})\right)$ contribute to $\xi^{-\Delta_1}(1+\mathcal{O}(\xi^{-1}))$ in $\mathcal{W}^{bulk}$, while $\xi^{-\Delta_2-i}\left(1+\mathcal{O}(\xi^{-1})\right)$ and $\xi^{-\frac{\Delta+\Delta_2-\Delta_1}{2}-i}\left(1+\mathcal{O}(\xi^{-1})\right)$ contribute to $\xi^{-\Delta_2}(1+\mathcal{O}(\xi^{-1}))$. On the other hand,  in the boundary channel decomposition $\hat{A}^{B,(1)}_0$ appears in  $\mathcal{W}^{bulk}$ as the coefficient of $\xi^{-\Delta_1}$. Therefore $\hat{A}^{B,(1)}_0$ can be obtained from (\ref{Wbulkcon}) and (\ref{Wbulkcontactaroundinfty}) by resumming all the $\xi^{-\Delta_1}$ coefficients. The result is
\begin{equation}\label{AhatBseed}\footnotesize
\begin{split}
\hat{A}^{B,(1)}_0={}&\sum_{i=0}^\infty P_i\,\frac{\pi ^{d/2} \Gamma (\Delta_2-\Delta_1) \Gamma \left(\frac{-d+2 i+\Delta_1+\Delta_2+1}{2}\right)}{\Gamma \left(\frac{-\Delta_1+\Delta_2+1}{2}\right) \Gamma (i+\Delta_2)}+\sum_{i=0}^\infty Q_i\,\frac{\pi ^{d/2} \Gamma (\Delta_2-\Delta_1) \Gamma \left(\frac{-d+2 i+\Delta +1}{2}\right)}{\Gamma \left(\frac{-\Delta_1+\Delta_2+1}{2}\right) \Gamma \left(\frac{\Delta -\Delta_1+\Delta_2+2i}{2}\right)}\\
={}&\frac{\pi ^{\frac{d}{2}} \Gamma (\Delta_2-\Delta_1) \Gamma \left(\frac{1}{2} (-d+\Delta_1+\Delta_2+1)\right)}{\Gamma (\Delta_2) (\Delta -\Delta_1-\Delta_2) \Gamma \left(\frac{1}{2} (-\Delta_1+\Delta_2+1)\right) (-d+\Delta +\Delta_1+\Delta_2)}\\
{}&\times {}_3F_2\left(1,\Delta_1,-\frac{d}{2}+\frac{\Delta_1}{2}+\frac{\Delta_2}{2}+\frac{1}{2};-\frac{\Delta }{2}+\frac{\Delta_1}{2}+\frac{\Delta_2}{2}+1,-\frac{d}{2}+\frac{\Delta }{2}+\frac{\Delta_1}{2}+\frac{\Delta_2}{2}+1;1\right)\\
{}&-\frac{\pi ^{\frac{d-3}{2}}  2^{\Delta_2-\Delta_1-3}\sin (\pi  \Delta_1)  \sin (\pi  \Delta_2) \sin \left(\frac{\pi  (d-2 \Delta )}{2}\right)\csc \left(\frac{\pi  (\Delta +\Delta_1-\Delta_2)}{2} \right)\csc \left(\frac{\pi  (-d+\Delta +\Delta_1+\Delta_2+2)}{2} \right)}{\Gamma \left(\frac{-\Delta -\Delta_1+\Delta_2+2}{2} \right) \Gamma \left(\frac{d-\Delta -\Delta_1-\Delta_2+2}{2}\right)}\\
{}&\times {}_2F_1\left(\frac{-d+\Delta +1}{2},\frac{\Delta +\Delta_1-\Delta_2}{2};-\frac{d}{2}+\Delta +1;1\right)   \Gamma \left(\frac{d-2 \Delta}{2}\right) \Gamma \left(\frac{-d+\Delta +1}{2}\right) \\
{}&\times \Gamma \left(\frac{\Delta_2-\Delta_1}{2}\right)  \Gamma \left(\frac{-\Delta +\Delta_1+\Delta_2}{2}\right)\Gamma (1-\Delta_1)\Gamma (1-\Delta_2)\;. 
\end{split}
\end{equation}
The seed coefficient $\hat{A}^{B,(2)}_0$ can be obtained from the above expression by exchanging $\Delta_1$ and $\Delta_2$. Moreover, as a consistency check, we can also reproduce other leading OPE coefficients in the bulk channel from  (\ref{Wbulkcon}). By expanding the contact diagrams in the bulk channel, it is clear that the leading bulk channel double-trace OPE coefficient $A^B_0$ is given by
\begin{equation}
A^B_0=P_0\,\mathcal{W}^{contact}_{\Delta_1,\Delta_2}(0)\;,
\end{equation}
while the single-trace OPE coefficient is given by 
\begin{equation}
A^B=Q_0\, \mathcal{W}^{contact}_{\frac{\Delta+\Delta_1-\Delta_2}{2},\frac{\Delta-\Delta_1+\Delta_2}{2}}(0)\;.
\end{equation}

We now consider the bulk mirror exchange Witten diagram (\ref{Wmirrorbulkcon}). In the bulk channel, the bulk mirror exchange diagram contains only double-trace blocks. In the bulk channel limit $\xi\to0$
\begin{equation}
\bar{\mathcal{W}}^{bulk}(\xi)=\bar{A}^B_0+\mathcal{O}(\xi)\;.
\end{equation}
Therefore, we can obtain the seed coefficient $\bar{A}^B_0$ from (\ref{Wmirrorbulkcon}) by simply setting $\xi=0$
\begin{equation}\label{AbarBseed}\footnotesize
\begin{split}
{}&\bar{A}^B_0=\sum_{i=0}^\infty P_i\,  \mathcal{W}^{contact}_{\Delta_1+i,\Delta_2+i}(0)+\sum_{i=0}^\infty Q_i\,  \mathcal{W}^{contact}_{\frac{\Delta+\Delta_1-\Delta_2}{2}+i,\frac{\Delta-\Delta_1+\Delta_2}{2}+i}(0)\\
{}&=\sum_{i=0}^\infty P_i\,\frac{\pi ^{d/2} \Gamma \left(\frac{1}{2} (-d+2 i+\Delta_1+\Delta_2+1)\right)}{\Gamma \left(\frac{1}{2} (2 i+\Delta_1+\Delta_2+1)\right)} +\sum_{i=0}^\infty Q_i\,\frac{\pi ^{d/2} \Gamma \left(\frac{1}{2} (-d+2 i+\Delta +1)\right)}{\Gamma \left(\frac{1}{2} (2 i+\Delta +1)\right)}\\
{}&=\frac{1}{4} \pi ^{d/2} \Gamma \left(\frac{1}{2} (-\Delta +\Delta_1+\Delta_2)\right)\times \Bigg[-\Gamma \left(\frac{1}{2} (-d+\Delta_1+\Delta_2+1)\right) \Gamma \left(\frac{1}{2} (-d+\Delta +\Delta_1+\Delta_2)\right)\\
{}&\times {} _4\tilde{F}_3\left(1,\Delta_1,\frac{-d+\Delta_1+\Delta_2+1}{2},\Delta_2;\frac{\Delta_1+\Delta_2+1}{2},\frac{-\Delta +\Delta_1+\Delta_2+2}{2},\frac{-d+\Delta +\Delta_1+\Delta_2+2}{2};1\right)\\
{}&-\frac{\pi   \Gamma (1-\Delta_1) \Gamma (1-\Delta_2) \Gamma \left(\frac{-d+\Delta +1}{2}\right) \csc \left(\frac{ \pi  (\Delta +\Delta_1-\Delta_2)}{2}\right) \csc \left(\frac{ \pi  (\Delta -\Delta_1+\Delta_2)}{2}\right)\sec \left(\frac{ \pi  (-d+\Delta +\Delta_1+\Delta_2+1)}{2}\right) }{\Gamma \left(\frac{-\Delta +\Delta_1-\Delta_2+2}{2}\right) \Gamma \left(\frac{-\Delta -\Delta_1+\Delta_2+2}{2}\right) \Gamma \left(\frac{d-\Delta -\Delta_1-\Delta_2+2}{2}\right)}\\
{}&\times \sin (\pi  \Delta_1)\sin (\pi  \Delta_2) {}_3\tilde{F}_2\left(\frac{-d+\Delta +1}{2},\frac{\Delta +\Delta_1-\Delta_2}{2} ,\frac{\Delta -\Delta_1+\Delta_2}{2};\frac{\Delta +1}{2},-\frac{d}{2}+\Delta +1;1\right)\Bigg]\;
\end{split}
\end{equation}
where ${}_3\tilde{F}_2$ and ${}_4\tilde{F}_3$ are the regularized hypergeometric functions. Let us also look at the decomposition of this diagram into the boundary channel. From (\ref{Wbulkcon}) and (\ref{Wmirrorbulkcon}), it is clear that to the leading order $\bar{\mathcal{W}}^{bulk}$ has the same large $\xi$ expansion as $\mathcal{W}^{bulk}$. Therefore the seed OPE coefficients of the two diagrams are the same
\begin{equation}
\bar{A}^{B,(i)}_0=A^{B,(i)}_0\;.
\end{equation}

Finally we consider the boundary exchange Witten diagram (\ref{Wbdrcon}). When decomposed in the bulk channel, the boundary exchange Witten diagram consists only double-trace operators. By taking $\xi\to 0$, we can isolate the seed OPE coefficient $A^b_0$ 
 \begin{equation}
 \mathcal{W}^{boundary}(\xi)=A^b_0+\mathcal{O}(\xi)\;.
 \end{equation}
Setting $\xi=0$ in (\ref{Wbdrcon}), we get
\begin{equation}\label{Abseed}\footnotesize
\begin{split}
A^b_0={}&\sum_{i=0}^\infty 2^{-2i} R_i\, \frac{\pi ^{d/2} \Gamma \left(\frac{1}{2} (-d+2 i+\Delta_1+\Delta_2+1)\right)}{\Gamma \left(\frac{1}{2} (2 i+\Delta_1+\Delta_2+1)\right)}+\sum_{i=0}^\infty 2^{-{\widehat{\Delta}}+\Delta_1-2i}S_i\,\frac{\pi ^{d/2} \Gamma \left(\frac{1}{2} (-d+2 i+{\widehat{\Delta}} +\Delta_2+1)\right)}{\Gamma \left(\frac{1}{2} (2 i+{\widehat{\Delta}} +\Delta_2+1)\right)}\\
={}&\frac{\pi ^{d/2}}{4}\times \Bigg[\frac{4 \Gamma \left(\frac{-d+\Delta_1+\Delta_2+1}{2} \right) \, }{({\widehat{\Delta}} -\Delta_1) (-d+{\widehat{\Delta}} +\Delta_1+1) \Gamma \left(\frac{1}{2} (\Delta_1+\Delta_2+1)\right)}\\
{}&\times {}_4F_3\left(1,\frac{\Delta_1}{2}+\frac{1}{2},\frac{\Delta_1}{2},-\frac{d}{2}+\frac{\Delta_1}{2}+\frac{\Delta_2}{2}+\frac{1}{2};-\frac{{\widehat{\Delta}} }{2}+\frac{\Delta_1}{2}+1,-\frac{d}{2}+\frac{{\widehat{\Delta}} }{2}+\frac{\Delta_1}{2}+\frac{3}{2},\frac{\Delta_1}{2}+\frac{\Delta_2}{2}+\frac{1}{2};1\right)\\
{}&+\frac{ 2^{\Delta_1-{\widehat{\Delta}} } \csc (\pi  {\widehat{\Delta}} ) \sin (\pi  \Delta_1) \Gamma (1-\Delta_1) \cos \left(\frac{\pi  (d-2 {\widehat{\Delta}} )}{2} \right) \Gamma \left(\frac{d-2 {\widehat{\Delta}} -1}{2}\right) \Gamma \left(\frac{\Delta_1-{\widehat{\Delta}} }{2}\right) \sec \left(\frac{\pi  (-d+{\widehat{\Delta}} +\Delta_1+2)}{2} \right)}{\Gamma (1-{\widehat{\Delta}} ) \Gamma \left(\frac{{\widehat{\Delta}} +\Delta_2+1}{2}\right) \Gamma \left(\frac{d-{\widehat{\Delta}} -\Delta_1+1}{2}\right)}\\
{}&\times \Gamma \left(\frac{-d+{\widehat{\Delta}} +\Delta_2+1}{2}\right) {}_3F_2\left(\frac{{\widehat{\Delta}} }{2}+\frac{1}{2},\frac{{\widehat{\Delta}} }{2},-\frac{d}{2}+\frac{{\widehat{\Delta}} }{2}+\frac{\Delta_2}{2}+\frac{1}{2};-\frac{d}{2}+{\widehat{\Delta}} +\frac{3}{2},\frac{{\widehat{\Delta}} }{2}+\frac{\Delta_2}{2}+\frac{1}{2};1\right)\Bigg]\;.
\end{split}
\end{equation}
A non-trivial consistency check of this formula is that (\ref{Abseed}) can be shown to be symmetric with respect to exchanging $\Delta_1$, $\Delta_2$ -- a property that is obvious from the definition of the Witten diagram. But this symmetry is totally obscured in the above expression by the asymmetric treatment of the two external legs in the method leading to (\ref{Wbdrcon}). We can also get $\hat{A}_0^{b,(1)}$ and $\hat{A}_0^{b,(2)}$ from (\ref{Wbdrcon}). It is easy to see there is only one term $R_0\, \mathcal{W}^{contact}_{\Delta_1,\Delta_2}(\xi)$ in (\ref{Wbdrcon}) contains $\xi^{-\Delta_1}$ in $1/\xi$ expansion, and we can get $\hat{A}_0^{b,(1)}$ by extracting the $\xi^{-\Delta_1}$ coefficient. On the other hand, every term in (\ref{Wbdrcon}) contributes to $\xi^{-\Delta_2}$. $\hat{A}_0^{b,(2)}$ can be obtained by resumming all the contributions, and reproduces our previous result.


\section{Formulae for Equal Weights}\label{SecEqualweights}
In this appendix, we consider the special case when the external conformal dimensions are degenerate, {\it i.e.}, $\Delta_1=\Delta_2=\Delta_\phi$. The boundary decomposition coefficients contain simple poles when $\Delta_1\to\Delta_2$.\footnote{More generally, this happens when $\Delta_1-\Delta_2\in2\mathbb{Z}$. But we will focus on the equal weight case for simplicity.}. By expanding in $\Delta_1-\Delta_2$, the singularities cancel at the cost of generating conformal blocks with derivatives in the boundary channel decomposition. 

 Let us start with the boundary channel conformal block decomposition (\ref{contactinboundary}) of the contact Witten diagram $\mathcal{W}^{contact}$. The coefficient $\hat{a}^{(1)}_m$ has a simple pole in $\Delta_1-\Delta_2$ because of the Gamma function $\Gamma(\frac{-2m-\Delta_1+\Delta_2}{2})$. The same simple pole appears in $\hat{a}^{(2)}_m$, but with an opposite sign. After expanding in $\Delta_1-\Delta_2$, it is easy to see that in the limit $\Delta_1,\Delta_2\to \Delta_\phi$, (\ref{contactinboundary}) becomes 
\begin{equation}\label{contactingbEW}
\mathcal{W}^{contact}(\xi)=\sum_{n}\hat{a}_{n}g^b_{\Delta_\phi+2n}(\xi)+\sum_{n}\hat{b}_{n}(\partial_{\Delta_\phi}g^b_{\Delta_\phi+2n})(\xi)\;.
\end{equation}
where the coefficients $\hat{a}_{n}$, $\hat{b}_{n}$ were given in (\ref{ahatbhat}).

Similarly, the equal-weight boundary exchange Witten diagram also contains simple poles in its boundary decomposition coefficients. In the boundary channel we find both the boundary channel conformal blocks and their derivatives
\begin{equation}
\mathcal{W}^{boundary}(\xi)=\hat{A}^{b}g^b_{\Delta}(\xi)+ \sum_{n}\hat{A}^{b}_ng^b_{\Delta_\phi+2n}(\xi)+\sum_{n}\hat{B}^{b}_n\partial_{\Delta_\phi}g^b_{\Delta_\phi+2n}(\xi)
\end{equation}
It is straightforward to get this expression from (\ref{boundaryinboundary}) by taking the equal weight limit, and we find that the coefficients $\hat{A}^{b}$, $\hat{A}^{b}_n$ and $\hat{B}^{b}_n$  are given by
\begin{equation}\small
\begin{split}
\hat{A}^{b}={}&\frac{\pi ^{\frac{d-1}{2}} 2^{-2 {\hat{\Delta}} +2 \Delta_\phi -3} \Gamma ({\hat{\Delta}} ) \Gamma \left(\frac{\Delta_\phi -{\hat{\Delta}} }{2}\right)^2 \Gamma \left(\frac{1}{2} (-d+{\hat{\Delta}} +\Delta_\phi +1)\right)^2}{\Gamma (\Delta_\phi )^2 \Gamma \left(-\frac{d}{2}+{\hat{\Delta}} +\frac{3}{2}\right)}\;,\\
\hat{A}^{b}_n={}&-\frac{\pi ^{\frac{d-1}{2}} 16^{-n} \Gamma (2 n+\Delta_\phi ) \Gamma \left(-\frac{d}{2}+n+\Delta_\phi +\frac{1}{2}\right)^2 }{(n!)^2 \Gamma (\Delta_\phi )^2 (-{\hat{\Delta}} +\Delta_\phi +2 n)^2 (-d+{\hat{\Delta}} +\Delta_\phi +2 n+1)^2 \Gamma \left(-\frac{d}{2}+2 n+\Delta_\phi +\frac{1}{2}\right)}\\
{}&\times \bigg[(-d+2 \Delta_\phi +4 n+1)+(-{\hat{\Delta}} +\Delta_\phi +2 n) (-d+{\hat{\Delta}} +\Delta_\phi +2 n+1) \bigg(-H_{-\frac{d}{2}+n+\Delta_\phi -\frac{1}{2}}\\
{}&\quad\quad\quad\quad\quad\quad\quad\quad\quad\quad+\psi\left[-\frac{d}{2}+2 n+\Delta_\phi +\frac{1}{2}\right]+H_n-\psi[2 n+\Delta_\phi ]+\log (4)\bigg)\bigg]\;,\\
\hat{B}^{b}_n={}&\frac{\pi ^{\frac{d-1}{2}} 16^{-n} \Gamma (2 n+\Delta_\phi ) \Gamma \left(-\frac{d}{2}+n+\Delta_\phi +\frac{1}{2}\right)^2}{(n!)^2 \Gamma (\Delta_\phi )^2 (-{\hat{\Delta}} +\Delta_\phi +2 n) (-d+{\hat{\Delta}} +\Delta_\phi +2 n+1) \Gamma \left(-\frac{d}{2}+2 n+\Delta_\phi +\frac{1}{2}\right)}\;.
\end{split}
\end{equation}
Here $\psi[z]=\Gamma'[z]/\Gamma[z]$ is the digamma function. 

Finally, let us consider the bulk exchange diagram $\mathcal{W}^{bulk}$ and the mirror bulk exchange diagram $\bar{\mathcal{W}}^{bulk}$. In working out the boundary channel decomposition for  generic external dimensions, we used a recursive method. To obtain the decomposition coefficients in the equal weight case
\begin{equation}
\mathcal{W}^{bulk}(\xi)=\sum_{n}\hat{A}^{B}_ng^b_{\Delta_1+n}(\xi)+\sum_{n}\hat{B}^{B}_n\partial_{\Delta_\phi}g^b_{\Delta_\phi+n}(\xi)\;.
\end{equation}
\begin{equation}
\bar{\mathcal{W}}^{bulk}(\xi)=\sum_n\hat{\bar{A}}^{B}_n g^b_{\Delta_\phi+n}(\xi)+\sum_n\hat{\bar{B}}^{B}_n \partial_{\Delta_\phi}g^b_{\Delta_\phi+n}(\xi)\;,
\end{equation}
we could also take the limit of (\ref{Wbulkintoboundary}) and (\ref{Wbulkmirrorintoboundary}) after we have recursively computed the unequal weight coefficients. In practice, however, it is more efficient to first find the recursion relations for the equal weight coefficients. This has the computational advantage that we only need to take the limit for the seed coefficients. Let us now derive the recursion relations, focusing only on $\mathcal{W}^{bulk}$. The result for $\bar{\mathcal{W}}^{bulk}$ is not needed since we know from Section \ref{Secbkmirrorinbdr} that the effect of adding $\bar{\mathcal{W}}^{bulk}$ to $\mathcal{W}^{bulk}$ is just to project out all the odd $n$ coefficients. We need the action $\mathbf{EOM}_B$ on $\partial_{\Delta_\phi}g^b_{\Delta_\phi+n}(\xi)$, and it can be obtained by computing the commutator of $\mathbf{EOM}_B$ with $\partial_{\Delta_\phi}$. The result is the following
\begin{equation}\label{EOMBEW}
\begin{split}
\mathbf{EOM}_B(\partial_{\Delta_\phi}g^b_{\Delta_\phi+n})(\xi)={}& \hat{{\alpha}}_n (\partial_{\Delta_\phi}g^b_{\Delta_\phi+n-1})(\xi)+\hat{{\beta}}_n (\partial_{\Delta_\phi}g^b_{\Delta_\phi+n})(\xi)+\hat{{\gamma}}_n(\partial_{\Delta_\phi}g^b_{\Delta_\phi+n+1})(\xi)\\
{}&+\hat{{\Theta}}_n g^b_{\Delta_\phi+n-1}(\xi)+\hat{{\Psi}}_n g^b_{\Delta_\phi+n}(\xi)+\hat{{\Omega}}_n g^b_{\Delta_\phi+n+1}(\xi)
\end{split}
\end{equation}
where $\hat{\alpha}_n$, $\hat{\beta}_n$, $\hat{\gamma}_n$ are $\hat{\alpha}_n^{(1)}$, $\hat{\beta}_n^{(1)}$, $\hat{\gamma}_n^{(1)}$ evaluated at $\Delta_1=\Delta_2=\Delta_\phi$ and 
\begin{equation}
\begin{split}
\hat{{\Theta}}_n={}&-8 n\;,\\
\hat{{\Psi}}_n={}&2 (-d+2 \Delta_\phi +2 n+1)\;,\\
\hat{{\Omega}}_n={}&\frac{2 (\Delta_\phi +n) (d-\Delta_\phi -n-2) (d-2 \Delta_\phi -n-1)}{(d-2 \Delta_\phi -2 n-3) (d-2 \Delta_\phi -2 n-1)}+d-2 \Delta_\phi -n-1\\
{}&+\frac{(d-3) (d-1) (n+2)^2}{4 (-d+2 \Delta_\phi +2 n+3)^2}-\frac{(d-3) (d-1) n^2}{4 (-d+2 \Delta_\phi +2 n+1)^2}\;.
\end{split}
\end{equation}
Using (\ref{EOMBEW}) and (\ref{contactingbEW}), we arrive at the following recursion equations for the decomposition coefficients
\begin{equation}
\hat{\gamma}_{n-1}\hat{B}^B_{n-1}+\hat{\beta}_{n}\hat{B}^B_{n}+\hat{\alpha}_{n+1}\hat{B}^B_{n+1}=\begin{cases}
\hat{b}_{\frac{n}{2}}\;,\quad n\text{ even}\;,\\
0\;,\quad\quad n\text{ odd}\;,
\end{cases}
\end{equation}
\begin{equation}
\hat{\gamma}_{n-1}\hat{A}^B_{n-1}+\hat{\beta}_{n}\hat{A}^B_{n}+\hat{\alpha}_{n+1}\hat{A}^B_{n+1}+\hat{\Omega}_{n-1}\hat{B}^B_{n-1}+\hat{\Psi}_{n}\hat{B}^B_{n}+\hat{\Theta}_{n+1}\hat{B}^B_{n+1}=\begin{cases}
\hat{a}_{\frac{n}{2}}\;,\quad n\text{ even}\;,\\
0\;,\quad\quad n\text{ odd}\;.
\end{cases}
\end{equation}
We should also impose the initial condition $\hat{A}^B_{-1}=\hat{B}^B_{-1}=0$. The seed coefficients $\hat{A}^B_{0}$, $\hat{B}^B_{0}$ can be obtained from taking the equal weight limit of 
\begin{equation}
\hat{A}^{B,(1)}_{0}g^b_{\Delta_1}(\xi)+\hat{A}^{B,(2)}_{0}g^b_{\Delta_2}(\xi)\;.
\end{equation}

\bibliography{bcftpolyakov} 
\bibliographystyle{utphys}

\end{document}